\documentclass{JFM_preprint}
\usepackage{graphicx}
\usepackage{amsmath}
\usepackage{hyperref}
\usepackage{subcaption}
\usepackage{orcidlink}
\usepackage[percent]{overpic}

\newcommand\Str{\mbox{\textit{St}}}  

\lefttitle{R. Klopsch, L. M. Fuchs, G. Rigas, K. Oberleithner and J. G. R. von Saldern}
\righttitle{Attached and separated turbulent flows over a Gaussian bump}

\title{Spectral analysis of attached and separated turbulent flows over a Gaussian-shaped bump}

\author{Roman Klopsch\aff{1,2}\orcidlink{0000-0003-0414-2471}, Lukas M. Fuchs\aff{2}\orcidlink{0009-0000-5068-7574}, Georgios Rigas\aff{1}\orcidlink{0000-0001-6692-6437}, Kilian Oberleithner\aff{2}\orcidlink{0000-0003-0964-872X} \and Jakob G. R. von Saldern\aff{2}\orcidlink{0000-0001-5003-8195}}

\affiliation{\aff{1}Department of Aeronautics, Imperial College London, London SW7 2AZ, United Kingdom
\aff{2}Laboratory for Flow Instabilities and Dynamics, Technische Universität Berlin, Berlin 10623, Germany}

\corresau{Roman Klopsch, \email{roman.klopsch24@imperial.ac.uk}}

\begin{document}
\maketitle

\begin{abstract}
We investigate the broadband turbulent dynamics of attached and separated flows over a Gaussian bump, focusing on the origin of low-frequency coherent structures. The analysis combines time-resolved experimental measurements with physics-based linear models, using mean fields previously assimilated from the same dataset as base flows. Spectral proper orthogonal decomposition reveals coherent dynamics in low- and medium-frequency regimes for both flows, with the low-frequency dynamics being substantially stronger in the separated case. In the separated flow, these dynamics are linked to a three-dimensional zero-frequency modal instability that generates large-scale streamwise-elongated structures downstream of the bump. A standing-wave model based on resolvent modes, incorporating finite-span effects, reproduces the experimentally observed spanwise structure of the dynamics and highlights the limitations of simulations with small spanwise extent and periodic boundary conditions. In the attached flow, similar low-frequency structures are identified. These are weaker, do not form a prominent standing-wave pattern, and cannot be definitively classified as either modal or non-modal. The three-dimensional zero-frequency instability and finite-span standing-wave dynamics are identified as the main drivers of low-frequency coherent structures in the separated flow. They offer an explanation for persistent discrepancies between simulations and experiments on the Gaussian bump, and provide guidance on spanwise domain size and boundary conditions for future simulations.

\end{abstract}


\section{Introduction}

Turbulent flows are fundamental to a wide range of industrial applications, including jets, combustion systems, and aerodynamic surfaces like aircraft wings and control surfaces, wind turbine rotor blades, as well as turbomachinery components like compressor and turbine blades. In wall-bounded flows, such as those over airfoils, flow separation is a critical phenomenon: When a boundary layer encounters a sufficiently strong adverse pressure gradient (APG) or a geometric discontinuity, it can detach from the surface and form a free shear layer. Under certain conditions, the flow may subsequently reattach downstream, forming a separation bubble. The onset of flow separation has significant implications for the aerodynamic performance of the flow, by increasing drag, decreasing lift, and introducing unsteady dynamics caused by instabilities of the free shear layer or the separation bubble \citep{simpson_Turbulent_1989}.

\subsection{Turbulent separation bubble dynamics}
Turbulent separation bubbles (TSBs) exhibit strong unsteadiness across a broad frequency range. This leads to fluctuating structural and thermal loads, as well as noise, in many engineering applications.
TSBs have been the subject of extensive study over the past five decades, across a wide range of geometries and flow configurations. They are typically classified as either geometry-induced or APG-induced. Geometry-induced TSBs arise at features such as backward-facing steps \citep{eaton_Low_1982} and rectangular leading edges \citep{cherry_Unsteady_1984, kiya_Structure_1983}. APG-induced TSBs occur, for instance, on airfoils \citep{wang_Unsteady_2022, sarras_Linear_2024}, backward-facing ramps \citep{kaltenbach_Study_1999, weiss_Spectral_2022}, or flat plates \citep{patrick_Flowfield_1987, na_Direct_1998, wu_Spatiotemporal_2020, cura_Lowfrequency_2024, abe_Reynoldsnumber_2017, mohammed-taifour_Unsteadiness_2016}. A related phenomenon is the appearance of stall cells, which are three-dimensional recirculation regions that occur near the trailing edge of two-dimensional airfoils \citep{winkelman_Flowfield_1980, sarras_Linear_2024}.
TSBs also play a key role in high-speed flows, particularly in shock–boundary layer interactions (SBLI), where shock waves induce separation and reattachment \citep{delery_Shock_1985, dussauge_Unsteadiness_2006, poggie_Spectral_2015, hao_Lowfrequency_2023}.

Over the years, different phenomena in TSBs have been associated with distinct frequency bands, which are typically referred to as \textit{shedding}, \textit{flapping} or \textit{breathing}. A dominant frequency was first identified by \citet{mabey_Analysis_1972}, who proposed the Strouhal scaling $\Str_\mathrm{sep} = f L_\mathrm{sep} / U_{\infty}$, where $L_\mathrm{sep}$ is the separation length and $U_\infty$ is the free-stream velocity. Shedding occurs at $\Str_\mathrm{sep} = 0.35-0.8$ and is linked to vortex roll-up in the shear layer \citep{eaton_Low_1982, kiya_Structure_1983, cherry_Unsteady_1984, weiss_Unsteady_2015}, often attributed to Kelvin–Helmholtz instability \citep{tenaud_Wall_2016}. 
Low-frequency dynamics ($\Str_\mathrm{sep} < 0.02$) with substantial energy content were already reported by \citet{eaton_Low_1982}, and low-frequency trailing-edge oscillations were first observed by \citet{zaman_Natural_1989}. These low-frequency motions are now commonly described as either flapping or breathing. Flapping refers to shear-layer oscillations at $\Str_\mathrm{sep} \approx 0.08-0.18$, typically observed in geometry-induced TSBs \citep{fang_Lowfrequency_2024, pearson_Turbulent_2013, largeau_Wall_2006}, whereas breathing occurs at $\Str_\mathrm{sep} \approx 0.01$ or below, and is often interpreted as a global expansion and contraction of the separation bubble \citep{weiss_Unsteady_2015, mohammed-taifour_Unsteadiness_2016, borgmann_ThreeDimensional_2024}.

The breathing motion has drawn particular attention due to its large-scale, low-frequency nature, which poses significant measurement and modelling challenges. Capturing these dynamics requires long observation times and sufficient spatial coverage. The origin of these dynamics is the subject of ongoing research.
Three key insights into the breathing mechanism have recently emerged: First, resolvent analyses show maximal amplification at finite spanwise wavenumbers, indicating a strong spanwise dependence of the breathing mode \citep{cura_Lowfrequency_2024, sarras_Linear_2024, fuchs_StandingWave_2025}. Second, global linear stability analysis reveals a stationary eigenmode with distinctly elevated growth rate at these spanwise wavenumbers \citep{cura_Lowfrequency_2024, sarras_Linear_2024, fuchs_StandingWave_2025} that is expected to be the origin of the low-frequency dynamics. Third, the eigenmode has been linked to a centrifugal instability \citep{barkley_Threedimensional_2002, rodriguez_Two_2013, savarino_Optimal_2025}.

\begin{figure}
    \centering
    \begin{subfigure}{0.5\textwidth}
        \includegraphics[width=\linewidth]{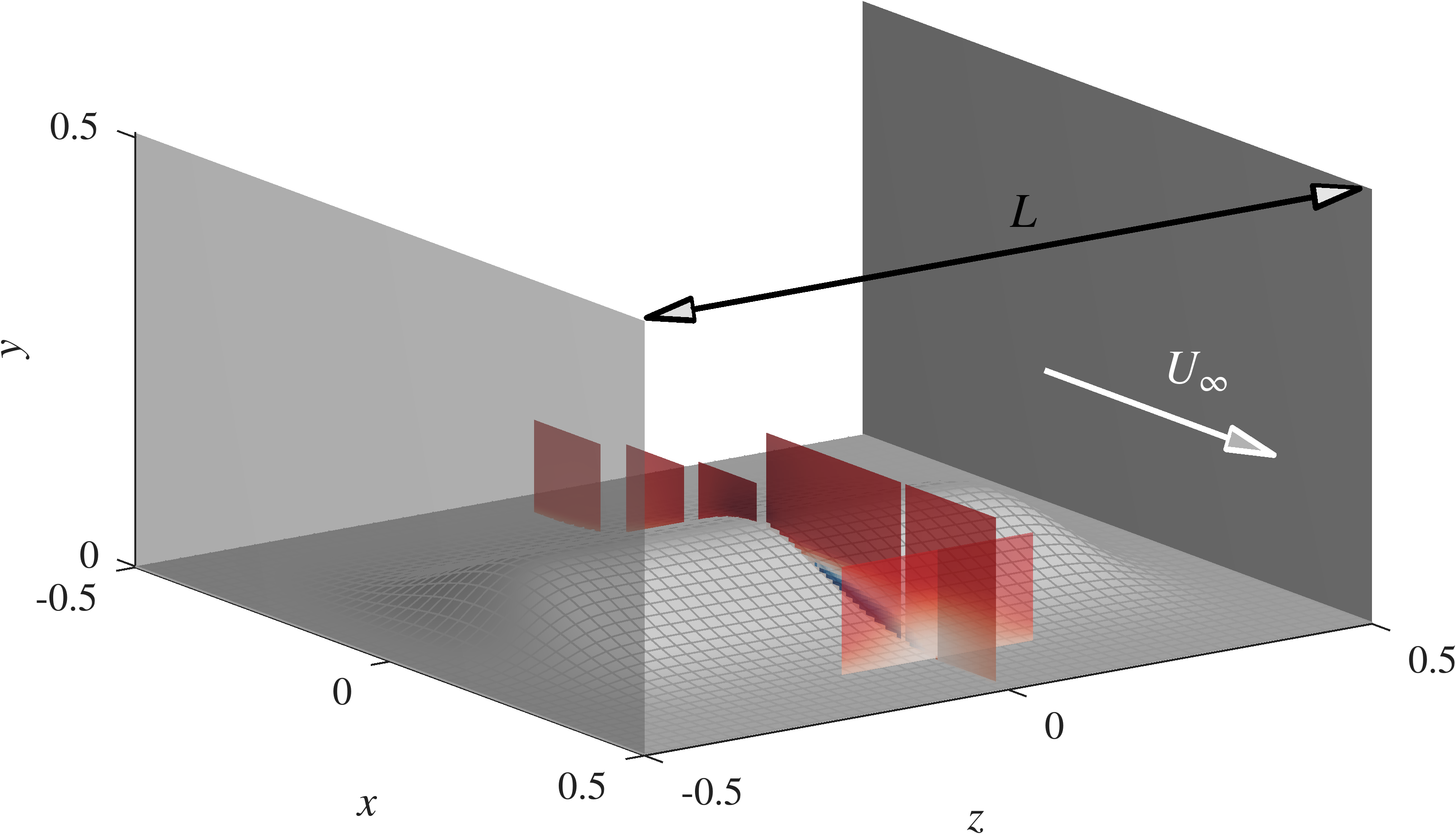}
        \vspace{0.25cm}
        \caption{}
        \label{fig:bump_3d}
    \end{subfigure}\begin{subfigure}{0.5\textwidth}
    \includegraphics[width=\linewidth]{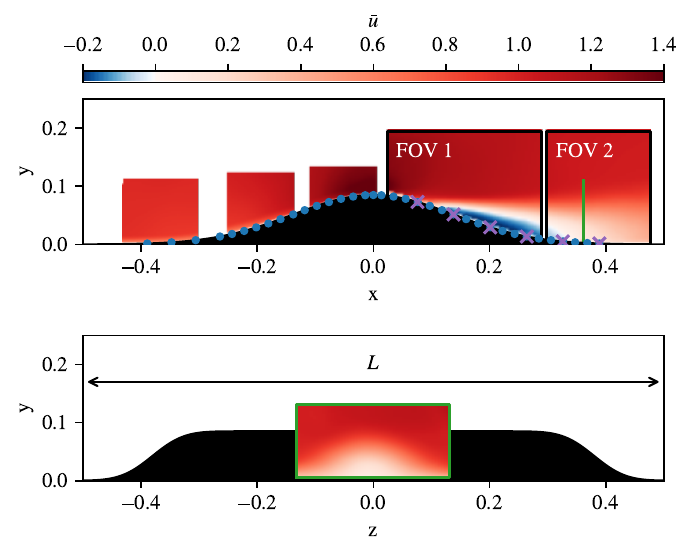}
    \caption{}
    \label{fig:bump_data_pos}
    \end{subfigure}

    \begin{subfigure}{0.5\textwidth}
        \includegraphics[width=0.9\linewidth]{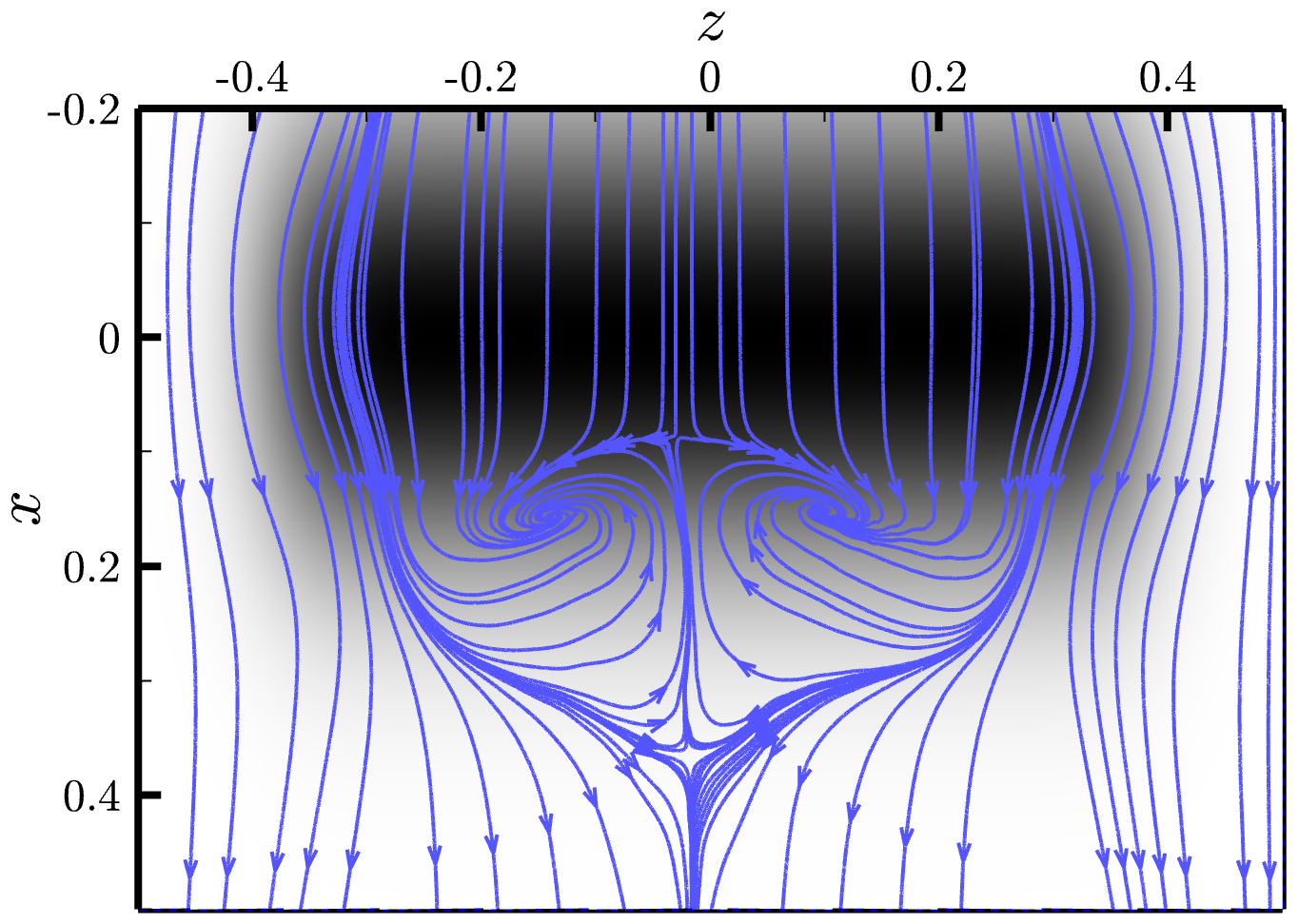}
        \caption{}
        \label{fig:bump_surf_stream}
    \end{subfigure}\begin{subfigure}{0.5\textwidth}
        \centering
        \begin{overpic}[width=0.9\textwidth]{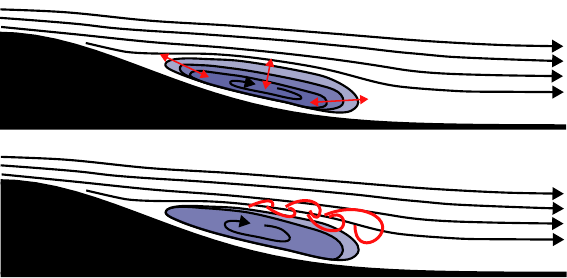}
            \put(70,45){\small breathing}
            \put(70,20){\small shedding}
        \end{overpic}
        \vspace{0.5cm}
        \caption{}
        \label{fig:bump_dynamics_sketch}
    \end{subfigure}
    \caption{Overview of the Gaussian bump test case. (a) Three-dimensional schematic of the bump mounted in the wind tunnel test section. Selected PIV interrogation windows are overlaid and coloured by the streamwise mean velocity component, $\bar{u}$. The wind tunnel side walls are located at $z=-0.5$ and $z=0.5$. (b) Geometry of the bump and measurement locations shown in the streamwise (top) and spanwise (bottom) principal views. Mean pressure measurement positions are indicated by blue dots and instantaneous pressure measurement positions by purple crosses. PIV measurement regions are coloured by $\bar{u}$ using the same colour scale as in (a). Instantaneous velocity data are available within the spanwise SPIV window (green border) and the streamwise PIV windows labelled ``FOV 1'' and ``FOV 2'' (black borders). (c) Wall skin friction streamlines obtained from a wall-modelled large-eddy simulation (LES) of the flow, adapted from \cite{iyer_Assessing_2023, iyer_Wallmodeled_2023} with permission. (d) Qualitative illustration of breathing and shedding dynamics. All data are shown for the separated flow at $\Rey = 2\times 10^6$.}
    \label{fig:bump}
\end{figure}

\subsection{The Gaussian bump benchmark experiment}
APG-induced flow separation occurring on smooth surfaces is sometimes termed smooth body separation (SBS). This process is particularly difficult to predict and model, often times requiring extensive numerical and experimental research. As part of the recent research efforts to improve the capability of computational fluid dynamics (CFD) to accurately capture SBS in turbulent flows, a new benchmark case, termed the \textit{Boeing Gaussian Bump} was introduced \citep{williams_Experimental_2020, gray_New_2021}.
Initial experiments using the geometry have been performed at the University of Washington by \citet{sarwas_Experimental_2019} and \citet{williams_Experimental_2020}, followed by an experimental campaign at the University of Notre Dame by \citet{gray_Experimental_2023a} and \citet{gray_New_2021, gray_Experimental_2022, gray_Benchmark_2022, gray_Turbulence_2023, gray_Experimental_2023}. The results of the latter are to a large extent archived on the ``NASA Turbulence Modeling Resource'' website. In this study we work with this dataset and, unless otherwise noted, ``experimental'' refers to the experiments performed at the University of Notre Dame. An overview of the three-dimensional geometry is provided in figure~\ref{fig:bump_3d}. Here, selected PIV interrogation windows are overlaid and coloured by the streamwise mean velocity component $\bar{u}$ from the experiment. All data shown in figure~\ref{fig:bump} are for $\Rey=2\times10^6$, where the flow is fully separated, with a TSB on the downstream face of the bump. Figure~\ref{fig:bump_data_pos} shows the streamwise and spanwise principal views, including the same PIV windows. Wall skin friction streamlines from a wall-modelled LES \citep{iyer_Assessing_2023, iyer_Wallmodeled_2023} are shown in figure~\ref{fig:bump_surf_stream} to provide an overview of the flow's three-dimensional structure, and figure~\ref{fig:bump_dynamics_sketch} shows a schematic of \textit{breathing} and \textit{shedding} dynamics in the flow.

A number of computational studies have been performed to benchmark the performance of various CFD approaches on the test case. \citet{williams_Experimental_2020} investigated the capability of two- and three-dimensional Reynolds-averaged Navier--Stokes (RANS) models to reproduce the results from their experiment. They found that all of their simulations failed to accurately reproduce the surface pressure coefficient in the separated flow region, whereas the pressure outside the separated region was generally well matched. They also found that the RANS models did not reflect the Reynolds number invariance of the surface pressure coefficient for $\Rey\geq 2\times 10^6$ (based on the wind tunnel width and free-stream velocity) that they observed in the experiment. \citet{gray_Experimental_2023} performed RANS and delayed detached eddy simulation (DDES) of the case at $\Rey=4\times 10^6$. They found that the RANS model failed to capture the pressure in the separated region by predicting almost no flow separation. DDES qualitatively captures the flow separation but quantitative deviations from the experiment remain. Their comparison also includes pressure coefficients from the experiment at the University of Washington at $\Rey=3.4\times10^6$ \citep{sarwas_Experimental_2019, williams_Experimental_2020}, which are almost indiscernible from the University of Notre Dame experiment at $\Rey = 4\times 10^6$, albeit the difference in Reynolds number, highlighting the Reynolds number invariance of the separated flow. \citet{zhou_Sensitivity_2024} investigated the sensitivities of wall-modelled large eddy simulation (LES) with respect to the specific modelling choices and mesh parameters. They found that especially the model for the subgrid-scale stresses significantly impacts the solution in the separated flow region, and that there is a significant dependence on the mesh resolution up to the point where the resolution approaches that of wall-resolved LES. Direct numerical simulations (DNS) of the flow have been performed at $\Rey=10^6$ \citep{uzun_Simulation_2021, balin_Direct_2021}, $\Rey=2\times 10^6$ \citep{uzun_HighFidelity_2022}, and $\Rey=4\times10^6$ \citep{uzun_Direct_2025}. For computational reasons, these studies do not take the three-dimensional geometry of the bump into account. Instead, the simulations employ a spanwise periodic domain where the profile of the bump corresponds to that at the spanwise centerline in the experimental configuration. They found evidence for relaminarization of the boundary layer in the acceleration zone at the upstream face of the bump at $\Rey=10^6$. At this Reynolds number, the flow undergoes only very weak separation in the deceleration region downstream of the bump apex. At $\Rey=2\times10^6$ and $4\times 10^6$, the relaminarization is suppressed and the flow undergoes much stronger separation. Comparison of the surface pressure- and skin friction coefficients from the DNS at these Reynolds numbers with experimental values from \citet{williams_Experimental_2020} shows relatively good agreement \citep{uzun_HighFidelity_2022, uzun_Direct_2025}. However, at $\Rey=4\times10^6$, the shear layer is tilted significantly more towards the wall in the DNS compared to the experiment, which the authors attribute to the spanwise periodic configuration of the DNS that neglects three-dimensionality and tunnel end-wall effects \citep{uzun_Direct_2025}. \citet{iyer_Wallmodeled_2023} performed wall-modeled LES of the case in both, a spanwise periodic and a fully three-dimensional configuration involving the wind tunnel side walls. They similarly found the shear layer to be tilted more towards the wall in the spanwise periodic simulation whereas the shear layer in the fully three-dimensional simulation shows good agreement with the experiment. This effect has been linked to the interaction of the shear layer with two counter-rotating vortices, which, at the spanwise centerline, helps to lift the shear layer away from the bump surface \citep{uzun_Direct_2025}. The effect becomes apparent from the surface streamline pattern of their three-dimensional LES, which is reproduced in figure \ref{fig:bump_surf_stream}. These studies highlight the challenges in accurately modelling the separated flow region downstream of the bump with reduced-order models and even with high-fidelity CFD, if the three-dimensional structure of the flow is not accounted for.

\subsection{Motivation and objectives}
In evaluating the performance of CFD computations, most studies of the Gaussian bump flow focus on mean-flow quantities such as time-averaged velocity fields, Reynolds stresses, or wall-pressure distributions, while investigations of dominant coherent structures are lacking. However, some of the challenges in accurately modelling the separated flow over the bump likely arise from the dynamics of the TSB that forms downstream of the bump. These dynamics generate large-scale, three-dimensional coherent structures that are difficult to capture in simulations \citep{mohammed-taifour_Unsteadiness_2016, manohar_Temporal_2023, borgmann_ThreeDimensional_2024, cura_Lowfrequency_2024}. In particular, the observation that unsteady simulations match experimental results significantly better when the full span of the wind tunnel is resolved motivates an investigation into the role of three-dimensional coherent structures in the flow.

We therefore address this gap by providing an extensive spectral characterisation of the coherent flow dynamics. The specific objectives of this study are to (i) identify the role of coherent structures in the broadband turbulent dynamics of the flow, (ii) compare their driving mechanisms between attached and fully separated flow conditions, and (iii) assess the role of the finite span and tunnel side walls on the dominant flow structures. In light of the discrepancies between CFD and experiments for this flow, the analysis is based entirely on experimental data.

\subsection{Structure}
Section~\ref{sec:database} provides a brief introduction to the \textit{Boeing Gaussian Bump} benchmark case and the available experimental database. Section~\ref{sec:methods} outlines the main methodologies used in the study, including spectral proper orthogonal decomposition (SPOD), linear stability analysis (LSA), and resolvent analysis (RA). In Section~\ref{sec:data_driven_results}, coherent structures are identified using SPOD, highlighting low-frequency streaky structures and medium-frequency vortex shedding in both flow configurations. In Section~\ref{sec:model_results}, LSA and RA are employed to model these coherent dynamics. In the separated case, the streaky structures are found to be linked to a three-dimensional zero-frequency global mode, whereas no evidence for a modal origin is found in the attached case. The possible driving mechanisms are then discussed and compared between both cases. Spanwise-standing wave dynamics resulting from the finite span of the wind tunnel are investigated in Section~\ref{sec:standing_waves}, and their implications regarding the domain size and boundary conditions of numerical simulations are discussed. Finally, Section~\ref{sec:conclusions} summarizes the main findings of the study.

\section{Database}
\label{sec:database}

The \textit{Boeing Gaussian Bump} was developed as a new benchmark test case for high Reynolds number flows undergoing SBS \citep{williams_Experimental_2020, gray_New_2021}. In the streamwise direction, the bump follows a Gaussian profile, where favourable and adverse pressure gradients are induced at the up- and downstream faces of the bump, respectively. In the spanwise direction, the bump is tapered according to an error function to minimize side wall interactions. The bump geometry is defined as

\begin{equation}
    y_\Gamma(x, z) = h\frac{1+\mathrm{erf}\left((\frac{1}{2}-2z_0 - |z|)/z_0\right)}{2}\exp\left(-\left(\frac{x}{x_0}\right)^2\right)\,,
\end{equation}
\noindent
where $(x, y, z)$ are the streamwise, vertical, and spanwise coordinates, respectively, and $y_\Gamma(x,\,z)$ is the bump surface. All variables in this paper are expressed in their respective non-dimensional form with the (spanwise) wind tunnel width $L$ and freestream velocity $U_\infty$ serving as integral reference scales. $h=0.085L$ is the bump height, erf is the (Gauss) error function, $z_0=0.06L$, and $x_0=0.195L$. Throughout this paper, unless otherwise noted, the Reynolds number is defined as $\Rey=U_\infty L/\nu$.

\subsection{Experimental database}
The bump geometry and measurement locations are shown in figure~\ref{fig:bump_data_pos}. Mean and instantaneous pressure measurements are available on the bump surface. The positions of the 6 pressure sensors where simultaneous time series were recorded are indicated in figure~\ref{fig:bump_data_pos} by purple crosses. The signals were sampled at 100~kHz for 20 seconds. In the streamwise plane, particle image velocimetry (PIV) measurements provide the streamwise and vertical mean velocity components, with instantaneous velocity data available in the black-bordered regions labeled ``FOV 1'' and ``FOV 2''. These measurements are located along the spanwise centerline at $z=0$. In the spanwise plane, downstream of the bump at $x=0.361$, mean and instantaneous velocity data for all three components are available from stereo PIV (SPIV) measurements. The pressure data and PIV mean fields are included in the aforementioned archive, and the PIV/SPIV snapshot data were additionally made available by Patrick Gray for use in this study. The PIV was recorded at 200~Hz for 3 intervals of 5 seconds each for the streamwise windows and for one continuous interval of 25 seconds for the spanwise SPIV window. It is noted that the PIV/SPIV measurements are not synchronised with the pressure measurements. The reader is referred to the ``NASA Turbulence Modeling Resource'' and the associated publications \citep{gray_Experimental_2023a, gray_New_2021, gray_Experimental_2022, gray_Benchmark_2022, gray_Turbulence_2023, gray_Experimental_2023} for a more detailed description of the experimental setup.

We consider two flow conditions with Reynolds numbers $\Rey=U_\infty L/\nu=10^6$ and $2\times 10^6$ based on the freestream velocity $U_\infty$ and wind tunnel width $L$. In the experimental configuration \citep{gray_Experimental_2023a}, the freestream velocities are $U_\infty=17\,\mathrm{m/s}$ and $34\,\mathrm{m/s}$, respectively, and the wind tunnel width is $L=0.91\,\mathrm{m}$. The Reynolds numbers based on the bump height are $8.5\times10^4$ and $1.7\times10^5$, and the Mach numbers are 0.05 and 0.1. The variation of the mean surface pressure coefficient for both cases is shown in figure \ref{fig:pinn_cp_validation}. The flow at $\Rey=10^6$ displays only intermittent or very weak separation with no reverse mean flow whereas the flow at $\Rey=2\times10^6$ is fully separated. This distinction is based on experimental observations \citep{gray_Experimental_2023} and consistent with results from DNS \citep{uzun_Simulation_2021, uzun_HighFidelity_2022}. In the following, we refer to these flow conditions as the \textit{attached} and \textit{separated} cases, respectively. It is worth noting, however, that these labels are based on the mean state of the flow, and that the instantaneous state at any specific time might be different.

\begin{figure}
    \centering
    \begin{subfigure}{0.5\textwidth}
        \includegraphics[width=\textwidth]{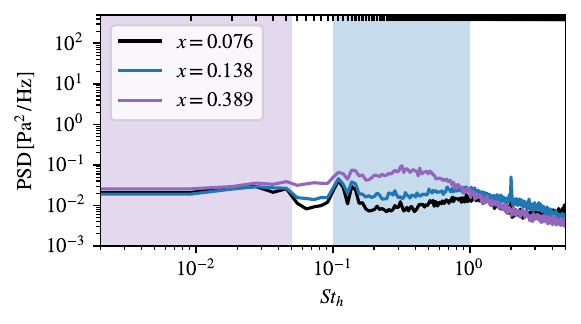}
        \caption{$\Rey=10^6$ (attached)}
    \end{subfigure}\begin{subfigure}{0.5\textwidth}
        \includegraphics[width=\textwidth]{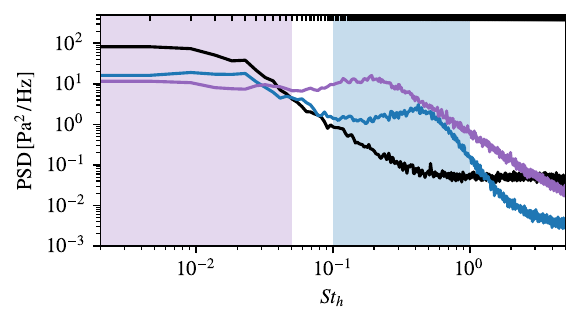}
        \caption{$\Rey=2\times 10^6$ (separated)}
    \end{subfigure}
    \caption{PSD of surface pressure measurements in the downstream region of the bump for the attached (a) and separated (b) cases. Tick marks at the top denote the frequency resolution.}
    \label{fig:surfp_psd}
\end{figure}

The difference in the flow dynamics between the two different cases is evident from figure~\ref{fig:surfp_psd}, which shows power spectral density (PSD) spectra of surface pressure measurements in the downstream region of the bump. The x-axis shows the Strouhal number $\Str_h=fh/U_\infty$, where $h=0.085L$ is the bump height. In the proposed scaling by \citet{mabey_Analysis_1972}, the Strouhal number is based on the separation length. For the fully separated flow over the bump, the separation length was reported as $L_\mathrm{sep}\approx0.3L$ \citep{gray_Experimental_2023a, uzun_HighFidelity_2022}. At $\Rey=10^6$, however, there is no flow separation and the length scale would be undefined. Therefore, a simple geometric length scale is used for both cases throughout this paper. For the separated case, the reader may convert to a separation-length-based scaling using $L_\mathrm{sep}=3.53h$. From figure~\ref{fig:surfp_psd}, two primary regions of elevated PSD are identified: There is a medium-frequency regime at $0.1 \lessapprox \Str_h \lessapprox 1$, indicated by the blue area. This regime is evident at both Reynolds numbers. For the separated case at $\Rey=2\times 10^6$, there is also a prominent low-frequency regime at $\Str_h \lessapprox 0.05$, indicated by the purple area. These spectra thus indicate shedding dynamics at both Reynolds numbers and pronounced low-frequency breathing dynamics in the separated case. In what follows, we identify coherent structures which govern the flow dynamics in these regimes.

\subsection{Data-assimilated mean flows}

To examine the dominant coherent structures within the broadband turbulent flow over the bump, we employ LSA and RA. These approaches require the construction of a linearized operator, which depends on the mean flow state, along with its spatial gradients, defined across the entire domain of interest.
However, the experimentally obtained mean velocity fields measured via PIV are limited to disjoint regions and do not provide full spatial coverage. The discrete nature of these measurements also complicates the accurate computation of gradients. Moreover, an eddy-viscosity field, typically used in linearized mean field methods to represent turbulence effects on coherent structures~\citep{reynolds_Mechanics_1972}, cannot be directly extracted from the sparse experimental data.

\begin{figure}
    \centering
    \begin{subfigure}{0.5\textwidth}
        \includegraphics[width=\linewidth]{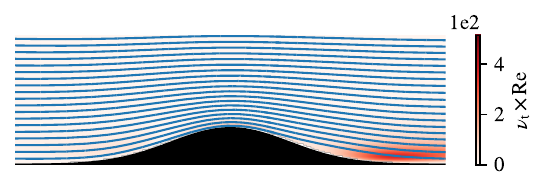}
        \caption{$\Rey=10^6$ (attached)}
    \end{subfigure}\begin{subfigure}{0.5\textwidth}
        \includegraphics[width=\linewidth]{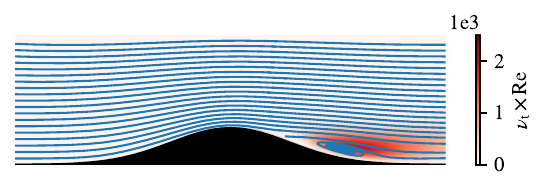}
        \caption{$\Rey=2\times10^6$ (separated)}
    \end{subfigure}
    \caption{Data-assimilated mean flow used in the approximation of the linear operators for the attached (a) and separated cases (b), with the eddy viscosity field shown as background contours and mean streamlines overlaid in blue. Based on data reported in \citep{klopsch_Enabling_2025}.}
    \label{fig:pinn_meanflows}
\end{figure}

\begin{figure}
    \centering
    \begin{subfigure}{0.5\textwidth}
        \includegraphics[width=\linewidth]{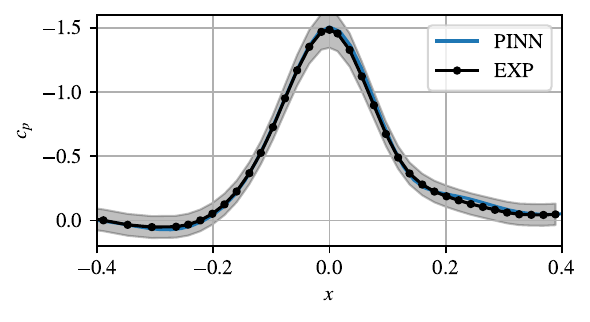}
        \caption{$\Rey=10^6$ (attached)}
    \end{subfigure}\begin{subfigure}{0.5\textwidth}
        \includegraphics[width=\linewidth]{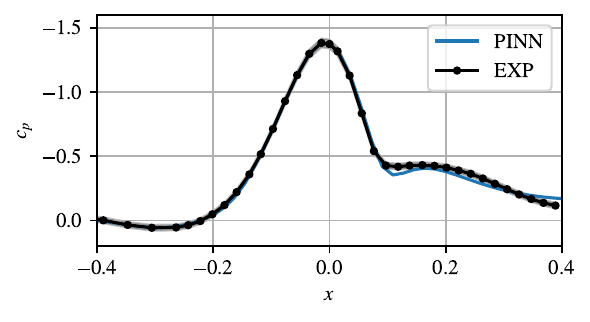}
        \caption{$\Rey=2\times10^6$ (separated)}
    \end{subfigure}
    \caption{Validation of the streamwise variation of mean surface pressure at $z=0$ for the attached (a) and separated cases (b), with the gray-shaded region indicating the 95\% confidence interval for the experimental data. The pressure coefficient is offset such that $c_p=0$ at $x=-0.4$. Based on data reported in \citep{klopsch_Enabling_2025}.}
    \label{fig:pinn_cp_validation}
\end{figure}

To address these issues, the mean flows have been data-assimilated using physics-informed neural networks (PINNs) in our previous study \citep{klopsch_Enabling_2025}. The data assimilation approach combines the measured velocity mean fields from PIV at the spanwise centerline of the bump \citep{gray_Experimental_2023a, gray_New_2021, gray_Experimental_2022, gray_Benchmark_2022, gray_Turbulence_2023, gray_Experimental_2023} with physical constraints in the form of the (two-dimensional) RANS- and continuity equations and a no-slip wall boundary condition. The PINN provides two-dimensional (automatically) differentiable velocity fields at the bump centerline, which cover the relevant region. Additionally, a corresponding mean-field-consistent eddy viscosity and pressure field are inferred in the process. The assimilated velocity- and eddy viscosity fields are shown in figure~\ref{fig:pinn_meanflows}. In order to validate the data assimilation approach against unseen data, the assimilated pressure at the bump surface is compared to experimental measurements in the previous study. The results of this validation step are shown in figure~\ref{fig:pinn_cp_validation}. Here, the offset is adjusted so that $c_p=0$ at $x=-0.4$. This is necessary because the RANS equations used in the data assimilation procedure only include the pressure gradient and provide no information on the absolute value. The gray-shaded region in figure~\ref{fig:pinn_cp_validation} indicates the $95\%$ confidence interval for the experimental data. For details regarding the uncertainty analysis, we refer to the Appendix of \citet{gray_Experimental_2023a}. The veracity of the eddy viscosity field was assessed through a breakdown of the RANS equations, which showed that the turbulent forces are represented with satisfactory accuracy (see figure 8 in \cite{klopsch_Enabling_2025}). Relatively small residuals in the force balance remain, which likely result from three-dimensionality of the mean flow, which is not accounted for in the two-dimensional equations used for the PINN, and from limitations of the Boussinesq hypothesis.
For a detailed description of the data assimilation methodology and validation of the assimilated mean fields, we refer to our previous studies~\citep{klopsch_Enabling_2025,vonsaldern_Mean_2022}.
The data-assimilated mean fields serve as foundation for approximating the linear operators in this study.

\section{Methodology}
\label{sec:methods}
This section outlines the main methods used in this study. We begin with an introduction to the SPOD methodology, which, throughout this study, serves as the primary data-driven approach for identifying coherent structures, followed by an introduction of the linearized Navier-Stokes equations, LSA, and the RA framework. 

\subsection{Spectral proper orthogonal decomposition}
To investigate the dominant dynamics in the broadband turbulent flow, SPOD is applied~\citep{lumley_Stochastik_1970,towne_Spectral_2018}. The method identifies structures of spatial and temporal coherence in a temporally and spatially resolved signal based on estimates of the corresponding cross-spectral density (CSD) matrix. In this work, SPOD is applied to both surface pressure measurements and time-resolved PIV snapshot data.

The SPOD algorithm involves several steps. First, the time series of a signal $\boldsymbol{q}'$ involving $M$ degrees of freedom is divided into $N$ (overlapping) blocks. Each block is then transformed into the frequency domain using a Fourier transform and the resulting coefficients of all blocks are partitioned into a data matrix

\begin{equation}
    \mathsfbi{\hat{Q}}_\omega = \left( \hat{\boldsymbol{q}}^1_\omega \quad \hat{\boldsymbol{q}}^2_\omega \quad \dots \quad \hat{\boldsymbol{q}}^N_\omega \right)
\end{equation}
\noindent
for each frequency $\omega$. Following Welch's method \citep{welch_Use_1967}, the CSD matrix at a given $\omega$ is estimated based on the Fourier modes of all $N$ blocks, $\mathrm{CSD}\approx \frac{\alpha}{N}\mathsfbi{\hat{Q}}_\omega\mathsfbi{\hat{Q}}_\omega^\mathsf{H}$. The superscript $\mathrm{H}$ denotes the complex-conjugate transpose, and the factor $\alpha = 1/\Delta f$ ensures that the eigenvalues represent modal PSD. Here $\Delta f$ is the discrete Fourier transform bin spacing of a windowed segment, $\Delta f = \frac{1}{N_w\,\Delta t}$, with time step $\Delta t$, and window length $N_w$. This choice is consistent with a forward normalization of the Fourier transform. If a taper is applied, $\alpha$ must be adjusted to compensate for the power loss.
An eigenvalue decomposition of the CSD matrix yields the SPOD eigenvalues and corresponding eigenmodes:

\begin{equation}
    \lambda_\omega^k\boldsymbol{\phi}_\omega^k = \frac{\alpha}{N}\mathsfbi{\hat{Q}}_\omega\mathsfbi{\hat{Q}}_\omega^\mathsf{H}\boldsymbol{\phi}_\omega^k
\end{equation}

This process is repeated for each frequency. For processing the PIV snapshot data, where $M > N$, we make use of the ``method of snapshots'', where

\begin{equation}
    \boldsymbol{\phi}^k_\omega = \sqrt{\frac{\alpha}{N \lambda_\omega^k}}\mathsfbi{\hat{Q}}_\omega \boldsymbol{\psi}_\omega^k\quad\text{with}\quad \lambda^k_\omega\boldsymbol{\psi}^k_\omega = \frac{\alpha}{N}\mathsfbi{\hat{Q}}_\omega^\mathsf{H}\mathsfbi{\hat{Q}}_\omega\boldsymbol{\psi}_\omega^k\,,
\end{equation}
which reduces the eigenvalue problem to be of size $N\times N$.
For a comprehensive overview of the method, we refer the reader to the SPOD guide by \citet{schmidt_Guide_2020}.

The method yields $N$ modes for each frequency bin that can be ranked according to their PSD, which is contained in the corresponding eigenvalue. If over certain frequency bands large parts of the overall PSD are associated with a small number of modes, one refers to low-rank dynamics. The leading modes are then considered as the dominant coherent structures of the flow in the respective frequency range.

For processing the PIV snapshot data, we make use of the \textit{PySPOD} package \citep{mengaldo_PySPOD_2021}.

\subsection{Linear stability analysis}
Linear stability analysis is based on a linearized formulation of the Navier--Stokes equations around the temporal mean flow. The governing equations can be derived by substituting the Reynolds decomposition $(\cdot)=\bar{(\cdot)}+(\cdot)'$ into the Navier--Stokes equations and subtracting the mean of the equations

\begin{equation} 
\label{eq:fluc}
    \frac{\partial\boldsymbol{u}'}{\partial t} = -(\boldsymbol{\bar{u}}\boldsymbol{\cdot}\nabla)\boldsymbol{u}' - (\boldsymbol{u'}\boldsymbol{\cdot}\nabla)\boldsymbol{\bar{u}} - \nabla p' + \frac{1}{\Rey} \nabla^2\boldsymbol{u}' - \nabla\boldsymbol{\cdot}\mathsfbi{R}'\,.
\end{equation}

The equations are given in non-dimensional form; $\boldsymbol{u}$ denotes the velocity, $t$ time, $p$ pressure, and $\Rey=U_\infty L/\nu$ the Reynolds number. $\bar{(\cdot)}$ indicates the (temporal) mean and $(\cdot)'$ is the fluctuating component. Incompressibility and constant density are assumed, allowing the non-dimensional density to be omitted in the equation. Due to the turbulent nature of the flow, the linearized equations include an unknown turbulent term $\boldsymbol{R}' = \boldsymbol{u'}\boldsymbol{u'} - \overline{\boldsymbol{u'}\boldsymbol{u'}}$ that is also referred to as the fluctuating Reynolds stress tensor~\citep{reynolds_Mechanics_1972}. This term represents the nonlinear interaction between different scales and is essential for their energy transfer~\citep{kuhn_Influence_2022,saldern_Role_2024}. 

A harmonic ansatz for all fluctuating quantities of the form

\begin{equation}
\label{eq:ansatz}
    \boldsymbol{q}'(x,y,z,t) = \int_\beta\int_\omega \hat{\boldsymbol{q}}(x,y) \exp (\mathrm{i} \beta z - \mathrm{i}\omega t) + \mathrm{c.c.}
\end{equation}

\noindent
is chosen, where $\beta$ denotes the spanwise wave number and $\omega$ the temporal frequency. Through $\beta$, we introduce a harmonic ansatz in the spanwise direction that reduces the three-dimensional problem to a two-dimensional problem for each $\beta$. This treatment of the spanwise coordinate allows to capture some three-dimensional effects while retaining the simplicity of two-dimensional computations. Implicitly, this approach assumes both the geometry and mean flow to be constant along the spanwise coordinate. Obviously, this is not the case in the configuration considered in this study. Nevertheless, it is shown in this paper, through extensive validation of the linear models against experimental data, that this modelling approach is sufficient to capture the key dynamics of the flow. By inserting this ansatz, the equation is transformed into frequency space,

\begin{equation}
\label{eq:hat_1}
   \mathrm{i}\omega\boldsymbol{\hat{u}} = (\boldsymbol{\bar{u}}\boldsymbol{\cdot}\nabla)\boldsymbol{\hat{u}} + (\boldsymbol{\hat{u}}\boldsymbol{\cdot}\nabla)\boldsymbol{\bar{u}} + \nabla\hat{p} - \frac{1}{\Rey} \nabla^2\boldsymbol{\hat{u}} + \nabla \boldsymbol{\cdot} \mathsfbi{\hat{R}},
\end{equation}

\noindent
where the coherent component of the Reynolds stress tensor $\mathsfbi{\hat{R}}$ represents the fluctuating Reynolds stresses in frequency space. In analogy to the Boussinesq model for the RANS equations, an eddy viscosity model is employed to represent the deviatoric component of the coherent Reynolds stress tensor,

\begin{equation}
\label{eq:hat}
   -\mathrm{i}\omega\boldsymbol{\hat{u}} + (\boldsymbol{\bar{u}}\boldsymbol{\cdot}\nabla)\boldsymbol{\hat{u}} + (\boldsymbol{\hat{u}}\boldsymbol{\cdot}\nabla)\boldsymbol{\bar{u}} + \nabla\hat{q} - \nabla\boldsymbol{\cdot} (\frac{1}{\Rey} + \nu_\mathrm{t})[(\nabla + \nabla^\mathsf{T})\boldsymbol{\hat{u}}] = \boldsymbol{\hat{f}}.
\end{equation}

The remaining spherical share is absorbed into the pressure term forming the modified harmonic pressure $\hat{q} = \hat{p} + 1/3 \text{Tr}(\mathsfbi{\hat{R}}) $, where $\text{Tr}$ is the trace operator. For the eddy viscosity, we follow the common approach and use the value consistent with the mean field equations~\citep{rukes_Assessment_2016,tammisola_Coherent_2016,saldern_Role_2024} that is available from the data assimilation. The vector $\boldsymbol{\hat{f}}$, also referred to as the nonlinear forcing vector, can be interpreted as the remaining share of  coherent Reynolds stresses that is not captured by the Boussinesq eddy viscosity model. Equation~\ref{eq:hat} is complemented by the continuity condition for the fluctuating velocity field $\nabla \boldsymbol{\cdot} \boldsymbol{\hat{u}} = 0$. In a compact form, the two governing equations can be written as

\begin{equation}
    \label{eq:linearized_system}
    -\mathrm{i}\omega\boldsymbol{\hat{q}} = \mathcal{L}\boldsymbol{\hat{q}}+\mathcal{B}\boldsymbol{\hat{f}}\,,
\end{equation}
where $\boldsymbol{\hat{q}}$ is the coherent state vector, including the velocity components and pressure, $\mathcal{L}$ is the linearized operator incorporating equation~\ref{eq:hat} and the continuity condition, and $\mathcal{B}$ is a restriction operator, constraining the forcing to the momentum equations. Considering $\boldsymbol{\hat{f}}=\boldsymbol{0}$, the eigenvalues and eigenvectors of this system are the linear stability modes. The real part of the eigenvalue $\Real(\omega_\mathrm{eig})$ represents the frequency and the imaginary part $\Imag(\omega_\mathrm{eig})$ the growth rate of the mode. The spatial shape of the mode is given by the corresponding eigenvector.

\subsection{Resolvent analysis}
To additionally investigate coherent structures emerging from non-modal mechanisms, resolvent analysis is applied~\citep{mckeon_Criticallayer_2010}. RA focuses on the forced system dynamics, equation~\ref{eq:linearized_system} with $\boldsymbol{\hat{f}}\neq\boldsymbol{0}$ that can be rearranged into an input-output transfer function

\begin{equation}
    \label{eq:RA_operator}
    \boldsymbol{\hat{q}} = \underbrace{(-\mathcal{L}-\mathrm{i}\omega\boldsymbol{I})^{-1}}_\mathcal{R}\mathcal{B}\boldsymbol{\hat{f}}\,,
\end{equation}

\noindent
where $\mathcal{R}$ is the Resolvent operator. The operator maps a given forcing to the corresponding response in terms of velocity and pressure fluctuations. Note that a separate resolvent operator is obtained for each $\beta$ and $\omega$. Since the true nonlinear forcings $\boldsymbol{\hat{f}}$ are generally not known, we analyze the operator in terms of its optimal input–output behavior. Specifically, we formulate

\begin{equation}
    \label{eq:RA_opt_problem}
    \sigma^2 = \max_{\boldsymbol{\hat{f}}}\frac{\boldsymbol{\hat{q}}^\mathsf{H}\mathsfbi{W}_r\boldsymbol{\hat{q}}}{\boldsymbol{\hat{f}}^\mathsf{H}\mathsfbi{W}_f\boldsymbol{\hat{f}}}\,,
\end{equation}

\noindent
which seeks forcing–response pairs that yield the maximum amplification $\sigma^2$. The matrices $\mathsfbi{W}_r$ and  $\mathsfbi{W}_f$ define the discrete turbulent kinetic energy (TKE) norms in which the output and input are measured, respectively.
This optimization problem is solved through a singular value decomposition (SVD) of the resolvent operator,

\begin{equation}
    \mathcal{R} = \mathsfbi{V} \boldsymbol{\Sigma} \mathsfbi{F}^\mathsf{H}\,,
\end{equation}

\noindent
where $\mathsfbi{F}$ and $\mathsfbi{V}$ contain the forcing and response modes, respectively. The corresponding amplification factors are found as the singular values $\sigma$ contained in the diagonal matrix $\boldsymbol{\Sigma}$. The leading (largest) singular value and its associated forcing and response modes represent the most amplified linear mechanism in the flow. Because of the strong linear amplification and the broadband excitation in turbulent flows, resolvent response modes associated with large gains have been shown to reliably model dominant coherent structures in turbulent shear flows~\citep{muller_Linear_2024, pickering_Liftup_2020, sarras_Linear_2024}. In the theoretical case of fully uncorrelated true nonlinear forcings, it can even be shown that the resolvent response modes correspond exactly to the SPOD modes \citep{towne_Spectral_2018, lesshafft_Resolventbased_2019}. 

Equation~\ref{eq:RA_operator} becomes singular when $\omega=\omega_\mathrm{eig}$, that is when the resolvent is computed at a frequency corresponding to an eigenvalue of the system. In this case, the optimization problem in equation~\ref{eq:RA_opt_problem} yields the eigenvector as response, the null vector as forcing, and infinite gain. In conventional RA, this manifests as sharp peaks in the gain spectrum near marginally stable (i.e. zero-growth-rate) eigenmodes. This problem can be addressed by employing discounted RA \citep{jovanovic_Modeling_2004, rolandi_Invitation_2024}, which is evaluated at a complex-valued $\omega$. Typically, an imaginary offset larger than the maximum growth rate of all eigenvalues is added to $\omega$ to ensure the resolvent operator is not evaluated at or very close to eigenvalues of the system. This way, an interpretable RA spectrum is obtained. This imaginary offset on the frequency acts as a temporal discounting parameter, enabling the system dynamics to be evaluated over a finite time horizon and separating shorter time scales from the exponential growth of the modal instability \citep{jovanovic_Modeling_2004, rolandi_Invitation_2024}.

\subsection{Computational domain}
\begin{figure}
    \centering
    \begin{subfigure}{0.5\textwidth}
        \includegraphics[width=\linewidth]{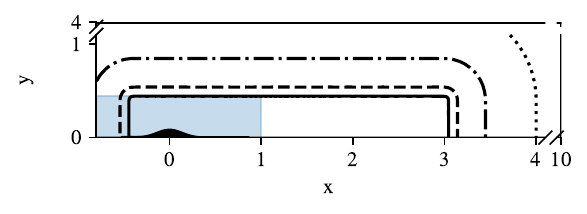}
        \caption{}
        \label{fig:domain_full}
    \end{subfigure}\begin{subfigure}{0.5\textwidth}
        \includegraphics[width=\linewidth]{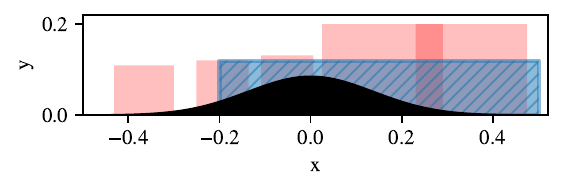}
        \caption{}
        \label{fig:domain_zoom}
    \end{subfigure}

    \begin{subfigure}{\textwidth}
        \includegraphics[width=\linewidth]{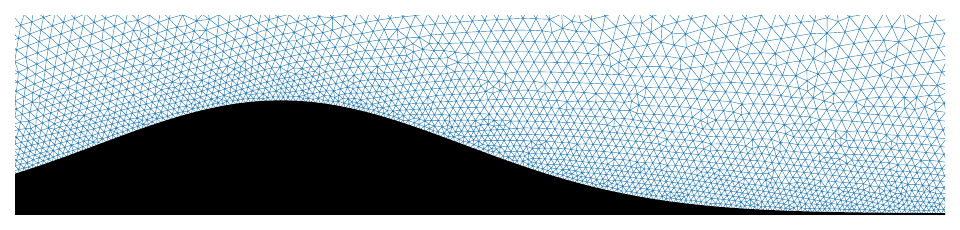}
        \caption{}
        \label{fig:domain_mesh}
    \end{subfigure}
    \caption{Sketch of the computational domain. (a) Full extent of the computational domain with the blue area indicating the PINN domain and black lines show contour levels of the sponge: 0.1 (solid), 1 (dashed), 10 (dash-dotted), and 50 (dotted). Axis breaks are used to highlight the most relevant part of the domain. (b) Zoomed view on the bump, where the red area denotes regions with PIV data and the blue area the response domain for RA. (c) Visualisation of the mesh near the bump surface.}
    \label{fig:domain}
\end{figure}

A sketch of the computational domain is shown in figure~\ref{fig:domain_full}. Here, the extent of the axes indicates the size of the computational domain used for LSA and RA. The extent of the data-assimilated mean flows is indicated by the blue area. Outside the blue area in the far field, nearest-neighbour extrapolation is applied to the mean field quantities. 
In order to suppress spurious free stream modes, the RA energy norm is constrained to the blue-coloured area in figure~\ref{fig:domain_zoom}. This is realized by using a corresponding weight matrix $\mathsfbi{W}_r$ in equation~\ref{eq:RA_opt_problem}. We do not apply a spatial restriction on the forcing.
To ensure numerical stability, sponging is applied, which absorbs and minimizes reflections from computational boundaries outside the region of interest \citep{Bodony2006_sponge}. The sponge level is zero within the rectangle defined by $-0.4\leq x\leq 3$, $0\leq y \leq 0.4$ and grows exponentially with the distance to the closest point on the rectangle. Contours of the sponge level are shown in figure~\ref{fig:domain_full}. The linear analyses are performed using \textit{FELiCS} \citep{kaiser_FELiCS_2023}, an open-source finite element solver for linearised mean field methods. Additional information regarding the implementation of the RA can be found in the documentation available at \url{https://felics.eu/}. The discretisation is performed on a triangular mesh. A close-up of the mesh near the bump surface is shown in figure~\ref{fig:domain_mesh}.

\subsection{Mode alignment}
In addition to a qualitative comparison of the modes through visualisation, we consider a quantitative measure, the alignment between the modes,

\begin{equation}
    A(\boldsymbol{\hat{q}}_1, \boldsymbol{\hat{q}}_2) = \frac{\boldsymbol{\hat{q}}_1^\mathsf{H} \mathsfbi{W}_A \boldsymbol{\hat{q}}_2}{\sqrt{\left(\boldsymbol{\hat{q}}_1^\mathsf{H} \mathsfbi{W}_A \boldsymbol{\hat{q}}_1\right)\left(\boldsymbol{\hat{q}}_2^\mathsf{H} \mathsfbi{W}_A \boldsymbol{\hat{q}}_2\right)}}\,,
\end{equation}
\noindent
based on an inner product defined by $\mathsfbi{W}_A$.
This measure is typically employed to assess the similarity between modes \citep{gudmundsson_Instability_2011, cavalieri_Wavepackets_2013, pickering_Optimal_2021} and is sometimes named correlation coefficient or normalized inner product. The alignment is bounded between 0 and 1, with 1 meaning ``parallel'' and 0 meaning ``orthogonal'' modes.

\section{Data-driven analysis of dominant flow structures}
\label{sec:data_driven_results}

In order to tackle objective (i) of this study, to identify the role of coherent structures in the broadband turbulent dynamics of the flow, we apply SPOD to identify coherent structures in the experimental flow data. In this section, we only consider the SPOD spectra. The respective mode shapes are shown in Section~\ref{sec:model_results}, where they are compared to the RA results.

\begin{figure}
    \centering
    \begin{subfigure}{0.5\textwidth}
        \includegraphics[width=\textwidth]{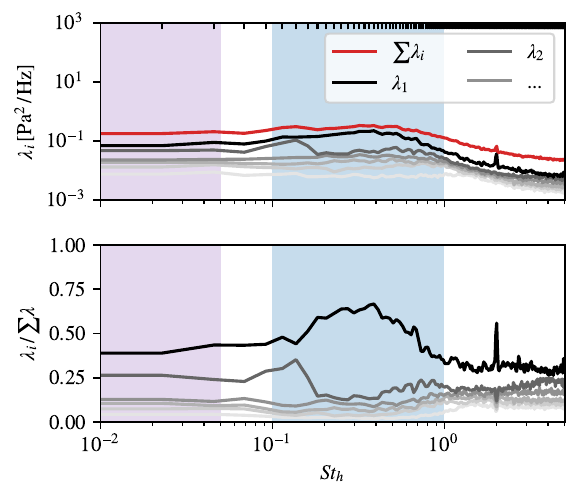}
        \caption{$\Rey=10^6$ (attached)}
    \end{subfigure}\begin{subfigure}{0.5\textwidth}
        \includegraphics[width=\textwidth]{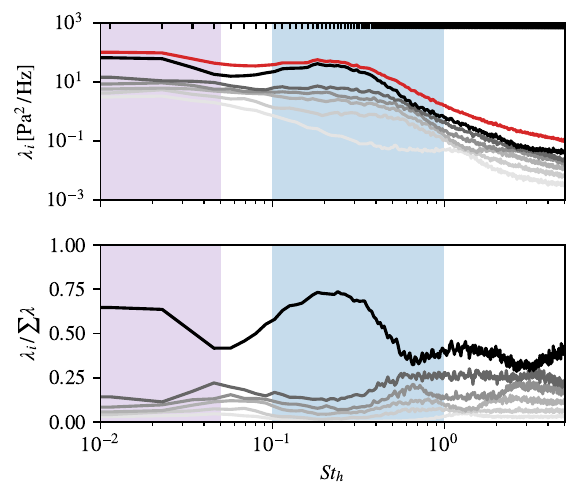}
        \caption{$\Rey=2\times10^6$ (separated)}
    \end{subfigure}
    
    \caption{Spectrum of the surface-pressure SPOD for the attached (a) and separated (b) cases, with the leading mode shown in black and subsequent modes in progressively lighter shades of gray. In the upper panels, the sum of all eigenvalues at each frequency is shown as a red line. In the lower panels, the eigenvalues are normalized by this value to indicate the PSD share of each mode. The blue-shaded region marks the medium-frequency regime, the purple-shaded region marks the low-frequency regime, and tick marks at the top denote the frequency resolution.}
    \label{fig:SPOD_spectrum}
\end{figure}

We first apply SPOD to the instantaneous pressure measurements from the downstream region of the bump. The block length of the SPOD was chosen as $2\times10^4$ snapshots which yields a frequency resolution of $\Delta f = 5\,\mathrm{Hz}$ ($\Delta\Str_h=0.023$ for the attached- and $\Delta\Str_h=0.011$ for the separated case). The minimum resolvable convection velocity for streamwise-travelling waves is set by the spatial Nyquist criterion, $u_\text{c,crit}=2 \Str\,\Delta x$. For the dominant shedding frequencies of $St=St_h/h=3.29$ and $St=2.12$ in the attached and separated cases, respectively, and the normalised sensor spacing of $\Delta x= 0.063$, this corresponds to minimum resolvable convection velocities of $u_\text{c,crit} = 0.41$ and $u_\text{c,crit} = 0.27$. In TSB flows, the expected convection velocities of coherent structures are typically in the range $0.4 < u_\text{c} < 0.6$ \citep{kiya_Structure_1983, cherry_Unsteady_1984, mohammed-taifour_Unsteadiness_2016}. Therefore, the expected coherent structures up to the dominant shedding frequency can be resolved using the pressure sensor data, especially in the separated case. The SPOD spectrum obtained from this dataset is shown in figure~\ref{fig:SPOD_spectrum}. In the upper panels, the sum over all eigenvalues at the respective frequency, $\sum\lambda_i$, is shown as a red line. Note that $\sum\lambda_i$ is equal to the sum of the PSDs of the individual sensor signals. It can be seen that the leading eigenvalue approaches the red line in the highlighted frequency regimes, identified in the PSDs of single sensors (see figure~\ref{fig:surfp_psd}). This is highlighted through the lower panels in figure~\ref{fig:SPOD_spectrum}, where the eigenvalues are divided by $\sum\lambda_i$. This representation shows the PSD share associated with the respective mode. Evidently, the prominent regimes identified so far also correspond to frequency ranges of low-rank dynamics. In the medium frequency regime, approximately 30 \% to 75 \% of the signal's PSD is represented by the leading SPOD mode in both cases. In the low-frequency regime, the leading mode accounts for approximately 40 \% of the PSD in the attached- and 40 \% to 65 \% in the separated case. This observation shows that, while low-frequency dynamics do not appear particularly prominent in the absence of flow separation ($\Rey=10^6$), the present dynamics nevertheless appear to be low-rank, indicating the presence of coherent structures.

\begin{figure}
    \centering
    \begin{subfigure}{0.5\textwidth}
        \includegraphics[width=\textwidth]{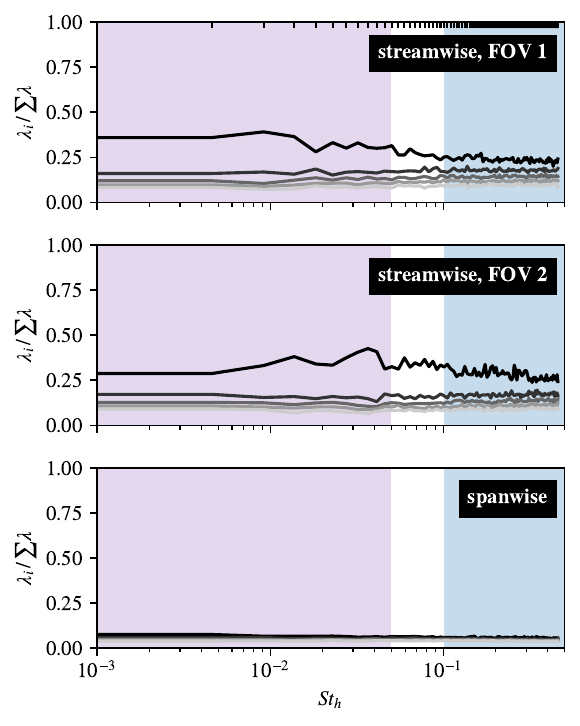}
        \caption{$\Rey=10^6$ (attached)}
    \end{subfigure}\begin{subfigure}{0.5\textwidth}
        \includegraphics[width=\textwidth]{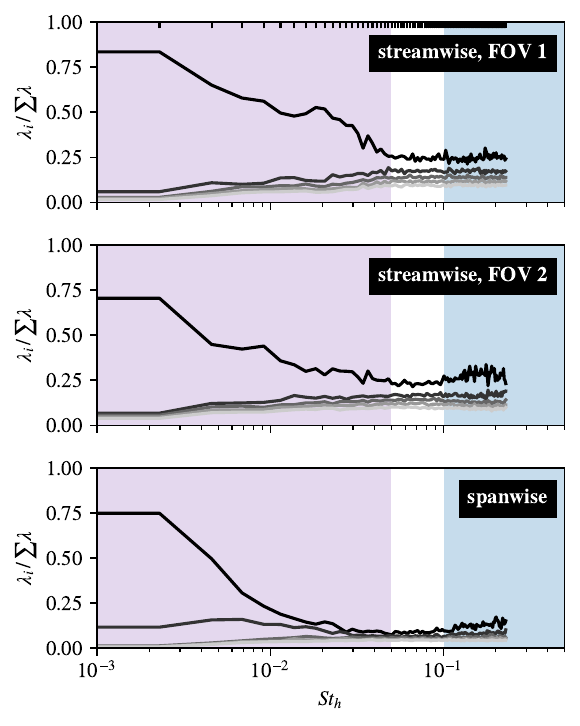}
        \caption{$\Rey=2\times10^6$ (separated)}
    \end{subfigure}
    
    \caption{SPOD spectra of the PIV data from the different measurement windows for the attached (a) and separated (b) cases, with the leading mode shown in black and subsequent modes in progressively lighter shades of gray. Eigenvalues are normalized by $\sum\lambda_i$ at each frequency to indicate the PSD share of each mode. The blue-shaded region marks the medium-frequency regime, the purple-shaded region marks the low-frequency regime, and tick marks at the top denote the frequency resolution.}
    \label{fig:piv_spod}
\end{figure}

SPOD is also performed on the PIV velocity time series in two streamwise and one spanwise measurement windows (PIV domains are shown in figure~\ref{fig:bump_data_pos}). Here, the block length was chosen as $200$ snapshots which yields a frequency resolution of $\Delta f=1\,\mathrm{Hz}$ ($\Delta\Str_h=0.005$ for the attached- and $\Delta\Str_h=0.002$ for the separated case). The Nyquist frequency is $100\,\mathrm{Hz}$, that is $\Str_h=0.46$ for the attached- and $\Str_h=0.23$ for the separated case, and as such, the medium-frequency regime cannot be fully resolved for the separated case. Instead, we focus on the low-frequency regime with this analysis.
Figure~\ref{fig:piv_spod} shows the resulting normalized eigenvalue spectra. For the separated case, these reveal very significant low-rank dynamics in the low-frequency regime. In all measurement windows, the leading mode approaches $\approx75\,\%$ of the PSD at the lowest resolvable frequency. Notably, the sub-leading mode in the spanwise window also shows clear separation from the subsequent modes. In the attached case however, no low-rank dynamics are observed in the spanwise window. In the streamwise windows, the separation between the eigenvalues is also less pronounced than in the separated case. Nevertheless, the leading mode accounts for more than $25\,\%$ of the PSD in the low-frequency regime, showing the presence of low-rank dynamics. This observation is consistent with the previous observation that pressure fluctuations at low-frequencies, in the absence of flow separation, are still low-rank. The second dominant region  with high gain separation in the medium-frequency regime (blue background), identified in the surface pressure SPOD spectra in figure~\ref{fig:SPOD_spectrum}, is absent in the PIV SPOD spectra (figure~\ref{fig:piv_spod}). We attribute this to the aforementioned Nyquist rate limitation of the PIV measurements.

Regarding objective (i), we draw the conclusion that the medium-frequency dynamics in both cases are characterized by low-rank dynamics driving coherent structures. The same holds for the low-frequency regime, especially in the separated case, and to a lesser degree also in the attached case.

\section{Physics-based model for coherent structures}
\label{sec:model_results}
In order to address objective (ii) of this study, to compare the driving mechanisms of the coherent structures between the cases, we analyse the coherent dynamics of the flow in a physics-based way by employing linear mean field methods, specifically LSA and RA. We first investigate the global stability of the mean flow using LSA and subsequently employ RA to analyse non-modal mechanisms.

\subsection{Linear stability analysis}
\begin{figure}
    \centering
    \begin{subfigure}{0.5\textwidth}
        \includegraphics[width=\textwidth]{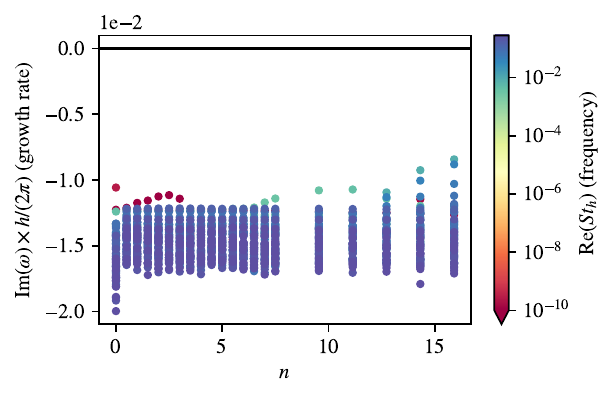}
        \caption{$\Rey=10^6$ (attached)}
    \end{subfigure}\begin{subfigure}{0.5\textwidth}
        \includegraphics[width=\textwidth]{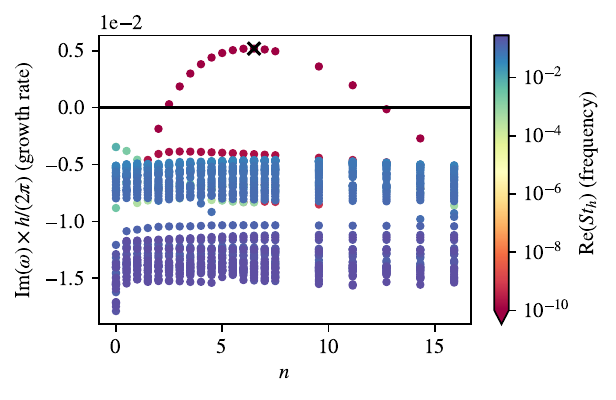}
        \caption{$\Rey=2\times10^6$ (separated)}
    \end{subfigure}
    \caption{LSA eigenvalue spectra over spanwise wavenumber  for the attached (a) and separated cases (b), coloured by $\Str_h$. The neutral stability line $\Imag(\omega)=0$ is shown as a black line. The most unstable (i.e. maximum growth rate) mode is marked with a black cross.}
    \label{fig:lsa_spectra}
\end{figure}

LSA is performed for each spanwise wavenumber $n=\frac{\beta}{2\pi}$ separately in the range of $0\leq n < 16$. Figure~\ref{fig:lsa_spectra} shows the resulting eigenvalue spectra. Eigenvalues with $\Imag(\omega)<0$ are stable, whereas those with $\Imag(\omega)>0$ are unstable (recall the harmonic ansatz in equation~\ref{eq:ansatz}). Evidently, the attached case does not exhibit any global instabilities: all eigenvalues have negative growth rates and are located at approximately the same distance from the neutral-stability line. We note that the branch that appears to be growing for high $n$ corresponds to spurious modes. In contrast, the separated case exhibits a distinct eigenvalue branch with numerically zero frequency ($\Str_h<10^{-10}$) and elevated growth rates over $2.5\leq n < 13$. The positive growth rate is unexpected for a stationary mean flow analysis and likely arises from model limitations, namely the spanwise-uniform mean-flow assumption and the eddy-viscosity modelling of the coherent Reynolds stresses.
Crucially, the isolation of this branch from the rest of the spectrum indicates that, for this spanwise wavenumber range, the low-frequency dynamics are governed by a global mode.

\begin{figure}
    \centering
    \includegraphics[width=0.5\linewidth]{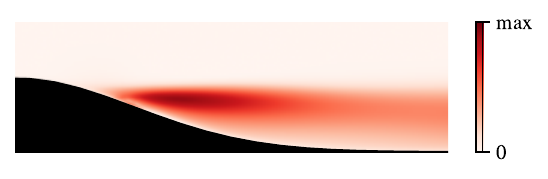}
    \caption{LSA eigenmode with maximum growth rate for the separated case. $n=6.5$ (marked in figure~\ref{fig:lsa_spectra} with a black cross).}
    \label{fig:lsa_mode}
\end{figure}

The corresponding mode shape for the largest growth rate (marked by the black cross) is shown in figure~\ref{fig:lsa_mode}. Note that the mode is real-valued due to the frequency being zero. In the following sections, we show that this globally unstable zero-frequency mode provides a good explanation for the low-frequency dynamics observed in the separated case.

\subsection{Resolvent analysis}

\begin{figure}
    \centering
    \begin{subfigure}{0.5\textwidth}
        \includegraphics[width=\textwidth]{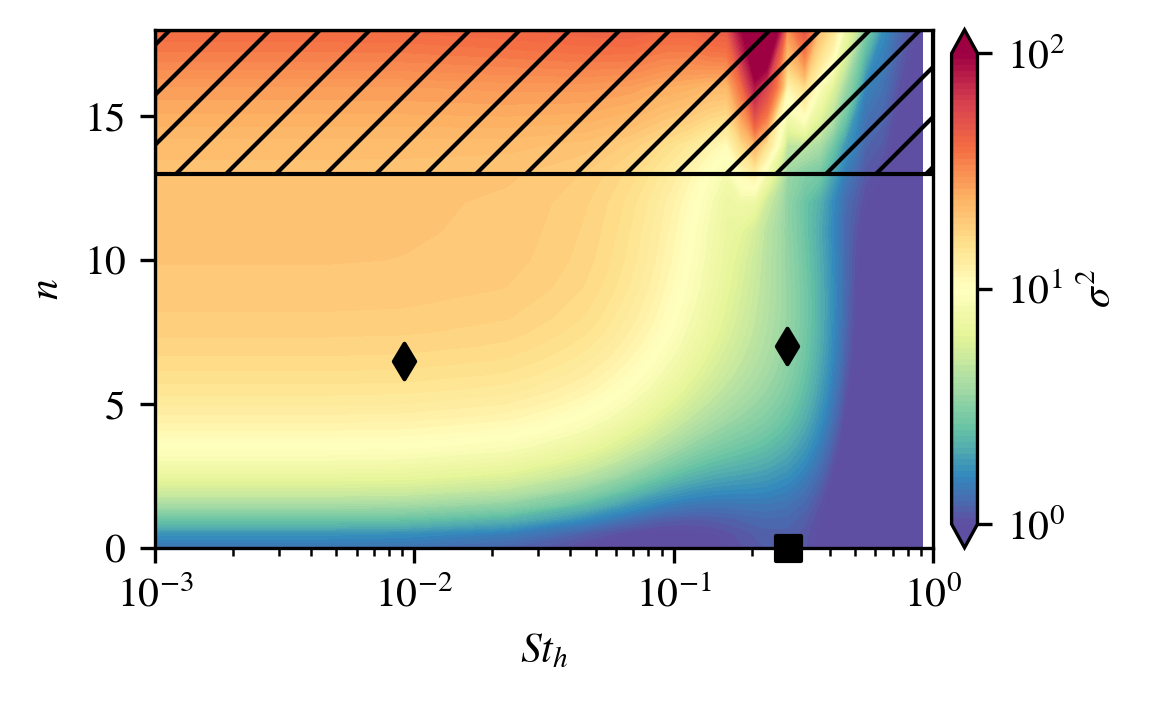}
        \caption{$\Rey=10^6$ (attached)}
        \label{fig:resolvent_gain_attached}
    \end{subfigure}\begin{subfigure}{0.5\textwidth}
        \includegraphics[width=\textwidth]{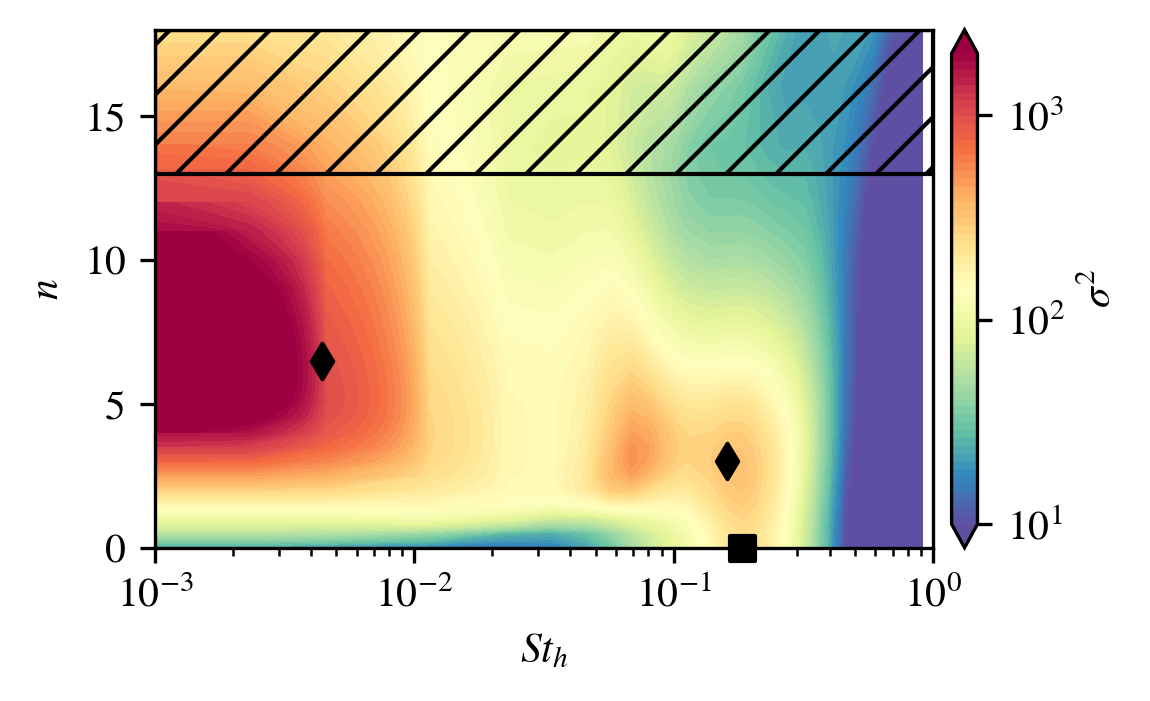}
        \caption{$\Rey=2\times10^6$ (separated)}
        \label{fig:resolvent_gain_separated}
    \end{subfigure}
    \caption{Resolvent gain as a function of Strouhal number and spanwise wavenumber for the attached (a) and separated cases (b), shown as a heatmap with a logarithmically scaled colorbar. Black markers indicate Strouhal number–spanwise wavenumber pairs discussed in the following sections.}
    \label{fig:resolvent_gain}
\end{figure}

The previous analysis with LSA did not reveal any distinc eigenmodes in the attached case, although low-rank dynamics have been identified in the data-driven analysis. We therefore employ RA in order to model coherent structures that originate from non-modal mechanisms.

RA is performed for both cases. In the separated case, however, the modal instability identified in the previous section would lead to singularities in the RA gain spectrum, which would obfuscate, for example, the preferential spanwise wavenumber of the low-frequency dynamics. One approach to obtain an interpretable gain spectrum would be to apply a scaling factor to the eddy viscosity in order to stabilise the system, similarly as in \cite{fuchs_StandingWave_2025}. However, we find that for this case eddy viscosity scaling deteriorates the linear models' performance to explain the data. We therefore take a different approach and apply discounted RA by adding an imaginary offset (discount factor) of $\gamma = \Imag(\omega) = 0.5$ to the RA frequencies in the separated case. The chosen value for $\gamma$ is slightly higher than the maximum growth rate of the unstable eigenvalues (see figure~\ref{fig:lsa_spectra}), effectively stabilising the system. It is important to note that for each value of $\gamma$, a different RA spectrum is obtained:  increasing the discount factor lowers the overall RA gains and flattens the spectrum \citep{rolandi_Invitation_2024, rolandi_Biglobal_2025}. In the present case, the corresponding RA mode shapes (in the region of interest) remain essentially unchanged for small discount factors, and only very large values lead to noticeable modifications. For the chosen $\gamma=0.5$, we demonstrate the negligible effect on the mode shapes in Appendix~\ref{sec:nondiscount_ra}. The discount factor is often interpreted as an effective time horizon $t_\text{eff}=\frac{1}{\gamma}$ over which the structures grow \citep{jovanovic_Modeling_2004, rolandi_Invitation_2024, rolandi_Biglobal_2025}. For $\gamma=0.5$ in the present case, this effective time horizon corresponds to $t_\text{eff}=2$ convective time scales based on $L$. Alternatively, the discounted RA spectrum can be interpreted as a slice through the pseudospectrum of the resolvent operator along a line in the complex-frequency plane specified by $\gamma$. We emphasize that discounting is applied solely to yield a finite, well-behaved spectrum, and the precise value chosen does not affect the qualitative interpretation of the results. In the attached case, all eigenvalues have a negative growth rate, so no discount factor is required. For a discussion of the relationship between the resolvent and the eigenvalues of the linear operator via the pseudospectrum, as well as the effect of discounting RA, we refer to Appendix~\ref{sec:nondiscount_ra}.

A grid search over $n$ and $St_h$ is performed and contours of the resulting resolvent gain are plotted in figure~\ref{fig:resolvent_gain}. Note the logarithmic colour scaling, which spans multiple orders of magnitude and is different for the two cases. At high spanwise wavenumbers ($n > 14$), the resolvent modes display increasing levels of numerical noise and the gain separation between the leading and sub-leading resolvent modes drops off (not shown). Furthermore, a different mode family is picked up, which is located on the upstream face of the bump, where no time-resolved data for validation are available. These modes are excluded from the analysis, as indicated by the hatched area in figure~\ref{fig:resolvent_gain}. The black markers in the figure mark parameter combinations of $\Str_h$ and $n$ that will be further discussed in the following sections.

In both cases, a medium-frequency regime is identified. The highest gain in this regime is observed for non-zero $n$, but the gain remains elevated as $n\to0$. The dominant frequency of this regime is $\Str_h\approx0.28$ for the attached case and slightly lower, at $\Str_h\approx0.18$ for the separated case (see square markers in figure~\ref{fig:resolvent_gain}), which is in line with the observations in the surface pressure spectra presented in figure~\ref{fig:surfp_psd} (or the surface pressure SPOD spectra in figure~\ref{fig:SPOD_spectrum}). The most notable difference between the two cases is seen at low-frequencies, $\Str_h\ll 0.1$, down to $\Str_h=0$. Here, we see the influence of the distinct eigenmode in the separated case, which manifests as a broad regime of very high RA gain around $n\approx6.5$.  In the attached case, the corresponding eigenvalues at $\Str_h=0$ are much more strongly damped, and the resolvent is therefore evaluated farther from any eigenvalue, so no comparable feature appears in the RA gain spectrum. However, there is still significant RA gain for $\Str_h<0.1$ and down to $\Str_h =0$ in this case, especially for elevated $n$.\\

\noindent
In this study, the resolvent gain is based on the TKE norm of the response mode and is thus indicative of the magnitude of the velocity fluctuations associated with the mode. In many experimental studies however, flow dynamics are investigated based on time-resolved wall pressure measurements. For a discussion regarding the coherent surface pressure fluctuations associated with the RA modes and their comparison with measurements from the experiment, we refer to Appendix~\ref{sec:surfp_ra}.

\subsection{Validation of the linear analyses}

\begin{figure}
    \centering
    \begin{subfigure}{0.5\textwidth}
        \includegraphics[width=\textwidth]{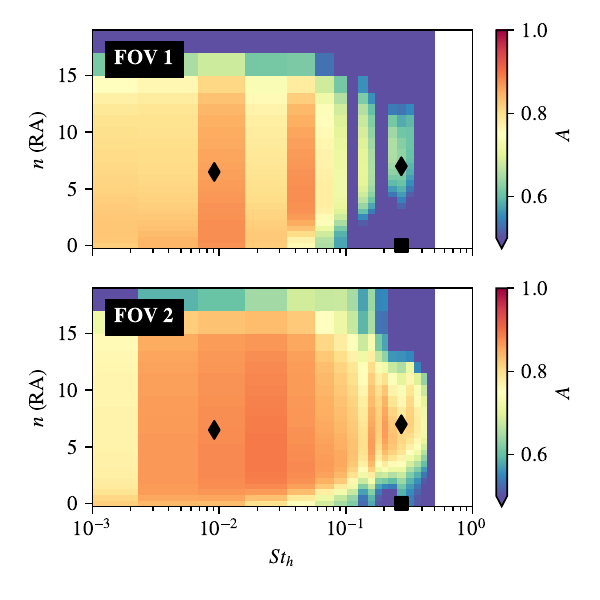}
        \caption{$\Rey=10^6$ (attached)}
    \end{subfigure}\begin{subfigure}{0.5\textwidth}
        \includegraphics[width=\textwidth]{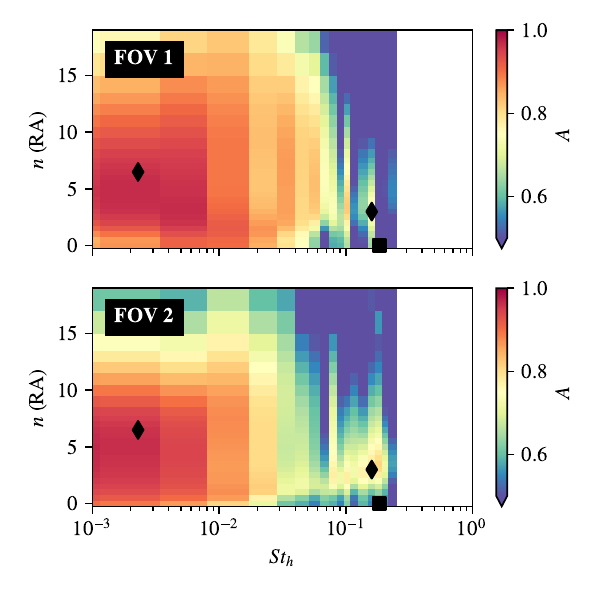}
        \caption{$\Rey=2\times10^6$ (separated)}
    \end{subfigure}
    
    \caption{Alignment between RA modes and streamwise SPOD modes at respective frequency as a function of Strouhal number and spanwise wavenumber for the attached (a) and separated (b) cases. Upper and lower panels show the alignment with the SPOD from FOV 1 and FOV 2, respectively. Black markers indicate Strouhal number–spanwise wavenumber pairs discussed in the following sections.}
    \label{fig:mode_alignment}
\end{figure}

In order to validate the linear analyses, we compute the alignment between the leading RA mode and the leading SPOD mode at the respective frequency. The alignment is shown in figure~\ref{fig:mode_alignment} as a heatmap over Strouhal number and spanwise wave number.
It should be noted that the SPOD computed from streamwise PIV measurements is not associated with a specific spanwise wavenumber $n$, but instead reflects contributions from a range of $n$. Consequently, strong alignment is not expected for all individual $n$ values. We consider the linear model valid if, at each frequency in the relevant regimes, at least one resolvent mode (over all $n$) has high alignment with the SPOD mode at this frequency. The maximum alignment at each frequency is expected at the dominant spanwise wavenumber of the dynamics in the flow. Because the PIV in FOV 1 and FOV 2 was not measured simultaneously, SPOD is performed for both areas separately. We therefore obtain two alignment values (one for FOV 1 and one for FOV 2) for each Strouhal number and spanwise wavenumber.

In the attached case, high values of alignment are observed for $\Str_h \lessapprox 0.1$ over a wide range of $n$. As the frequency exceeds $\Str_h= 0.1$, high alignment becomes limited to higher $n$. In the separated case, very high alignment is observed for $\Str_h < 0.01$ with a preferential $n$ around $5$. In the medium-frequency regime, around $\Str_h=0.15$, there is a rather confined regime of high alignment for $n \approx 3$. 
Generally, there is a trend that the alignment is high where the RA gain is high, confirming that the dominant coherent structures in the flow originate from linear dynamics. However, for the low-frequency modes, there is a broad region of high alignment around the dominant spanwise wavenumbers, which reflects the fact that the corresponding mode shapes in the plane at $z=0$ are relatively insensitive to $n$. For the medium-frequency modes, the alignment is confined to a narrower band of wavenumbers, especially for the separated case. We attribute this mostly to the higher sensitivity of the planar mode shapes to $n$ in this regime, combined with the observation that the underlying mechanism appears to preferentially amplify these wavenumbers (see figure~\ref{fig:resolvent_gain_separated}).

\begin{figure}
    \centering
    \includegraphics[width=0.5\linewidth]{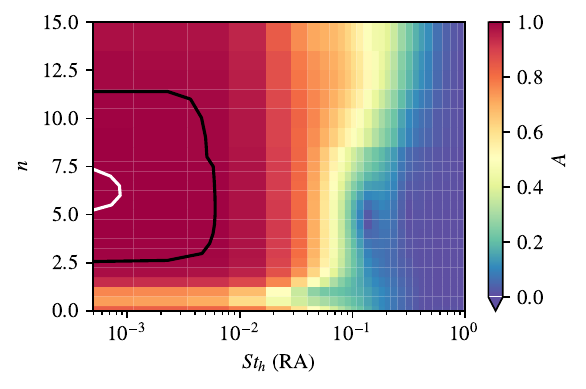}
    \caption{Alignment between the zero-frequency LSA eigenmode and RA modes. Both modes share the same $n$ but the RA mode frequency is indicated on the x-axis whereas the LSA mode is always at $\Str_h=0$. Contour lines show $A=0.99$ (black) and $A=0.999$ (white).}
    \label{fig:LSA_RA_alignment}
\end{figure}

For the separated case, we further compute the alignment between the RA modes and the $\Str_h=0$ LSA mode with maximum growth rate at the respective spanwise wavenumber. This alignment is shown as a heatmap in figure~\ref{fig:LSA_RA_alignment}. Evidently, as $\Str_h \to 0$, the RA- and LSA modes converge in the range of spanwise wavenumbers where the eigenmode has a distinct growth rate. Figure~\ref{fig:LSA_RA_alignment} thus demonstrates that, in the separated case, RA reflects the modal instability at $\Str_h=0$ throughout the low-frequency regime, with little contribution from non-modal effects.

For the attached case, comparing all LSA modes with $\Str_h\approx0$ against the RA mode at $\Str_h=0$ and the respective $n$ confirms that there is no significant alignment. The reader is referred to Appendix~\ref{sec:all_lsa_vs_ra} for details.

\begin{figure}
    \centering
    \begin{subfigure}{0.5\textwidth}
        \centering
        \includegraphics[width=0.9\textwidth]{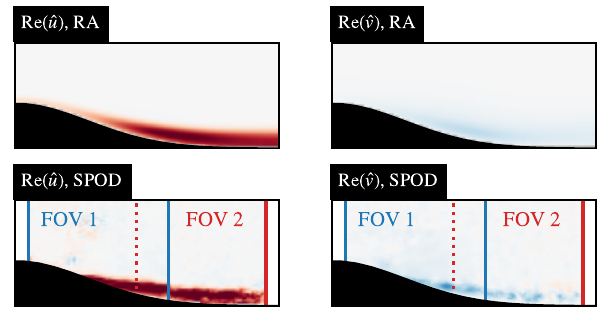}
        \includegraphics[width=0.8\textwidth]{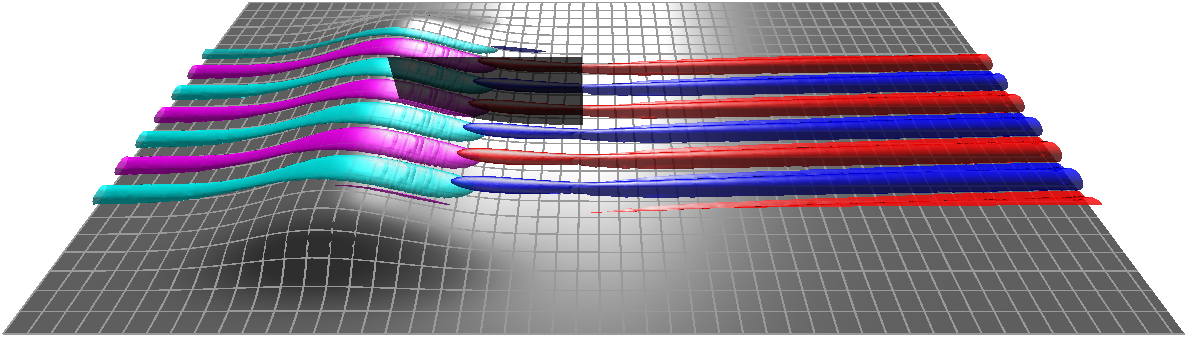}
        \caption{$\Rey=10^6$ (attached), $\Str_h=0.01$, $n=6.5$,\\ $A_\text{FOV1}=0.84$, $A_\text{FOV2}=0.88$}
        \label{fig:spod_ra_compare_attached_low}
    \end{subfigure}\begin{subfigure}{0.5\textwidth}
        \centering
        \includegraphics[width=0.9\textwidth]{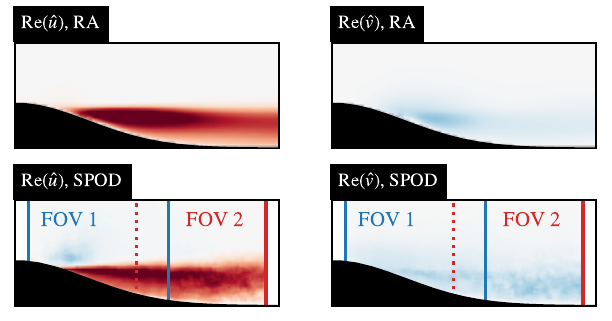}
        \includegraphics[width=0.8\textwidth]{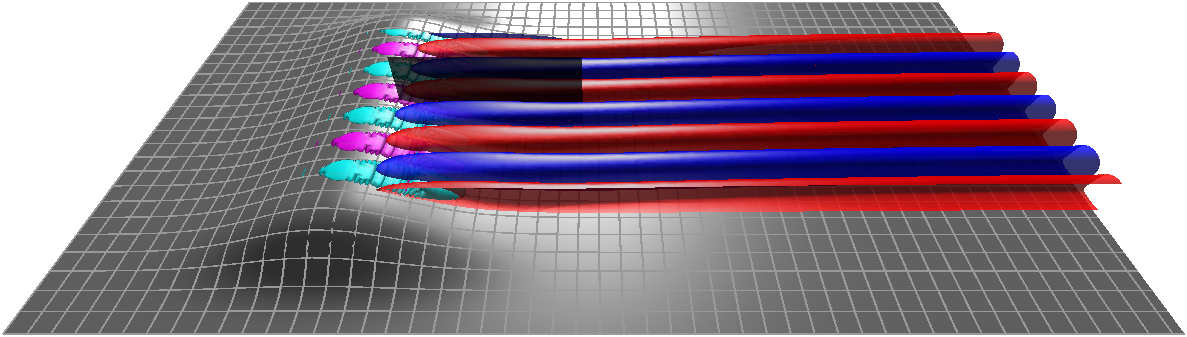}
        \caption{$\Rey=2\times10^6$ (separated), $\Str_h=0.002$, $n=6.5$,\\ $A_\text{FOV1}=0.95$, $A_\text{FOV2}=0.96$}
        \label{fig:spod_ra_compare_sep_low}
    \end{subfigure}

    \begin{subfigure}{0.5\textwidth}
        \centering
        \includegraphics[width=0.9\textwidth]{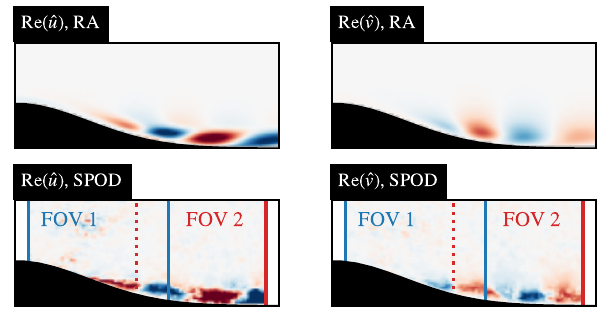}
        \includegraphics[width=0.8\textwidth]{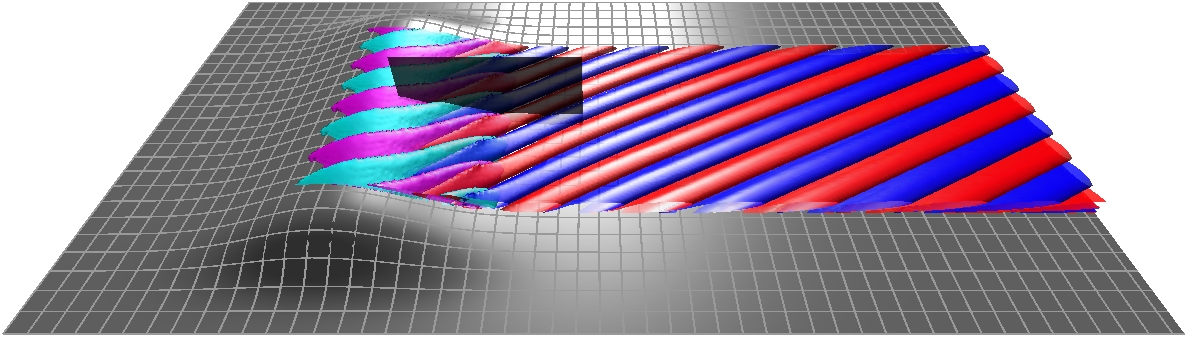}
        \caption{$\Rey=10^6$ (attached), $\Str_h=0.27$, $n=7$,\\ $A_\text{FOV1}=0.70$, $A_\text{FOV2}=0.80$}
        \label{fig:spod_ra_compare_attached_med}
    \end{subfigure}\begin{subfigure}{0.5\textwidth}
        \centering
        \includegraphics[width=0.9\textwidth]{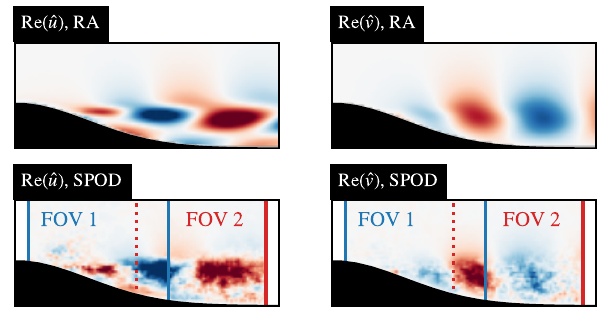}
        \includegraphics[width=0.8\textwidth]{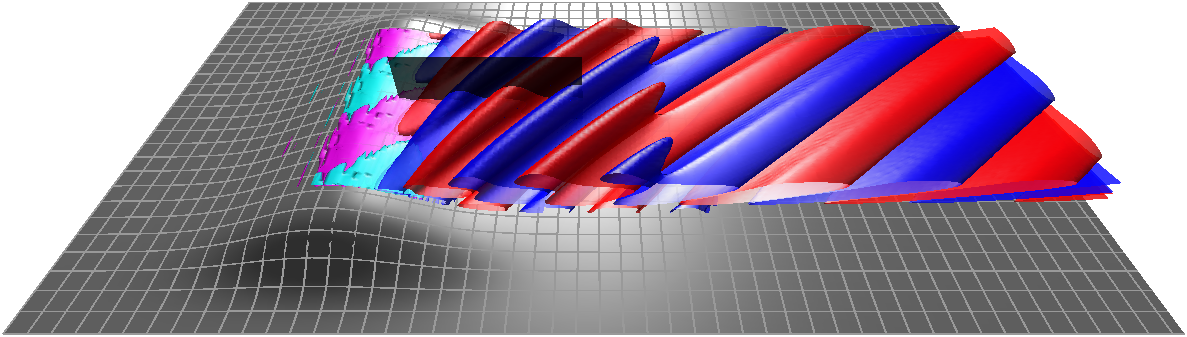}
        \caption{$\Rey=2\times10^6$ (separated), $\Str_h=0.16$, $n=3$,\\ $A_\text{FOV1}=0.78$, $A_\text{FOV2}=0.78$}
        \label{fig:spod_ra_compare_sep_med}
    \end{subfigure}
    
    \caption{Mode shape comparison between RA and SPOD at selected frequencies and spanwise wavenumbers for attached (a, c) and separated (b, d) cases. SPOD modes from both fields of view (independent measurements) are shown on the same axis, and visually framed with coloured lines for clarity. Three-dimensional isosurfaces of the RA forcing- (cyan, magenta) and response (blue, red) modes are included for reference, with the black plane indicating the region used for the SPOD–RA comparison. Frequencies and spanwise wavenumbers are indicated in figures~\ref{fig:resolvent_gain},~\ref{fig:mode_alignment}, and~\ref{fig:resolvent_surfp} through diamond markers.}
    \label{fig:spod_ra_compare}
\end{figure}

For representative frequencies and spanwise wavenumbers of the low- and medium-frequency regimes, the RA modes are shown in figure~\ref{fig:spod_ra_compare} and compared to the respective SPOD modes of the streamwise PIV windows in the downstream region of the bump. In this section, we only focus on SPOD modes of the streamwise PIV, which include the streamwise and vertical velocity components ($u, v$). A comparison of the RA mode with the SPOD modes from the spanwise SPIV data requires consideration of the three-dimensional structure of the modes, which will be addressed in Section~\ref{sec:standing_waves}.
The selected frequencies and spanwise wavenumbers are indicated in the RA gain spectrum (figure~\ref{fig:resolvent_gain}) and the RA-SPOD alignment heatmaps (figure~\ref{fig:mode_alignment}) through diamond markers.
Note that the SPOD was performed for both windows separately, as they were not measured simultaneously. For presentation purposes, the SPOD modes from both windows are still shown on the same axes, and visually framed with coloured lines for clarity.
The low-frequency modes ($\Str_h=0.01$ and $\Str_h =0.002$, for the attached and separated cases, respectively), shown in figures~\ref{fig:spod_ra_compare_attached_low} and~\ref{fig:spod_ra_compare_sep_low}, resemble large-scale, streamwise-elongated structures. We use the term ``streaky structures'' in the remainder of this manuscript in order to avoid confusion with the term ``streaks'', which is often used specifically to refer to structures associated with the lift-up mechanism. Interestingly, we find qualitatively similar low-frequency modes in the attached and separated cases. In the attached case, the structure follows the attached shear layer along the wall, whereas the structure in the separated case is located within the free shear layer further away from the wall. This is notable because low-frequency dynamics are often attributed to the presence of a TSB, which is not the case in the mean flow of the attached case.
The medium-frequency modes ($\Str_h=0.27$ and $\Str_h =0.16$, for the attached and separated cases, respectively), shown in figures~\ref{fig:spod_ra_compare_attached_med} and~\ref{fig:spod_ra_compare_sep_med}, are likewise similar between the attached and separated cases. These modes resemble vortices with relatively small streamwise extent. In this case, the similarity is expected because these medium-frequency dynamics are typically associated with vortex shedding in the shear layer, which is present across both cases.

Based on these qualitative and quantitative comparisons between the LSA, RA, and SPOD modes, we conclude that the linear modelling results qualitatively capture the relevant trends and mode shapes of the flows' coherent dynamics with satisfactory accuracy.  In Appendix~\ref{sec:surfp_ra}, we further show that RA accurately captures the coherent surface pressure fluctuations on the downstream side of the bump.

\subsection{Physical mechanisms}
As the model is validated, we now turn back to objective (ii) and compare the physical mechanisms leading to the formation of coherent structures in both cases. Here, we focus on the low-frequency dynamics.

For the separated case, the low-frequency dynamics are associated with a distinct ``steady'' (i.e., $\Str_h=0$), three-dimensional (i.e., $n\neq 0$) eigenmode. We therefore speak of modal dynamics. In the study of laminar separation bubbles, such a mode was attributed to a centrifugal instability \citep{gallaire_Threedimensional_2007, rodriguez_Two_2013, savarino_Optimal_2024}. In TSB flows, this type of mode was also previously identified \citep{sarras_Linear_2024, cura_Lowfrequency_2024, fuchs_StandingWave_2025} and it was suggested that a similar centrifugal mechanism is responsible.
In the attached case, on the other hand, we did not identify any distinct eigenmode corresponding to a physically relevant feature. Yet, RA correctly captures the dynamics, with $A>0.75$ throughout the low-frequency regime (see figure~\ref{fig:mode_alignment}). There are two possible interpretations for the dynamics in the attached case:

\begin{enumerate}
    \item The dynamics are modal but associated with a highly stable eigenvalue, which the LSA fails to isolate. In this case, the driving mechanism might be a similar centrifugal instability as in the separated case, acting on the geometry-induced curvature of the mean flow streamlines.
    \item The dynamics are non-modal; that is, they are not associated with a specific eigenmode but rather emerge from the non-orthogonality of several eigenmodes. A likely explanation for non-modal dynamics would be the lift-up mechanism that generates streamwise-elongated low-frequency streaks through the transport of momentum across a shear layer, which is facilitated by the action of streamwise vorticity \citep{landahl_Note_1980, brandt_Liftup_2014}.
\end{enumerate}

\noindent
Note that the coherent structures generated through both of these mechanisms are very similar and are subsequently both described as low-frequency streaky structures, as introduced above. 
From the aforementioned results, we cannot draw a definite conclusion on whether a weakened version of the modal instability gives rise to the streaky structures in the attached case, or whether these originate through a qualitatively different mechanism like the lift-up effect. 
However, the absence of a distinct branch in the eigenvalue spectrum favours the latter explanation.

Here, we compare the cases at $\Rey=10^6$ (attached) and $\Rey =2\times10^6$ (separated). We find that a further increase of the Reynolds number to $4\times10^6$ does not qualitatively change the dynamics of the separated flow. This analysis is available in Appendix~\ref{sec:reynolds_invariance}.

\section{Finite-span effects on the low-frequency dynamics}
\label{sec:standing_waves}

\begin{figure}
    \centering
    \begin{subfigure}{0.5\textwidth}
        \includegraphics[width=\textwidth]{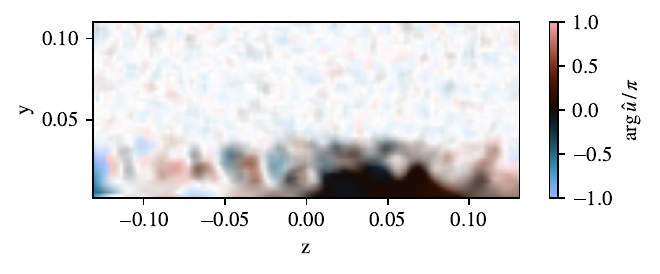}
        \caption{$\Rey=10^6$ (attached), $\Str_h=0.005$, leading}
        \label{fig:standing_wave_demo_attached_leading}
    \end{subfigure}\begin{subfigure}{0.5\textwidth}
        \includegraphics[width=\textwidth]{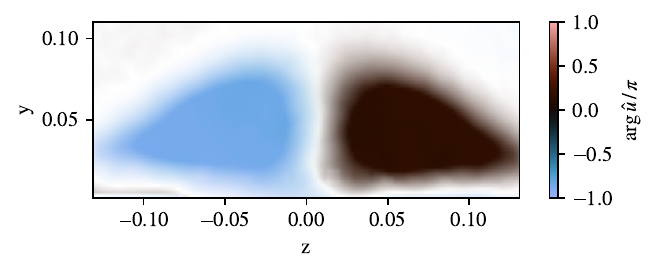}
        \caption{$\Rey=2\times10^6$ (separated), $\Str_h=0.002$, leading}
        \label{fig:standing_wave_demo_separated_leading}
    \end{subfigure}

    \begin{subfigure}{0.5\textwidth}
        \includegraphics[width=\textwidth]{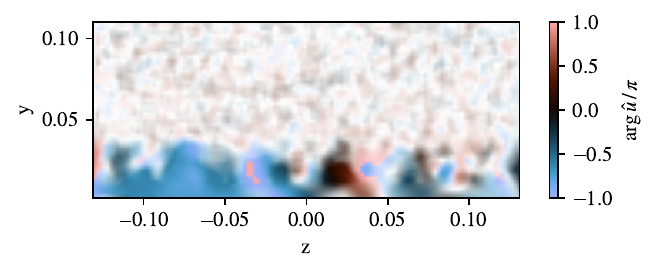}
        \caption{$\Rey=10^6$ (attached), $\Str_h=0.005$, sub-leading}
        \label{fig:standing_wave_demo_attached_subleading}
    \end{subfigure}\begin{subfigure}{0.5\textwidth}
        \includegraphics[width=\textwidth]{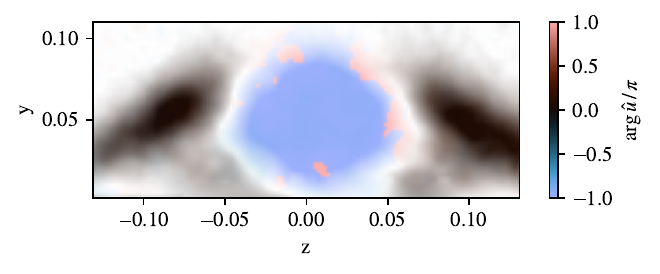}
        \caption{$\Rey=2\times10^6$ (separated), $\Str_h=0.002$, sub-leading}
        \label{fig:standing_wave_demo_separated_subleading}
    \end{subfigure}
    
    \caption{Phase angle of the $\hat{u}$-component of the leading (a, b) and sub-leading (c, d) low-frequency SPOD modes in the spanwise plane for the attached (a, c) and separated cases (b, d). Transparency is set according to the magnitude of the mode with full transparency at zero magnitude and no transparency at half the maximum value.}
    \label{fig:standing_wave_demo}
\end{figure}

In order to address objective (iii) of this study, to assess the role of the finite span and tunnel side walls on the dominant flow structures, we analyse the three-dimensional structure of the low-frequency dynamics. To this end, we consider SPOD of SPIV measurements in the spanwise plane as shown in figure~\ref{fig:bump}.

The phase angle of the $\hat{u}$-component of the leading and sub-leading SPOD modes at $f=1\,\mathrm{Hz}$ (the lowest non-zero frequency bin) are shown in figure~\ref{fig:standing_wave_demo}. The phase fields are masked by the mode's magnitude, with the transparency set inversely proportional to the absolute value of the mode. This depiction reveals the wave structure of the mode. In the case of travelling waves, the phase evolves continuously in space. Standing waves, on the other hand, are characterized by nodes, where the oscillation amplitude remains zero, and antinodes, where it reaches a maximum. Between nodes, the phase remains constant, whereas discontinuities occur at the nodes. In the attached case (a, c), the SPOD modes are very noisy and the mode cannot be as clearly extracted as from the side view data. In the separated case~\mbox{(b, d)}, on the other hand, the modes show clear evidence of a standing wave structure. Here, the leading mode~(b) corresponds to a standing wave pattern with a node on the centerline, whereas the sub-leading mode (d) has an antinode at this position.

The observation of spanwise-standing waves constitutes a challenge for modelling these structures with RA because, by periodically expanding the RA mode in spanwise direction according to the ansatz in equation~\ref{eq:ansatz}, a spanwise-travelling wave is obtained. In order to model the experimentally observed dynamics, we therefore apply a spanwise-standing wave assumption \citep{fuchs_StandingWave_2025}, where the RA mode is expanded as

\begin{equation}
    \begin{split}
        \boldsymbol{\hat{u}}^\mathrm{SW}_{\beta,\omega}(x,y,z,t) &= \boldsymbol{\hat{u}}_{\beta,\omega}(x,y)\exp(\mathrm{i}\beta [z-0.5L_\text{eff}^\text{SW}] - \mathrm{i}\omega t) \\ &+ \boldsymbol{\hat{u}}_{-\beta,\omega}(x,y)\exp(-\mathrm{i}\beta[z-0.5L_\text{eff}^\text{SW}] - \mathrm{i}\omega t)\,.
    \end{split}
\end{equation}

\noindent
Here, $\boldsymbol{\hat{u}}_{\beta, \omega}$ denotes the velocity field associated with the leading resolvent response mode, and $\boldsymbol{\hat{u}}_{-\beta,\omega}$ is its spanwise-symmetric counterpart travelling in the opposite direction. The two modes are identical in all components except the spanwise velocity, which has the opposite sign due to the symmetry condition. An effective spanwise length scale for the standing wave dynamics $L_\text{eff}^\text{SW}$ is introduced, and the term $-0.5L_\text{eff}^\text{SW}$ shifts the spanwise coordinate, which is convenient for imposing boundary conditions. We impose slip wall boundary conditions ($\hat{w}=0$) at $z=\pm0.5L_\text{eff}^\text{SW}$ \citep{fuchs_StandingWave_2025}, which yields permissible wavenumbers $\beta=\frac{2\pi n^\text{SW}}{L_\text{eff}^\text{SW}}$ with

\begin{equation}
    n^\text{SW} = \{0.5, 1, 1.5, 2, 2.5, \dots\}\,.
\end{equation}

\noindent
This model assumes that the hydrodynamic waves are reflected at the slip-wall boundaries without phase lag and resonate in the channel. We note that this model cannot satisfy no-slip boundary conditions, due to the phase relationships between the different components \citep{fuchs_StandingWave_2025}. In \citep{fuchs_StandingWave_2025}, the wind tunnel width is the natural choice for the effective spanwise length scale, as their backward facing ramp geometry is spanwise homogeneous and bounded by the wind tunnel side walls. In the present case, however, the spanwise profile of the bump and the highly three-dimensional flow enable multiple plausible choices for $L_\text{eff}^\text{SW}$. For the results presented here, we chose $L_\mathrm{eff}^\text{SW}=L$ as in \citet{fuchs_StandingWave_2025}, as the wind tunnel width represents the least ambiguous option. This leads to the slip wall boundary conditions being placed at the side walls. However, we do not intend to imply with this choice that the wind tunnel sidewalls are necessarily the dynamically relevant features governing the formation of the standing wave pattern. Alternative candidates include the bump shoulders or specific features of the mean flow, as discussed below. We note that for $L_\text{eff}^\text{SW}=L$ the wavenumber convention used in the previous sections ($n=\frac{\beta}{2\pi}$) is recovered.

\begin{figure}
    \centering
    \begin{subfigure}{0.5\textwidth}
        \includegraphics[width=\linewidth]{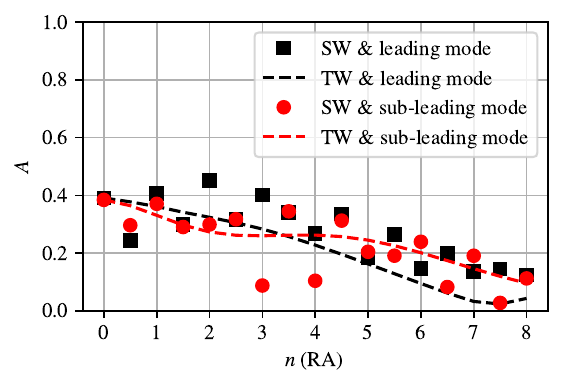}
        \caption{$\Rey=10^6$ (attached), $\Str_h=0.005$}
        \label{fig:SW_model_alignment_attached}
    \end{subfigure}\begin{subfigure}{0.5\textwidth}
        \includegraphics[width=\linewidth]{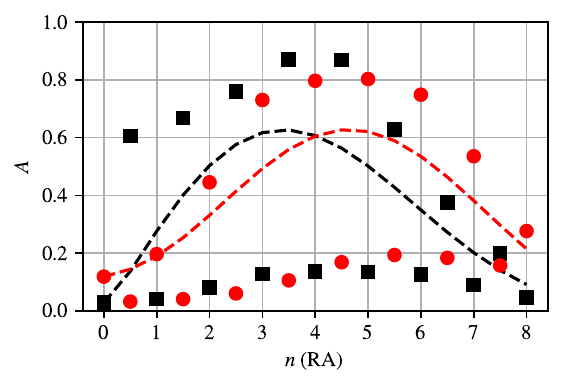}
        \caption{$\Rey=2\times10^6$ (separated), $\Str_h=0.002$}
        \label{fig:SW_model_alignment_sep}
    \end{subfigure}
    \caption{Alignment between the RA model and the low-frequency SPOD mode in the spanwise plane for the attached (a) and separated (b) cases. The markers show the alignment between standing wave (SW) model and SPOD mode, and the dashed lines show the alignment of the traveling wave (TW) mode for comparison. Black markers and curves show the alignment with the leading, and red markers and curves with the sub-leading SPOD mode.}
    \label{fig:SW_model_alignment}
\end{figure}

The alignment of both the travelling and standing wave models with the leading and sub-leading SPOD modes in the low-frequency regime is shown in figure~\ref{fig:SW_model_alignment} for different spanwise wavenumbers. Half-integer wavenumbers $n=\{0.5, 1.5, 2.5, \dots\}$ correspond to standing waves with a node for $\hat{u}$ on the centerline, whereas integer wavenumbers $n=\{1, 2, 3,\dots\}$ have an antinode at this position. At $n=0$, both the travelling wave and the standing wave reduce to a spanwise-constant mode.

In the attached case (see figure~\ref{fig:SW_model_alignment_attached}), the results are inconclusive and both the travelling wave modes and the standing wave model display low alignments with the spanwise SPOD modes. For modelling the leading SPOD mode, the standing wave model with wavenumbers $n=\{2, 3\}$ provides some improvement over the travelling wave modes. Yet, the respective alignments are still low, which is an expected result given the unclear structure of the SPOD modes (see figures~\ref{fig:standing_wave_demo_attached_leading} and \ref{fig:standing_wave_demo_attached_subleading}) and the low separation between the leading- and sub-leading modes in the respective SPOD spectrum (see figure~\ref{fig:piv_spod}).
In the separated case, however, the standing wave model achieves far higher alignment with the SPOD mode compared to the traveling wave modes, provided that the wavenumber corresponds to the same node position (i.e. either integer or half-integer $n$). Both the standing wave pattern with a node at the centerline, observed in the leading SPOD mode, as well as the pattern with an antinode at the centerline, observed in the sub-leading SPOD mode, are well captured through the RA standing wave model.

The respective mode shapes, at the wavenumbers corresponding to the highest alignments, are displayed and compared to the respective SPOD modes in figure~\ref{fig:SW_SPOD_compare}. The reader is referred to the animated version of this figure in the supplementary materials. This visualisation demonstrates the capability of the model to capture the main characteristics of the mode shape. This is particularly the case for the $\hat{u}$ component that contains most of the modes' energy. The dashed lines show how a standing wave structure with the selected spanwise wave number fulfils the slip wall condition at the side walls. However, there are some differences between the RA and SPOD mode shapes: Most notably, the $\hat{u}$ component of the SPOD mode displays a roughly triangular shape with vertices $(z,y)$ around $(\pm 0.125, 0.025)$ and $(0, 0.1)$, which is antisymmetric with respect to the node at the centerline but not symmetric with respect to the antinodes. The RA modes, on the other hand, are both antisymmetric with respect to the nodes and symmetric with respect to the antinodes, which is a consequence of the spanwise periodic ansatz for the RA modes (see equation~\ref{eq:ansatz}). Yet, alignments of $A=0.87$ and $0.80$, for the leading and sub-leading modes, respectively, are obtained with this model, demonstrating its applicability to the present flow.

The relevant spanwise length scale for the standing wave dynamics remains unclear, as the available data is insufficient to directly determine this value. If a different effective spanwise length scale is considered, the modes are simply represented by different wavenumbers based on that scale. Nevertheless, the modal structure provides a speculative argument suggesting that, if the standing waves result from a single effective spanwise length scale, this value would have to be relatively large. From the mode shapes in figures~\ref{fig:standing_wave_demo_separated_leading} and~\ref{fig:standing_wave_demo_separated_subleading}, we know that the leading mode requires a half-integer wavenumber, whereas the sub-leading mode requires an integer wavenumber in order to satisfy the node and antinode at $z=0$, respectively, under the assumed standing-wave ansatz. We can also deduce that the ratio of spanwise wavelengths (leading to sub-leading) is roughly $1.5$. If the effective spanwise length scale was close to the spanwise extend of the separation bubble, approximately $0.18L$, the leading mode would best be represented by $n=0.5$ and the sub-leading mode by $n=1$ (the wavelength of the sub-leading mode approximately coincides with the bubble width). This would give a wavelength ratio of $1/0.5=2$, which is in conflict with the data. Hence, the modes more likely correspond to higher wavenumbers on a larger length scale. For example, $n=1.5$ and $n=2$ gives a wavelength ratio of  $1.33$, which is in much better agreement with the data. This observation suggests that the effective spanwise length scale is at least two times the spanwise wavelength of the sub-leading mode, that is $L_\text{eff}^\text{SW}\geq 0.36$.

The prominence of spanwise standing-wave patterns in the low-frequency regime, similar to the findings by \citet{fuchs_StandingWave_2025}, has broader implications for numerical simulations of separated flows. Many computational studies, for reasons of efficiency, assume periodic spanwise boundary conditions and simulate only a thin spanwise section of the flow (e.g. \citet{uzun_Direct_2025, balin_Direct_2021}). These modelling choices can inadvertently filter out low-wavenumber modes, particularly those associated with large-scale breathing behavior. Moreover, periodic boundary conditions restrict admissible spanwise wavenumbers to integer multiples of the fundamental wavenumber, excluding half-integer standing-wave patterns, which may arise from reflection of hydrodynamic waves from the channel side walls in configurations with finite span, and in the present case, account for up to 75 \% of the PSD at low-frequencies as measured in the spanwise SPIV (see figure~\ref{fig:piv_spod}). As a result, the low-frequency dynamics of TSBs may be underrepresented or misrepresented in such simulations, potentially affecting both qualitative understanding and quantitative model validation.

\begin{figure}
    \centering
    \begin{subfigure}{\textwidth}
        \centering
        \includegraphics[width=0.8\linewidth]{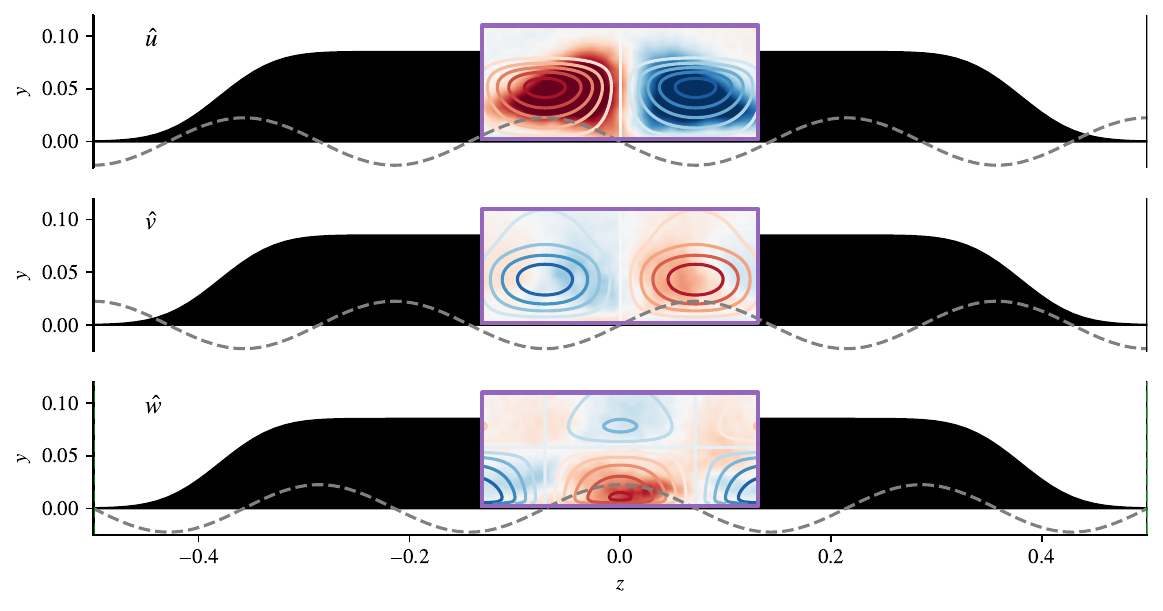}
        \caption{RA mode at $n=3.5$, leading SPOD mode, $A=0.87$}
    \end{subfigure}\vspace{0.5cm}\\
    \begin{subfigure}{\textwidth}
        \centering
        \includegraphics[width=0.8\linewidth]{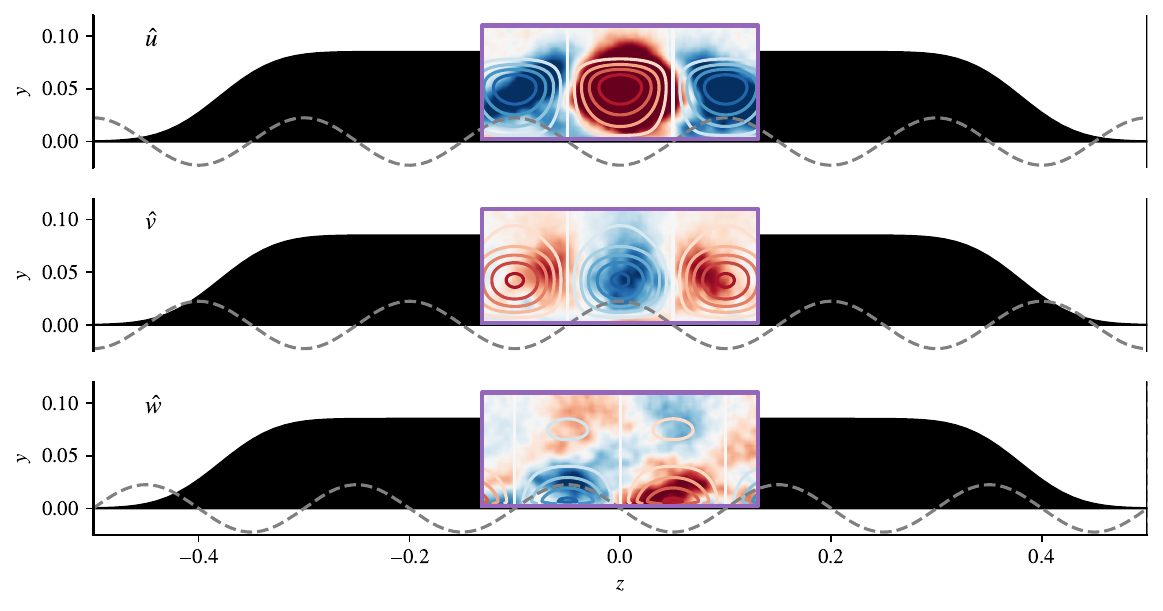}
        \caption{RA mode at $n=5$, sub-leading SPOD mode, $A=0.80$}
    \end{subfigure}
    \caption{Comparison of the standing wave RA model (contour lines) and SPOD mode (image behind contours) in the spanwise view of the bump. The leading SPOD mode is shown in (a) and the sub-leading SPOD mode is shown in (b). The colour-scale for the SPOD modes is clipped at $0.25$ times the respective maximum absolute value of $\hat{u}$ to highlight the structure of the other components, which are much smaller in comparison. Gray lines visualise how the standing wave systems fulfils slip-wall boundary conditions at the sidewalls: $\hat{u}$ and $\hat{v}$ have an antinode at sidewalls and $\hat{w}$ has a node at the sidewalls. $\Str=0.002$, $\Rey=2\times 10^6$ (separated). The reader is referred to the animated version of the figure.}
    \label{fig:SW_SPOD_compare}
\end{figure}

\section{Conclusions}
\label{sec:conclusions}
We have investigated the dynamics of attached and separated turbulent flows over the Boeing Gaussian Bump using both data-driven and physics-based approaches. Both flows exhibit coherent low-frequency dynamics characterized by streamwise-elongated streaky structures. Notably, these low-frequency structures are not a unique signature of fully separated flow, challenging the conventional view that they arise solely from separation bubble \textit{breathing}. Instead, they are already present in an attached flow state, so they may be a precursor or indicator of incipient separation.

In the separated flow, the dynamics are clearly modal (in the global sense), associated with a three-dimensional zero-frequency eigenmode that likely corresponds to a centrifugal instability. The instability produces a prominent spanwise-standing-wave pattern, which dominates the very-low-frequency spectral content. In contrast, the attached flow's dynamics show no evidence of a modal origin, and a non-modal amplification mechanism underlying the coherent streaky structures seems more likely. This delineation connects the Gaussian bump to recent work on APG-induced TSB breathing mechanisms. However, our results do not justify a definitive classification of the attached flow's low-frequency dynamics as either modal or non-modal. Compared to the separated flow, these dynamics are weaker, and no standing-wave structure emerges.

The strong modal dynamics in the separated case help explain persistent challenges in simulating this flow, even with high-fidelity computational fluid dynamics. The most prominent low-frequency dynamics are associated with spanwise wavelengths corresponding to roughly $20\,\%$ to $30\,\%$ of the wind tunnel width, exceeding the periodic span of many simulations. Additionally, the use of spanwise-periodic boundary conditions further restricts the resolved spanwise wavenumbers to integer multiples of the fundamental wavenumber. Small spanwise extents and periodic boundary conditions can therefore exclude the dominant low-frequency modes, especially half-integer standing-wave patterns resulting from sidewall reflections. These findings offer an explanation for discrepancies between simulations on spanwise-periodic domains and experimental measurements, contributing to a better understanding of smooth-body separation and turbulent separation bubble dynamics. This study also highlights the potential of combining data assimilation with linear mean field methods to investigate flow dynamics using mean flow data from limited regions, an approach that can be applied to other flows. Open questions for future research include the mechanisms driving low-frequency streaky structures in the attached flow, the influence of spanwise confinement, and the full three-dimensional structure of the dynamics.

\backsection[Supplementary movies]{\label{SupMat}[link to supplementary movies]}

\backsection[Acknowledgements]{The authors gratefully acknowledge Patrick Gray for providing the PIV snapshot data from the Smooth Body Separation Experiment, which was essential for validating the present analysis. The authors also thank Prahladh S. Iyer for providing the wall skin friction data from their three-dimensional LES.}

\backsection[Funding]{\begin{sloppy}
Funded by the Deutsche Forschungsgemeinschaft (DFG, German Research Foundation) - 506170981, 504349109. This work has been supported by the German Academic Exchange Service (DAAD).
\end{sloppy}}

\backsection[Declaration of interests]{The authors report no conflict of interest.}

\backsection[Data availability statement]{The data from the experiment at the University of Notre Dame are openly available in the NASA Turbulence Modeling Resource at \url{https://tmbwg.github.io/turbmodels/Other_exp_Data/speedbump_sep_exp.html}. The data-assimilated mean flows are available at \url{https://git.tu-berlin.de/pida-group}. The code used for the linearized analyses is available at \url{https://gitlab.com/felics-group/FELiCS}.}

\backsection[Author ORCIDs]{\\
R. Klopsch, \url{https://orcid.org/0000-0003-0414-2471}\\
L. Fuchs, \url{https://orcid.org/0009-0000-5068-7574}\\
G. Rigas, \url{https://orcid.org/0000-0001-6692-6437}\\
K. Oberleithner, \url{https://orcid.org/0000-0003-0964-872X}\\
J.G.R. von Saldern, \url{https://orcid.org/0000-0001-5003-8195}
}

\backsection[Author contributions]{This work builds upon R.K.’s Master’s thesis. R.K.: methodology, investigation, analysis, software, visualization, writing – original draft, writing – review \& editing. L.F.: methodology, analysis, writing – review \& editing. G.R.: supervision, writing – review \& editing. K.O.: supervision, funding acquisition, writing – review \& editing. J.v.S.: conceptualization, methodology, supervision, funding acquisition, writing – original draft, writing – review \& editing. All authors approved the final manuscript.}

\backsection[Use of artificial intelligence (AI) tools]{Artificial intelligence tools (DeepL and ChatGPT) were used solely for language refinement.}

\appendix
\section{Effects of discounting resolvent analysis for the separated case}
\label{sec:nondiscount_ra}
\begin{figure}
    \centering
    \begin{subfigure}{0.5\textwidth}
        \includegraphics[width=\textwidth]{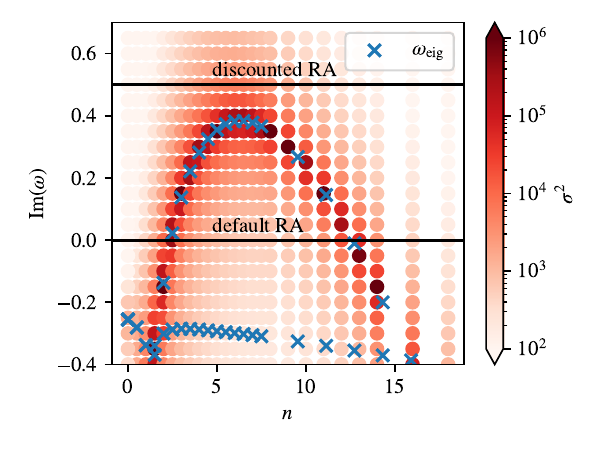}
        \caption{Pseudospectrum at $\Str_h=0$}
        \label{fig:separated_pseudospectrum}
    \end{subfigure}\begin{subfigure}{0.5\textwidth}
        \includegraphics[width=\textwidth]{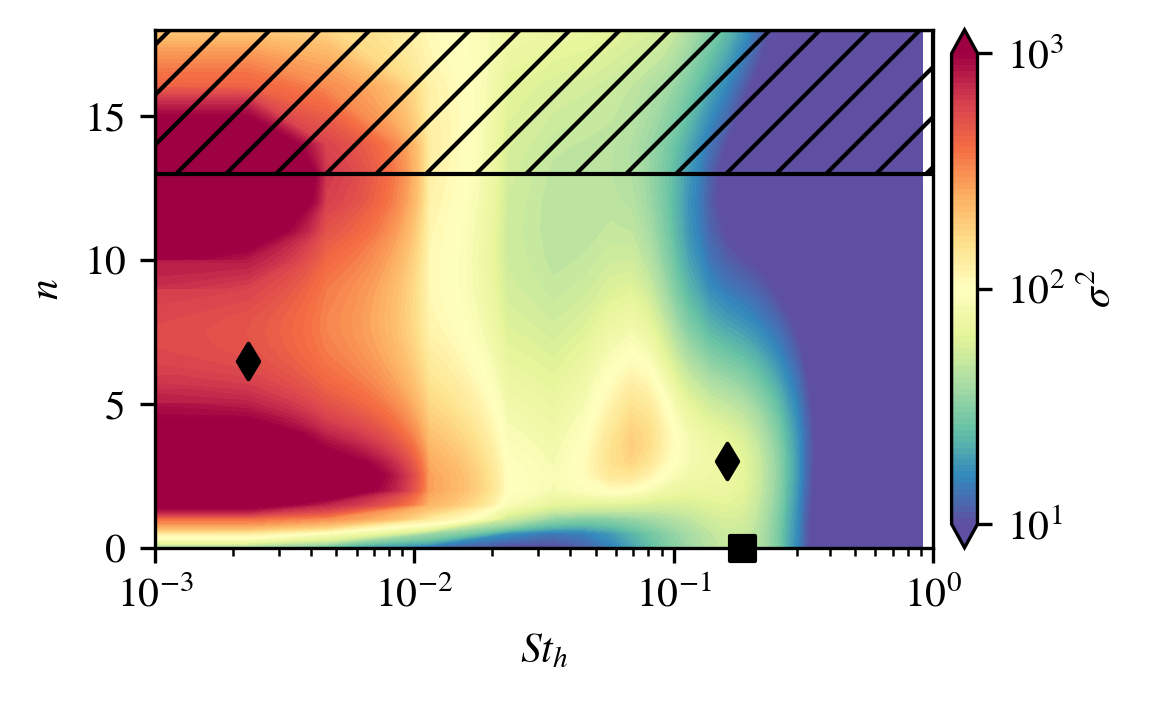}
        \caption{Non-discounted spectrum for $\mathrm{Im}(\omega)=0$}
        \label{fig:resolvent_gain_separated_non-discount}
    \end{subfigure}
    \caption{Relationship between resolvent gain and eigenvalues of the linear operator for $\Rey=2\times10^6$ (separated). Slice through the pseudospectrum at $St_h=0$ (a) and gain heatmap at $\mathrm{Im}(\omega)=0$ (b).}
\end{figure}

To highlight the relationship between RA gain and the eigenvalues of the linearized operator, and to demonstrate how discounting modifies the resulting spectrum, we compute a pseudospectrum slice (at a fixed frequency) of the resolvent operator for the separated case. In particular, we examine the slice at $\Str_h=0$, where the unstable eigenvalues are located. This pseudospectrum slice is shown in figure~\ref{fig:separated_pseudospectrum}. The default RA gain spectrum corresponds to evaluating the resolvent along the real axis (i.e. at $\mathrm{Im}(\omega)=0$) for each $\Str_h$. This spectrum is shown as a heatmap in figure~\ref{fig:resolvent_gain_separated_non-discount}, and clearly shows the imprint of singularities at zero frequency and at spanwise wavenumbers $n\approx2.5$ and $\approx12.5$, where the resolvent is evaluated very close to eigenvalues of the linear operator. These peaks obscure the underlying dependence on spanwise wavenumber at low frequencies. In contrast, the discounted RA gain spectrum shown in figure~\ref{fig:resolvent_gain_separated} is obtained by evaluating the resolvent at $\mathrm{Im}(\omega)=0.5$, which shifts the resolvent away from the eigenvalues and removes these singularities. The resulting gain distribution reveals a trend over $n$ that closely matches the LSA growth rate.

\begin{figure}
    \centering
    \begin{subfigure}{0.5\textwidth}
        \includegraphics[width=\linewidth]{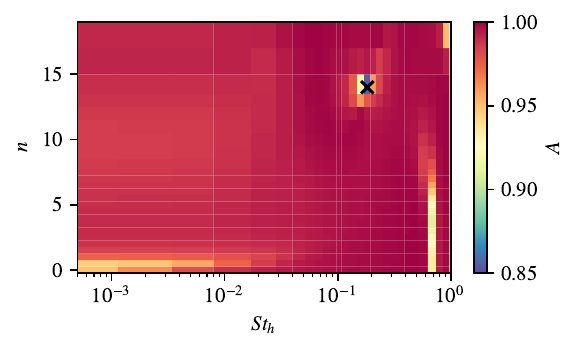}
        \caption{alignment heatmap}
        \label{fig:RA_discount_nondiscount_alignment_heatmap}
    \end{subfigure}\begin{subfigure}{0.5\textwidth}
        \includegraphics[width=\linewidth]{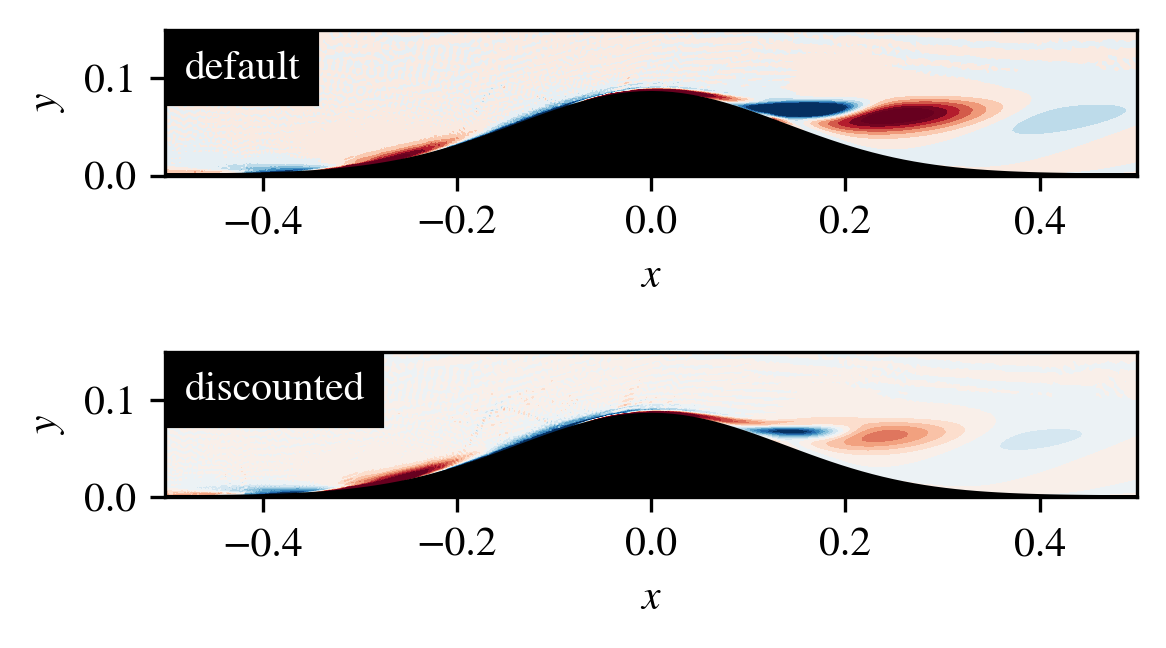}
        \caption{$\hat{u}$-component of mode pair with lowest alignment}
        \label{fig:RA_discount_nondiscount_mode_compare}
    \end{subfigure}
    \caption{Alignment between the modes from discounted ($\gamma=0.5$) and default ($\gamma=0$) RA (a) and comparison of the mode shapes with the lowest alignment (b). $\Rey=2\times10^6$ (separated).}
\end{figure}

Although discounting dramatically affects the RA gain spectrum, the mode shapes remain very similar. This is demonstrated through figure~\ref{fig:RA_discount_nondiscount_alignment_heatmap}, where the alignment between the modes from discounted ($\gamma=0.5$) and default ($\gamma=0$) RA is shown. The alignment was evaluated in the region $0\leq x\leq 0.5$ and $0\leq y \leq 0.2$, which corresponds to the primary region of interest in this study. It should be noted, that different values would be obtained, should the alignment be computed over a different region. For most $\Str_h$ and $n$, $A>0.98$, but there are some modes with lower values, down to $A\approx0.85$. Notably, these modes are not associated with a particularly high gain, and are therefore not of special interest. The mode pair with the lowest alignment (black marker in figure~\ref{fig:RA_discount_nondiscount_alignment_heatmap}) is shown in figure~\ref{fig:RA_discount_nondiscount_mode_compare} for comparison. Although some differences between the mode shapes are clearly identified, these would likely not change the interpretation of the mode, and we therefore conclude that the effect of discounting on the RA mode shapes is negligible in the present case.

\section{Surface pressure signature of resolvent modes}
\label{sec:surfp_ra}

In the experimental study of TSB flows, instantaneous surface pressure measurements are often applied to gain insight into the flow dynamics (e.g. \citet{gray_Experimental_2023a, weiss_Unsteady_2015, mohammed-taifour_Unsteadiness_2016}). We therefore assess to what extent coherent structures in the flow translate into surface pressure fluctuations. This is not immediately obvious: For example, we observe an additional regime of elevated gain in the separated case, around $n=4$ and $\Str_h=0.08$, which is slightly below half the characteristic frequency of the medium frequency regime in this case (see figure~\ref{fig:resolvent_gain_separated}). We refer to this as the intermediate regime. In this regime, the surface pressure SPOD yields comparatively low separation between the leading- and sub-leading modes (see figure~\ref{fig:SPOD_spectrum}); therefore, the observation of high RA gain is surprising.

\begin{figure}
    \centering
    \begin{subfigure}{0.5\textwidth}
        \includegraphics[width=\textwidth]{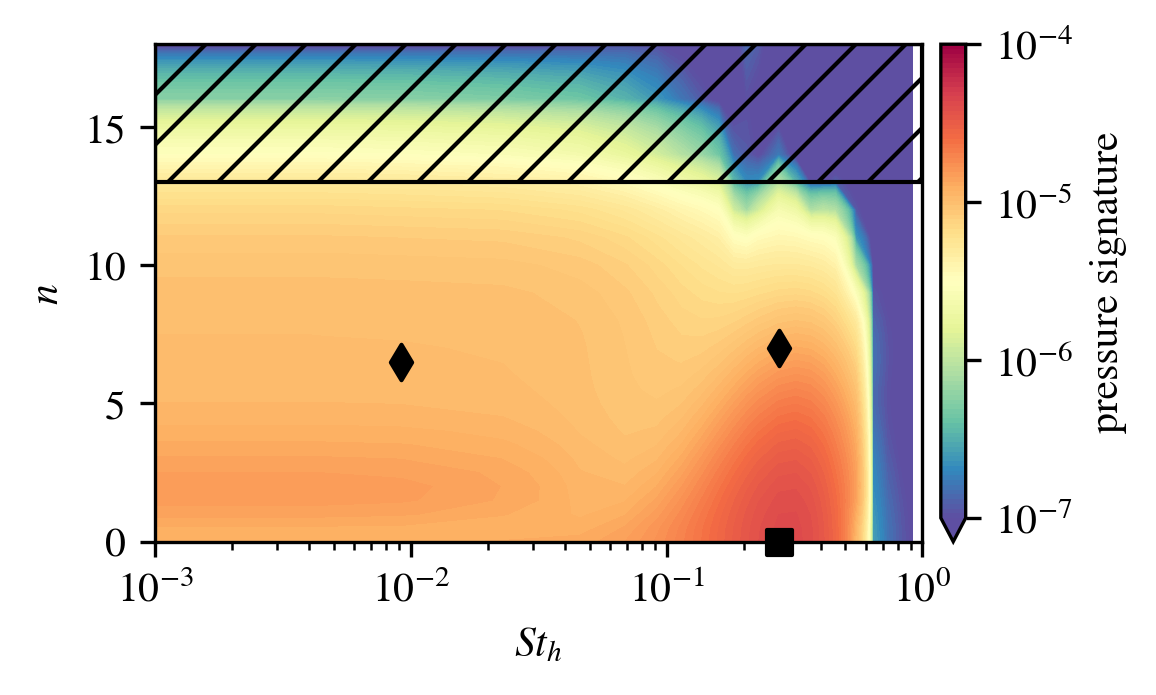}
        \caption{$\Rey=10^6$ (attached)}
    \end{subfigure}\begin{subfigure}{0.5\textwidth}
        \includegraphics[width=\textwidth]{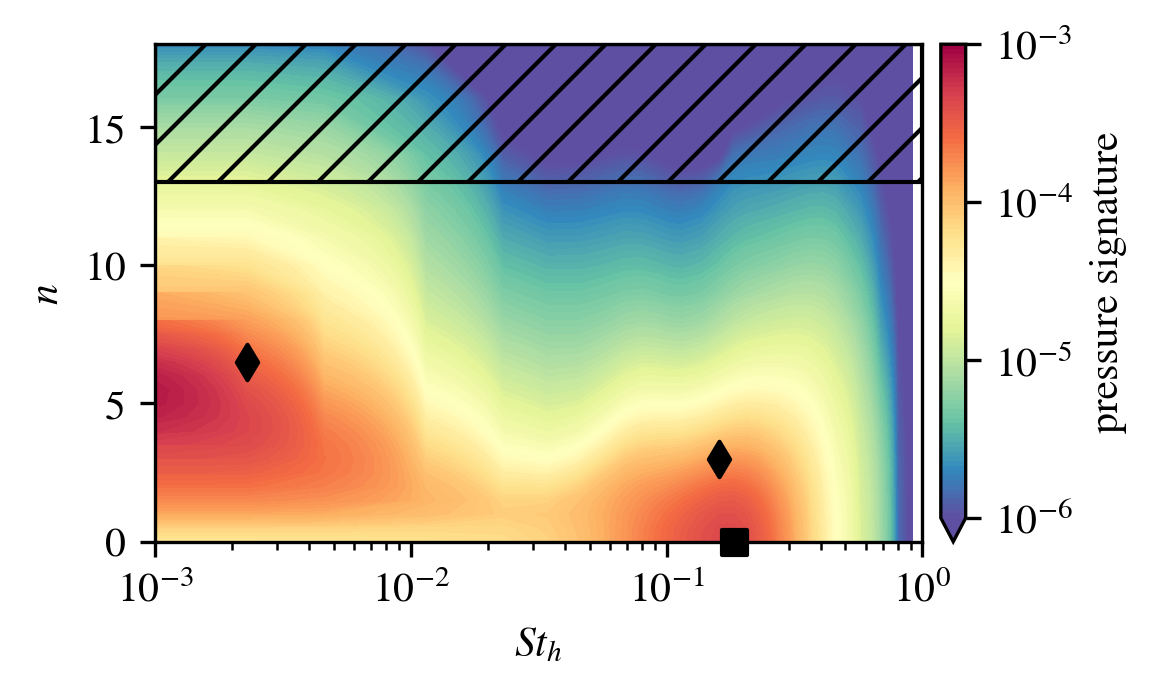}
        \caption{$\Rey=2\times10^6$ (separated)}
    \end{subfigure}
    \caption{Pressure signature of the resolvent modes at the bump surface as a function of Strouhal number and spanwise wavenumber for the attached (a) and separated cases (b), shown as a heatmap with a logarithmically scaled colourbar. Black markers indicate Strouhal number–spanwise wavenumber pairs discussed in the previous sections.}
    \label{fig:resolvent_surfp}
\end{figure}

In order to facilitate a direct comparison of the RA spectrum with the surface pressure SPOD spectra (figure ~\ref{fig:SPOD_spectrum}), the pressure signature of the RA modes is computed by extracting the squared magnitude of the pressure component at the $N$ sensor locations from the RA mode, and averaging over the number of sensors,

\begin{equation}
    \tilde{p} = \frac{\sum_{i=1}^N|\hat{p}|^2(\boldsymbol{x}_i)}{N}
\end{equation}
\noindent
where $|\hat{p}|$ is the magnitude of the pressure component of the resolvent mode. Note that the TKE norm of the resolvent mode corresponds to the gain value, $\sigma$, at the respective $\Str_h$ and $n$. The resulting pressure signature is shown in figure~\ref{fig:resolvent_surfp} as contours over $\Str_h$ and $n$. In the attached case, the pressure signature is most significant in the medium-frequency regime. Interestingly, whereas the RA gain increases with $n$ at this frequency, the response at lower $n$ induces the strongest pressure signature at the sensor locations.
In the separated case, the pressure signature displays pronounced low- and medium-frequency regimes. In the low-frequency regime, the pressure signature displays a similar dependence on the spanwise wavenumber as the resolvent gain, with a preferential $n\approx6.5$, and vanishing for $n\to0$. In the medium-frequency regime, the highest pressure signature is observed for $n=0$, a similar observation as in the attached case. Remarkably, the intermediate regime identified in the resolvent gain spectrum does not produce a distinct pressure imprint at the sensor locations.
This highlights a critical point: the modes most amplified (in terms of the resolvent gain, based on TKE) are not necessarily those most observable in the pressure near the wall, and thus may not always be captured through surface measurements alone.

\begin{figure}
    \centering
    \begin{subfigure}{0.5\textwidth}
        \includegraphics[width=\textwidth]{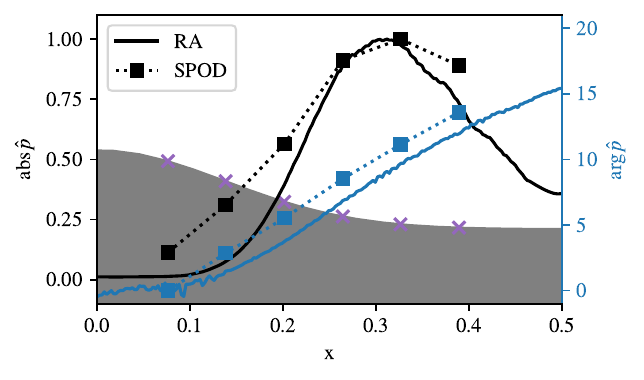}
        \caption{$\Rey=10^6$ (attached), $\Str_h=0.28$, $A=0.98$}
    \end{subfigure}\begin{subfigure}{0.5\textwidth}
        \includegraphics[width=\textwidth]{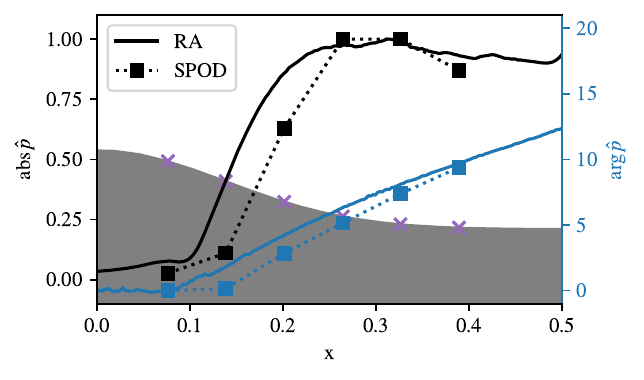}
        \caption{$\Rey=2\times10^6$ (separated), $\Str_h=0.18$, $A=0.9$}
    \end{subfigure}
    \caption{Comparison of coherent surface-pressure fluctuations from measurements extracted using SPOD and from RA modelling for the attached (a) and separated (b) cases. Black plots and the left axis show the magnitude, normalized by its respective maximum value, and blue plots and the right axis show the unrolled phase angle. The bump geometry and positions of the pressure sensors (purple crosses) are included for reference. RA results correspond to a spanwise wavenumber of $n=0$.}
    \label{fig:surf_pressure_modes}
\end{figure}

As a further step, we compare the shape of the modelled pressure fluctuation and the experimentally observed structures. As shown in figure~\ref{fig:resolvent_surfp}, the largest pressure response is in the medium-frequency regime and associated with low spanwise wavenumbers. We therefore focus on the two-dimensional limit $n=0$ for this analysis. Representative frequencies for the comparison were chosen according to the highest RA gain in the medium-frequency regime at $n=0$ for the respective case. These are indicated in the RA gain- and surface pressure spectra (figures~\ref{fig:resolvent_gain},~\ref{fig:resolvent_surfp}) through square markers.
Figure~\ref{fig:surf_pressure_modes} shows the comparison of the RA pressure response mode, extracted at the bump surface, with the leading mode from the surface pressure SPOD.
For presentation purposes, the amplitude is scaled by its maximum value and the phase angle is unwrapped and set to zero at the position of the first sensor. 
The figure demonstrates the qualitative similarity of the mode shapes.
A quantitative measure for the similarity of the modes is provided through the mode alignment.
The pressure modes shown in figure~\ref{fig:surf_pressure_modes} have alignment values of $A=0.98$ for the attached and $A=0.90$ for the separated case. These high alignments demonstrate the capability of RA to model the dominant coherent structure at this frequency.

\section{Alignment between zero-frequency resolvent- and eigenmodes}
\label{sec:all_lsa_vs_ra}

\begin{figure}
    \centering
    \begin{subfigure}{0.5\textwidth}
        \includegraphics[width=\textwidth]{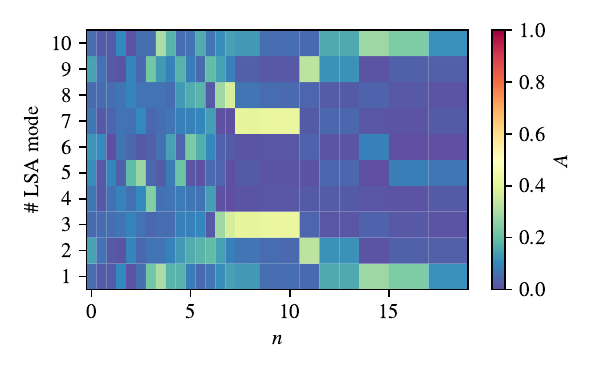}
        \caption{$\Rey=10^6$ (attached)}
    \end{subfigure}\begin{subfigure}{0.5\textwidth}
        \includegraphics[width=\textwidth]{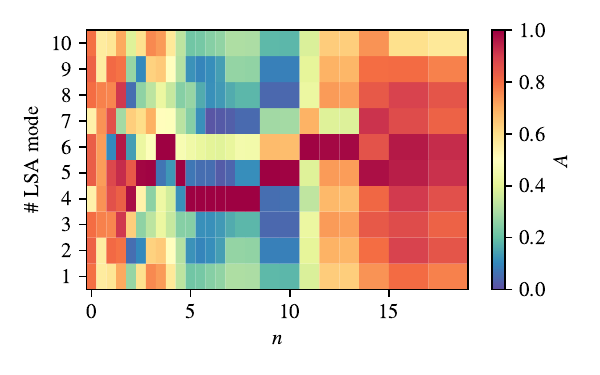}
        \caption{$\Rey=2\times10^6$ (separated)}
    \end{subfigure}
    \caption{Alignment between all LSA modes at $\Str_h\approx0$ and the RA mode at $\Str_h=0$ for the attached (a) and separated cases (b), shown as a heatmap over the spanwise wave number and LSA mode number.}
    \label{fig:zero-freq-ra-lsa-alignment}
\end{figure}

In the LSA spectrum of the attached case, we did not identify any eigenvalue branch with distinctly elevated growth rates. This suggests that the dynamics identified with RA for this case are non-modal. To further support this hypothesis, we assess whether the RA modes are associated with a specific eigenmode.

The LSA solver computes 10 eigenmodes with close-to-zero frequencies for each $n$. In order to assess their similarity with the RA mode computed at $\Str_h=0$, we compare the respective modes based on their alignment. Specifically, at each $n$, we compute the alignment between all LSA modes $(\Str_h\approx0,n)$ with the RA mode $(\Str_h=0, n)$. The result is shown in figure~\ref{fig:zero-freq-ra-lsa-alignment}. In the attached case, the alignment for all pairs is very low. We therefore conclude that the $\Str_h=0$ RA mode identified in this case does not correspond to an eigenmode of the system. In the separated case on the other hand, much higher alignment values are observed. In the range of approximately $2<n<15$, a single LSA mode is very highly aligned with the RA mode at the respective $n$. This range of spanwise wavenumbers corresponds to the same range where the corresponding eigenvalue branch has a distinctly higher growth rate compared to the remaining eigenvalues (see figure~\ref{fig:lsa_spectra}). For $n<2$ or $n>15$, there is no single eigenvalue with very high alignment. Instead, all eigenmodes are moderately aligned with the RA mode. We note that for these $n$, the growth rate of the respective eigenvalue is no longer elevated compared to the mode cloud (see figure~\ref{fig:lsa_spectra}).

\section{Effect of Reynolds number on the separated flow}
\label{sec:reynolds_invariance}
In their initial study, \citet{williams_Experimental_2020} reported a regime of approximate Reynolds number invariance of the separated mean flow. Later experimental \citep{gray_Experimental_2023a} and computational \citep{uzun_Direct_2025, iyer_Assessing_2023} studies mostly confirmed this observation. However, whereas the separation point and separated flow region are largely independent of the Reynolds number, differences in the boundary layer development have been reported \citep{uzun_Direct_2025}. In order to assess the influence of the Reynolds number on the separated flow dynamics, we consider an additional flow case at $\Rey=4\times10^6$. The data-assimilated mean flow for this additional case was obtained through the same procedure as for the other cases. However, it was not included in our previous publication \citep{klopsch_Enabling_2025} due to scope limitations.

\begin{figure}
    \centering
        \begin{subfigure}{0.5\textwidth}
            \includegraphics[width=\textwidth]{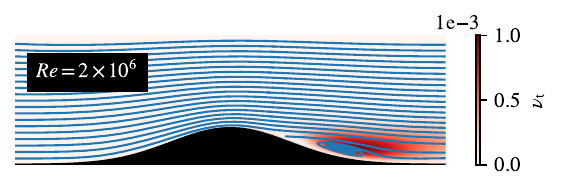}
            \includegraphics[width=\textwidth]{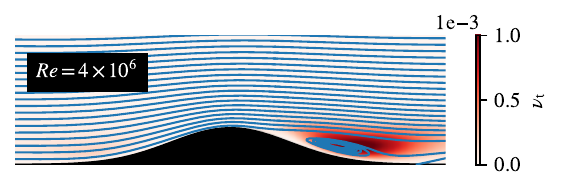}
            \caption{Comparison of mean flow streamlines and eddy viscosity contours between $\Rey=2\times10^6$ and $4\times10^6$.}
            \label{fig:reynolds_invariance_meanflow}
        \end{subfigure}\begin{subfigure}{0.5\textwidth}
            \includegraphics[width=\textwidth]{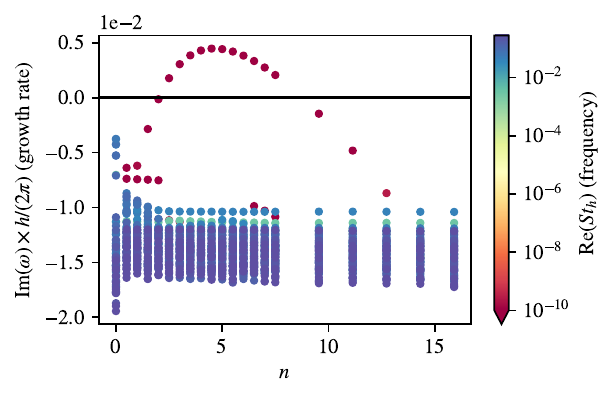}
            \caption{LSA eigenvalue spectrum for $\Rey=4\times10^6$.}
            \label{fig:reynolds_invariance_lsa}
        \end{subfigure}
        \begin{subfigure}{0.5\textwidth}
            \includegraphics[width=\textwidth]{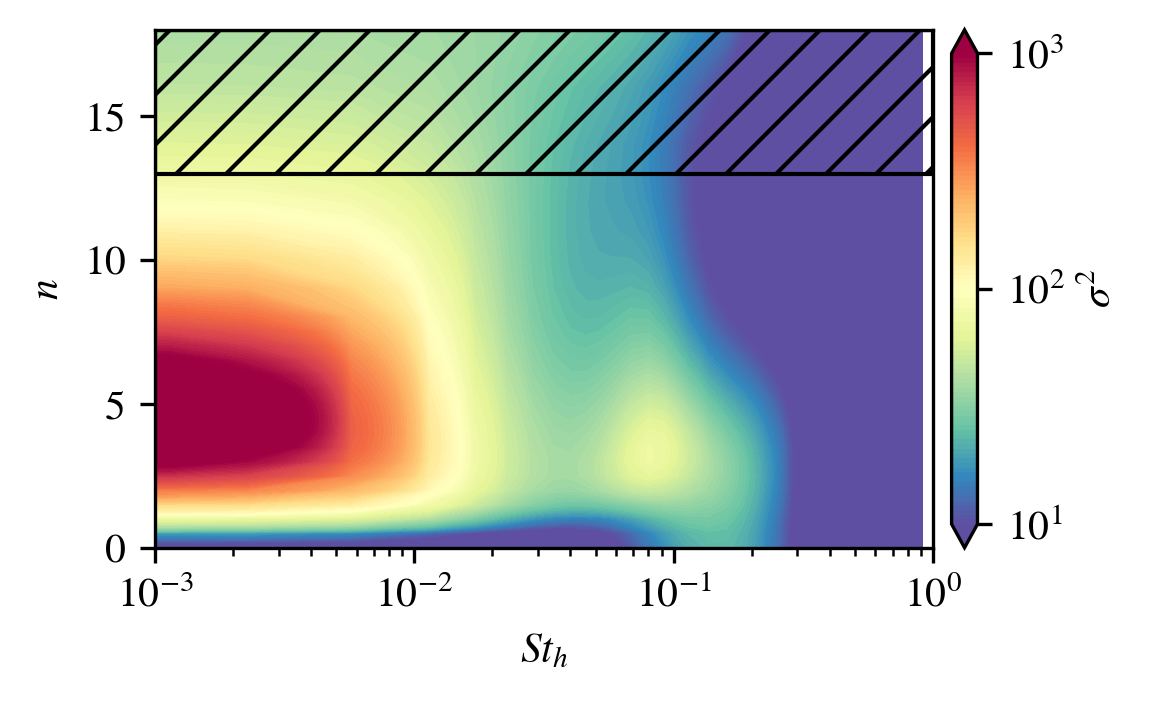}
            \caption{Resolvent gain heatmap for $\Rey=4\times10^6$.}
            \label{fig:reynolds_invariance_ra}
        \end{subfigure}\begin{subfigure}{0.5\textwidth}
            \includegraphics[width=\textwidth]{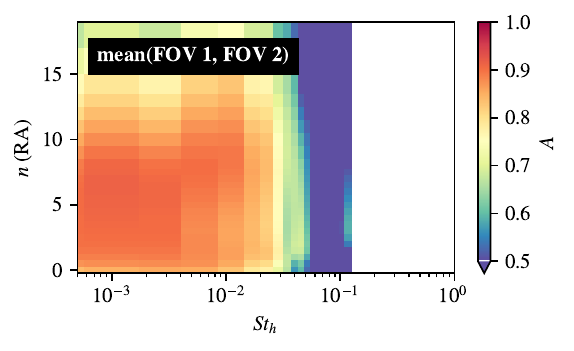}
            \caption{RA-SPOD alignment at $\Rey=4\times10^6$.}
            \label{fig:reynolds_invariance_alignment}
        \end{subfigure}

    \caption{Effect of increasing Reynolds number on the separated flow. (a) Comparison of mean flow streamlines and eddy viscosity contours between $\Rey=2\times10^6$ and $4\times10^6$, (b) LSA eigenvalue spectrum for $\Rey=4\times 10^6$, (c) Resolvent gain heatmap for $\mathrm{Re=4\times10^6}$, (d) and alignment between RA modes and SPOD modes at $\Rey=4\times10^6$.}
    \label{fig:reynolds_invariance}
\end{figure}

Mean flow streamlines and eddy viscosity contours of this additional case and the separated case at \mbox{$\Rey=2\times10^6$} are compared in figure~\ref{fig:reynolds_invariance_meanflow}. Note that the same contour levels for the eddy viscosity are chosen for both cases. Evidently, the mean flows are very similar, but there are some differences: In the additional case, the flow in the separated region is slightly less tilted towards the wall. Moreover, the extent of the region with high eddy viscosity is larger in this case. Nevertheless, the mean flows are very similar, supporting the approximate Reynolds number independence of the mean flow (outside the boundary layer) in this Reynolds number regime. We proceed by performing LSA and RA on this additional mean flow in order to investigate whether these small differences change the dynamics of the flow predicted by the linear models. The LSA eigenvalue spectrum is shown in figure~\ref{fig:reynolds_invariance_lsa}. Here we observe the same distinct branch of $\Str_h=0$ eigenvalues as in the separated case. However, the maximum growth rate occurs at slightly lower spanwise wavenumbers. The RA gain spectrum is shown in figure~\ref{fig:reynolds_invariance_ra}. The spectrum is very similar to the separated case, but the highest gain in the low-frequency regime is shifted to slightly lower spanwise wavenumbers, reflecting the trend observed in the LSA. Nevertheless, the results of the linear analyses are remarkably similar between this additional case and the separated case. To validate the RA modes for this additional case, we perform SPOD on streamwise PIV snapshot data for this case as well, and compute the alignment between the RA and respective SPOD modes. The resulting alignment is shown in figure~\ref{fig:reynolds_invariance_alignment}. For conciseness, the mean over both FOVs is shown. Very high alignments are observed in the low-frequency regime, demonstrating that this approximate Reynolds number independence is not an artifact of the linear models. Note that the Nyquist frequency of the SPOD is $\Str_h=0.11$ in this case, which excludes most of the medium-frequency regime. The validation is therefore limited to the low-frequency regime.  Based on this analysis we conclude that the dominant dynamics of the separated flow in the range $2\times10^6\leq\Rey\leq4\times10^6$ do not significantly depend on the Reynolds number.

\bibliographystyle{jfm}  
\bibliography{refs}

@inproceedings{iyer_Assessing_2023,
  title = {Assessing {{Reynolds Number Effects}} for Flow over a {{Gaussian}} Bump Using {{Wall-modeled LES}}},
  booktitle = {{{AIAA AVIATION}} 2023 {{Forum}}},
  author = {Iyer, Prahladh S. and Malik, Mujeeb R.},
  year = 2023,
  month = jun,
  publisher = {{American Institute of Aeronautics and Astronautics}},
  address = {San Diego, CA and Online},
  doi = {10.2514/6.2023-3982},
  urldate = {2026-02-05},
  isbn = {978-1-62410-704-7},
  langid = {english}
}

@article{rolandi_Biglobal_2025,
  title = {Biglobal Resolvent Analysis of Separated Flow over a {{NACA0012}} Airfoil},
  author = {Rolandi, Laura Victoria and Smith, Luke and Amitay, Michael and Theofilis, Vassilis and Taira, Kunihiko},
  year = 2025,
  month = oct,
  journal = {Journal of Fluid Mechanics},
  volume = {1021},
  pages = {A53},
  issn = {0022-1120, 1469-7645},
  doi = {10.1017/jfm.2025.10728},
  urldate = {2026-03-03},
  langid = {english}
}

@article{abe_Reynoldsnumber_2017,
  title = {Reynolds-Number Dependence of Wall-Pressure Fluctuations in a Pressure-Induced Turbulent Separation Bubble},
  author = {Abe, Hiroyuki},
  year = 2017,
  month = dec,
  journal = {Journal of Fluid Mechanics},
  volume = {833},
  pages = {563--598},
  issn = {0022-1120, 1469-7645},
  doi = {10.1017/jfm.2017.694},
  urldate = {2025-07-29},
  copyright = {https://www.cambridge.org/core/terms},
  langid = {english}
}

@article{balin_Direct_2021,
  title = {Direct Numerical Simulation of a Turbulent Boundary Layer over a Bump with Strong Pressure Gradients},
  author = {Balin, Riccardo and Jansen, K.E.},
  year = 2021,
  month = jul,
  journal = {Journal of Fluid Mechanics},
  volume = {918},
  pages = {A14},
  issn = {0022-1120, 1469-7645},
  doi = {10.1017/jfm.2021.312},
  urldate = {2025-03-27},
  langid = {english}
}

@article{barkley_Threedimensional_2002,
  title = {Three-Dimensional Instability in Flow over a Backward-Facing Step},
  author = {Barkley, Dwight and Gomes, M. Gabriela M. and Henderson, Ronald D.},
  year = 2002,
  month = dec,
  journal = {Journal of Fluid Mechanics},
  volume = {473},
  pages = {167--190},
  issn = {0022-1120, 1469-7645},
  doi = {10.1017/S002211200200232X},
  urldate = {2025-10-23},
  copyright = {https://www.cambridge.org/core/terms},
  langid = {english}
}

@article{borgmann_ThreeDimensional_2024,
  title = {Three-{{Dimensional Nature}} of {{Low-Frequency Unsteadiness}} in a {{Turbulent Separation Bubble}}},
  author = {Borgmann, David and Cura, Carolina and Weiss, Julien and Little, Jesse},
  year = 2024,
  month = sep,
  journal = {AIAA Journal},
  pages = {1--15},
  issn = {0001-1452, 1533-385X},
  doi = {10.2514/1.J064167},
  urldate = {2025-01-14},
  langid = {english}
}

@article{brandt_Liftup_2014,
  title = {The Lift-up Effect: {{The}} Linear Mechanism behind Transition and Turbulence in Shear Flows},
  shorttitle = {The Lift-up Effect},
  author = {Brandt, Luca},
  year = 2014,
  month = sep,
  journal = {European Journal of Mechanics - B/Fluids},
  volume = {47},
  pages = {80--96},
  issn = {09977546},
  doi = {10.1016/j.euromechflu.2014.03.005},
  urldate = {2025-10-02},
  langid = {english}
}

@article{cavalieri_Wavepackets_2013,
  title = {Wavepackets in the Velocity Field of Turbulent Jets},
  author = {Cavalieri, Andr{\'e} V. G. and Rodr{\'i}guez, Daniel and Jordan, Peter and Colonius, Tim and Gervais, Yves},
  year = 2013,
  month = sep,
  journal = {Journal of Fluid Mechanics},
  volume = {730},
  pages = {559--592},
  issn = {0022-1120, 1469-7645},
  doi = {10.1017/jfm.2013.346},
  urldate = {2025-08-01},
  copyright = {https://www.cambridge.org/core/terms},
  langid = {english}
}

@article{cherry_Unsteady_1984,
  title = {Unsteady Measurements in a Separated and Reattaching Flow},
  author = {Cherry, N. J. and Hillier, R. and Latour, M. E. M. P.},
  year = 1984,
  month = jul,
  journal = {Journal of Fluid Mechanics},
  volume = {144},
  pages = {13--46},
  issn = {0022-1120, 1469-7645},
  doi = {10.1017/S002211208400149X},
  urldate = {2025-04-22},
  copyright = {https://www.cambridge.org/core/terms},
  langid = {english}
}

@article{cura_Lowfrequency_2024,
  title = {On the Low-Frequency Dynamics of Turbulent Separation Bubbles},
  author = {Cura, C. and Hanifi, A. and Cavalieri, A.V.G. and Weiss, J.},
  year = 2024,
  month = jul,
  journal = {Journal of Fluid Mechanics},
  volume = {991},
  pages = {A11},
  issn = {0022-1120, 1469-7645},
  doi = {10.1017/jfm.2024.532},
  urldate = {2025-04-22},
  langid = {english}
}

@article{delery_Shock_1985,
  title = {Shock Wave/Turbulent Boundary Layer Interaction and Its Control},
  author = {Delery, Jean M.},
  year = 1985,
  journal = {Progress in Aerospace Sciences},
  volume = {22},
  number = {4},
  pages = {209--280},
  issn = {03760421},
  doi = {10.1016/0376-0421(85)90001-6},
  urldate = {2025-07-29},
  copyright = {https://www.elsevier.com/tdm/userlicense/1.0/},
  langid = {english}
}

@article{dussauge_Unsteadiness_2006,
  title = {Unsteadiness in Shock Wave Boundary Layer Interactions with Separation},
  author = {Dussauge, Jean-Paul and Dupont, Pierre and Debi{\`e}ve, Jean-Fran{\c c}ois},
  year = 2006,
  month = mar,
  journal = {Aerospace Science and Technology},
  volume = {10},
  number = {2},
  pages = {85--91},
  issn = {12709638},
  doi = {10.1016/j.ast.2005.09.006},
  urldate = {2025-07-29},
  copyright = {https://www.elsevier.com/tdm/userlicense/1.0/},
  langid = {english}
}

@inproceedings{eaton_Low_1982,
  title = {Low {{Frequency Unsteadyness}} of a {{Reattaching Turbulent Shear Layer}}},
  booktitle = {Turbulent {{Shear Flows}} 3},
  author = {Eaton, John K. and Johnston, James P.},
  editor = {Bradbury, Leslie J. S. and Durst, Franz and Launder, Brian E. and Schmidt, Frank W. and Whitelaw, James H.},
  year = 1982,
  pages = {162--170},
  publisher = {Springer},
  address = {Berlin, Heidelberg},
  doi = {10.1007/978-3-642-95410-8_16},
  isbn = {978-3-642-95410-8},
  langid = {english}
}

@article{fang_Lowfrequency_2024,
  title = {On the Low-Frequency Flapping Motion in Flow Separation},
  author = {Fang, Xingjun and Wang, Zhan},
  year = 2024,
  month = apr,
  journal = {Journal of Fluid Mechanics},
  volume = {984},
  pages = {A76},
  issn = {0022-1120, 1469-7645},
  doi = {10.1017/jfm.2024.280},
  urldate = {2025-07-29},
  langid = {english}
}

@article{fuchs_StandingWave_2025,
    title={Standing-wave dynamics in low-frequency breathing of a turbulent separation bubble}, volume={1030},
    DOI={10.1017/jfm.2026.11191},
    journal={Journal of Fluid Mechanics},
    author={Fuchs, Lukas M. and Steinfurth, Ben and von Saldern, Jakob G.R. and Weiss, Julien and Oberleithner, Kilian}, year={2026}, pages={A34}}

@article{gallaire_Threedimensional_2007,
  title = {Three-Dimensional Transverse Instabilities in Detached Boundary Layers},
  author = {Gallaire, Fran{\c c}ois and Marquillie, Matthieu and Ehrenstein, Uwe},
  year = 2007,
  month = jan,
  journal = {Journal of Fluid Mechanics},
  volume = {571},
  pages = {221--233},
  issn = {0022-1120, 1469-7645},
  doi = {10.1017/S0022112006002898},
  urldate = {2024-12-19},
  copyright = {https://www.cambridge.org/core/terms},
  langid = {english}
}

@inproceedings{gray_Benchmark_2022,
  title = {Benchmark {{Characterization}} of {{Separated Flow Over Smooth Gaussian Bump}}},
  booktitle = {{{AIAA AVIATION}} 2022 {{Forum}}},
  author = {Gray, Patrick D. and Gluzman, Igal and Thomas, Flint O. and Corke, Thomas C. and Lakebrink, Matthew T. and Mejia, Kevin},
  year = 2022,
  month = jun,
  publisher = {{American Institute of Aeronautics and Astronautics}},
  address = {Chicago, IL \& Virtual},
  doi = {10.2514/6.2022-3342},
  urldate = {2024-11-21},
  isbn = {978-1-62410-635-4},
  langid = {english}
}

@inproceedings{gray_Experimental_2022,
  title = {Experimental {{Characterization}} of {{Smooth Body Flow Separation Over Wall-Mounted Gaussian Bump}}},
  booktitle = {{{AIAA SCITECH}} 2022 {{Forum}}},
  author = {Gray, Patrick D. and Gluzman, Igal and Thomas, Flint O. and Corke, Thomas C.},
  year = 2022,
  month = jan,
  publisher = {{American Institute of Aeronautics and Astronautics}},
  address = {San Diego, CA \& Virtual},
  doi = {10.2514/6.2022-1209},
  urldate = {2024-11-21},
  isbn = {978-1-62410-631-6},
  langid = {english}
}

@inproceedings{gray_Experimental_2023,
  title = {Experimental and {{Computational Evaluation}} of {{Smooth-Body Separated Flow}} over {{Boeing Bump}}},
  booktitle = {{{AIAA AVIATION}} 2023 {{Forum}}},
  author = {Gray, Patrick D. and Lakebrink, Matthew T. and Thomas, Flint O. and Corke, Thomas C. and Gluzman, Igal and Straccia, Joseph},
  year = 2023,
  month = jun,
  publisher = {{American Institute of Aeronautics and Astronautics}},
  address = {San Diego, CA and Online},
  doi = {10.2514/6.2023-3981},
  urldate = {2024-11-21},
  isbn = {978-1-62410-704-7},
  langid = {english}
}

@phdthesis{gray_Experimental_2023a,
  title = {An {{Experimental Investigation}} of {{Smooth-Body Separation}} over a {{Tapered Gaussian Bump}}},
  author = {Gray, Patrick},
  year = 2023,
  school = {University of Notre Dame}
}

@inproceedings{gray_New_2021,
  title = {A {{New Validation Experiment}} for {{Smooth-Body Separation}}},
  booktitle = {{{AIAA AVIATION}} 2021 {{FORUM}}},
  author = {Gray, Patrick D. and Gluzman, Igal and Thomas, Flint and Corke, Thomas and Lakebrink, Matthew and Mejia, Kevin},
  year = 2021,
  month = aug,
  publisher = {{American Institute of Aeronautics and Astronautics}},
  address = {VIRTUAL EVENT},
  doi = {10.2514/6.2021-2810},
  urldate = {2024-11-21},
  isbn = {978-1-62410-610-1},
  langid = {english}
}

@techreport{gray_Turbulence_2023,
  title = {Turbulence {{Model Validation Through Joint Experimental}} /{{Computational Studies}} of {{Separated Flow Over A Three-Dimensional Tapered Bump}}: {{Part I}} - {{Experimental Investigation}}},
  shorttitle = {Turbulence {{Model Validation Through Joint Experimental}} /{{Computational Studies}} of {{Separated Flow Over A Three-Dimensional Tapered Bump}}},
  author = {Gray, Patrick and Corke, Thomas and Thomas, Flint and Gluzman, Igal and Straccia, Joseph},
  year = 2023,
  month = jan,
  urldate = {2024-08-30},
  langid = {english}
}

@article{gudmundsson_Instability_2011,
  title = {Instability Wave Models for the Near-Field Fluctuations of Turbulent Jets},
  author = {Gudmundsson, K. and Colonius, Tim},
  year = 2011,
  month = dec,
  journal = {Journal of Fluid Mechanics},
  volume = {689},
  pages = {97--128},
  issn = {0022-1120, 1469-7645},
  doi = {10.1017/jfm.2011.401},
  urldate = {2025-08-01},
  copyright = {https://www.cambridge.org/core/terms},
  langid = {english}
}

@article{hao_Lowfrequency_2023,
  title = {On the Low-Frequency Unsteadiness in Shock Wave--Turbulent Boundary Layer Interactions},
  author = {Hao, Jiaao},
  year = 2023,
  month = sep,
  journal = {Journal of Fluid Mechanics},
  volume = {971},
  pages = {A28},
  issn = {0022-1120, 1469-7645},
  doi = {10.1017/jfm.2023.687},
  urldate = {2025-07-29},
  langid = {english}
}

@inproceedings{iyer_Wallmodeled_2023,
  title = {Wall-Modeled {{LES}} of the {{Three-dimensional Speed Bump Experiment}}},
  booktitle = {{{AIAA SCITECH}} 2023 {{Forum}}},
  author = {Iyer, Prahladh S. and Malik, Mujeeb R.},
  year = 2023,
  month = jan,
  publisher = {{American Institute of Aeronautics and Astronautics}},
  address = {National Harbor, MD \& Online},
  doi = {10.2514/6.2023-0253},
  urldate = {2025-07-31},
  isbn = {978-1-62410-699-6},
  langid = {english}
}

@phdthesis{jovanovic_Modeling_2004,
  title = {Modeling, {{Analysis}}, and {{Control}} of {{Spatially Distributed Systems}}},
  author = {Jovanovic, Mihailo},
  year = 2004,
  month = jan
}

@inproceedings{kaiser_FELiCS_2023,
  title = {{{FELiCS}}: {{A Versatile Linearized Solver Addressing Dynamics}} in {{Multi-Physics Flows}}},
  shorttitle = {{{FELiCS}}},
  booktitle = {{{AIAA AVIATION}} 2023 {{Forum}}},
author = {Thomas L. Kaiser and Simon Demange and Jens S. Müller and Sophie Knechtel and Kilian Oberleithner},
  year = 2023,
  month = jun,
  publisher = {{American Institute of Aeronautics and Astronautics}},
  address = {San Diego, CA and Online},
  doi = {10.2514/6.2023-3434},
  urldate = {2024-11-20},
  isbn = {978-1-62410-704-7},
  langid = {english}
}

@article{kaltenbach_Study_1999,
  title = {Study of Flow in a Planar Asymmetric Diffuser Using Large-Eddy Simulation},
  author = {Kaltenbach, H.-J. and Fatica, M. and Mittal, R. and Lund, T. S. and Moin, P.},
  year = 1999,
  month = jul,
  journal = {Journal of Fluid Mechanics},
  volume = {390},
  pages = {151--185},
  issn = {0022-1120, 1469-7645},
  doi = {10.1017/S0022112099005054},
  urldate = {2025-07-28},
  copyright = {https://www.cambridge.org/core/terms},
  langid = {english}
}

@article{kiya_Structure_1983,
  title = {Structure of a Turbulent Separation Bubble},
  author = {Kiya, Masaru and Sasaki, Kyuro},
  year = 1983,
  month = dec,
  journal = {Journal of Fluid Mechanics},
  volume = {137},
  pages = {83--113},
  issn = {0022-1120, 1469-7645},
  doi = {10.1017/S002211208300230X},
  urldate = {2025-04-22},
  copyright = {https://www.cambridge.org/core/terms},
  langid = {english}
}

@inproceedings{klopsch_Enabling_2025,
  title = {Enabling {{Resolvent Analysis Through Assimilation}} of {{Experimental Mean Flows}} with {{Physics-Informed Neural Networks}}: {{A Case Study}} on the {{Boeing Gaussian Bump}}},
  shorttitle = {Enabling {{Resolvent Analysis Through Assimilation}} of {{Experimental Mean Flows}} with {{Physics-Informed Neural Networks}}},
  booktitle = {{{AIAA AVIATION FORUM AND ASCEND}} 2025},
  author = {Klopsch, Roman and Fuchs, Lukas M. and Rigas, Georgios and Oberleithner, Kilian and Von Saldern, Jakob G.},
  year = 2025,
  month = jul,
  publisher = {{American Institute of Aeronautics and Astronautics}},
  address = {Las Vegas, Nevada},
  doi = {10.2514/6.2025-3604},
  urldate = {2025-07-28},
  isbn = {978-1-62410-738-2},
  langid = {english}
}

@inproceedings{kuhn_Influence_2022,
  title = {Influence of {{Eddy Viscosity}} on {{Linear Modeling}} of {{Self-Similar Coherent Structures}} in the {{Jet Far Field}}},
  booktitle = {{{AIAA SCITECH}} 2022 {{Forum}}},
  author = {Kuhn, Phoebe and M{\"u}ller, Jens S. and Knechtel, Sophie and Soria, Julio and Oberleithner, Kilian},
  year = 2022,
  month = jan,
  publisher = {{American Institute of Aeronautics and Astronautics}},
  address = {San Diego, CA \& Virtual},
  doi = {10.2514/6.2022-0460},
  urldate = {2025-05-16},
  isbn = {978-1-62410-631-6},
  langid = {english}
}

@article{landahl_Note_1980,
  title = {A Note on an Algebraic Instability of Inviscid Parallel Shear Flows},
  author = {Landahl, M. T.},
  year = 1980,
  month = may,
  journal = {Journal of Fluid Mechanics},
  volume = {98},
  number = {2},
  pages = {243--251},
  issn = {0022-1120, 1469-7645},
  doi = {10.1017/S0022112080000122},
  urldate = {2025-10-02},
  copyright = {https://www.cambridge.org/core/terms},
  langid = {english}
}

@article{largeau_Wall_2006,
  title = {Wall Pressure Fluctuations and Topology in Separated Flows over a Forward-Facing Step},
  author = {Largeau, J. F. and Moriniere, V.},
  year = 2006,
  month = dec,
  journal = {Experiments in Fluids},
  volume = {42},
  number = {1},
  pages = {21--40},
  issn = {0723-4864, 1432-1114},
  doi = {10.1007/s00348-006-0215-9},
  urldate = {2025-07-29},
  copyright = {http://www.springer.com/tdm},
  langid = {english}
}

@article{lesshafft_Resolventbased_2019,
  title = {Resolvent-Based Modeling of Coherent Wave Packets in a Turbulent Jet},
  author = {Lesshafft, Lutz and Semeraro, Onofrio and Jaunet, Vincent and Cavalieri, Andr{\'e} V. G. and Jordan, Peter},
  year = 2019,
  month = jun,
  journal = {Physical Review Fluids},
  volume = {4},
  number = {6},
  pages = {063901},
  issn = {2469-990X},
  doi = {10.1103/PhysRevFluids.4.063901},
  urldate = {2025-10-24},
  langid = {english}
}

@book{lumley_Stochastik_1970,
  title = {Stochastik {{Tools}} in {{Turbulence}}},
  author = {Lumley, John L.},
  year = 1970,
  series = {Applied {{Mathematics}} and {{Mechanics}}},
  number = {12},
  publisher = {Academic Press},
  address = {New York}
}

@article{mabey_Analysis_1972,
  title = {Analysis and {{Correlation}} of {{Data}} on {{Pressure Fluctuations}} in {{Separated Flow}}},
  author = {Mabey, Dennis G.},
  year = 1972,
  month = sep,
  journal = {Journal of Aircraft},
  volume = {9},
  number = {9},
  pages = {642--645},
  issn = {0021-8669, 1533-3868},
  doi = {10.2514/3.59053},
  urldate = {2025-07-29},
  langid = {english}
}

@article{manohar_Temporal_2023,
  title = {Temporal Super-Resolution Using Smart Sensors for Turbulent Separated Flows},
  author = {Manohar, Kevin H. and Williams, Owen and Martinuzzi, Robert J. and Morton, Chris},
  year = 2023,
  month = may,
  journal = {Experiments in Fluids},
  volume = {64},
  number = {5},
  pages = {101},
  issn = {0723-4864, 1432-1114},
  doi = {10.1007/s00348-023-03639-2},
  urldate = {2024-12-19},
  langid = {english}
}

@article{mckeon_Criticallayer_2010,
  title = {A Critical-Layer Framework for Turbulent Pipe Flow},
  author = {McKeon, B. J. and Sharma, A. S.},
  year = 2010,
  month = sep,
  journal = {Journal of Fluid Mechanics},
  volume = {658},
  pages = {336--382},
  issn = {0022-1120, 1469-7645},
  doi = {10.1017/S002211201000176X},
  urldate = {2024-12-18},
  copyright = {https://www.cambridge.org/core/terms},
  langid = {english}
}

@article{mengaldo_PySPOD_2021,
  title = {{{PySPOD}}: {{A Python}} Package for {{Spectral Proper Orthogonal Decomposition}} ({{SPOD}})},
  shorttitle = {{{PySPOD}}},
  author = {Mengaldo, Gianmarco and Maulik, Romit},
  year = 2021,
  month = apr,
  journal = {Journal of Open Source Software},
  volume = {6},
  number = {60},
  pages = {2862},
  issn = {2475-9066},
  doi = {10.21105/joss.02862},
  urldate = {2025-01-13},
  langid = {english}
}

@article{mohammed-taifour_Unsteadiness_2016,
  title = {Unsteadiness in a Large Turbulent Separation Bubble},
  author = {{Mohammed-Taifour}, Abdelouahab and Weiss, Julien},
  year = 2016,
  month = jul,
  journal = {Journal of Fluid Mechanics},
  volume = {799},
  pages = {383--412},
  issn = {0022-1120, 1469-7645},
  doi = {10.1017/jfm.2016.377},
  urldate = {2025-01-23},
  copyright = {https://www.cambridge.org/core/terms},
  langid = {english}
}

@article{muller_Linear_2024,
  title = {Linear Amplification of Inertial-Wave-Driven Swirl Fluctuations in Turbulent Swirling Pipe Flows: A Resolvent Analysis Approach},
  shorttitle = {Linear Amplification of Inertial-Wave-Driven Swirl Fluctuations in Turbulent Swirling Pipe Flows},
  author = {M{\"u}ller, J.S. and Von Saldern, J.G.R. and Kaiser, T.L. and Oberleithner, K.},
  year = 2024,
  month = dec,
  journal = {Journal of Fluid Mechanics},
  volume = {1000},
  pages = {A91},
  issn = {0022-1120, 1469-7645},
  doi = {10.1017/jfm.2024.679},
  urldate = {2025-09-19},
  langid = {english}
}

@article{na_Direct_1998,
  title = {Direct Numerical Simulation of a Separated Turbulent Boundary Layer},
  author = {Na, Y. and Moin, P.},
  year = 1998,
  month = nov,
  journal = {Journal of Fluid Mechanics},
  volume = {374},
  pages = {379--405},
  issn = {0022-1120, 1469-7645},
  doi = {10.1017/S002211209800189X},
  urldate = {2025-07-29},
  copyright = {https://www.cambridge.org/core/terms},
  langid = {english}
}

@techreport{patrick_Flowfield_1987,
  title = {Flowfield Measurements in a Separated and Reattached Flat Plate Turbulent Boundary Layer},
  author = {Patrick, William P.},
  year = 1987,
  month = mar,
  number = {NAS 1.26:4052},
  urldate = {2025-07-29}
}

@article{pearson_Turbulent_2013,
  title = {Turbulent Separation Upstream of a Forward-Facing Step},
  author = {Pearson, D. S. and Goulart, P. J. and Ganapathisubramani, B.},
  year = 2013,
  month = jun,
  journal = {Journal of Fluid Mechanics},
  volume = {724},
  pages = {284--304},
  issn = {0022-1120, 1469-7645},
  doi = {10.1017/jfm.2013.113},
  urldate = {2025-07-29},
  copyright = {https://www.cambridge.org/core/terms},
  langid = {english}
}

@article{pickering_Liftup_2020,
  title = {Lift-up, {{Kelvin}}--{{Helmholtz}} and {{Orr}} Mechanisms in Turbulent Jets},
  author = {Pickering, Ethan and Rigas, Georgios and Nogueira, Petr{\^o}nio A. S. and Cavalieri, Andr{\'e} V. G. and Schmidt, Oliver T. and Colonius, Tim},
  year = 2020,
  month = aug,
  journal = {Journal of Fluid Mechanics},
  volume = {896},
  pages = {A2},
  issn = {0022-1120, 1469-7645},
  doi = {10.1017/jfm.2020.301},
  urldate = {2024-12-19},
  copyright = {https://www.cambridge.org/core/terms},
  langid = {english}
}

@article{pickering_Optimal_2021,
  title = {Optimal Eddy Viscosity for Resolvent-Based Models of Coherent Structures in Turbulent Jets},
  author = {Pickering, Ethan and Rigas, Georgios and Schmidt, Oliver T. and Sipp, Denis and Colonius, Tim},
  year = 2021,
  month = jun,
  journal = {Journal of Fluid Mechanics},
  volume = {917},
  pages = {A29},
  issn = {0022-1120, 1469-7645},
  doi = {10.1017/jfm.2021.232},
  urldate = {2024-12-19},
  langid = {english}
}

@article{poggie_Spectral_2015,
  title = {Spectral {{Characteristics}} of {{Separation Shock Unsteadiness}}},
  author = {Poggie, Jonathan and Bisek, Nicholas J. and Kimmel, Roger L. and Stanfield, Scott A.},
  year = 2015,
  month = jan,
  journal = {AIAA Journal},
  volume = {53},
  number = {1},
  pages = {200--214},
  issn = {0001-1452, 1533-385X},
  doi = {10.2514/1.J053029},
  urldate = {2025-07-29},
  langid = {english}
}

@article{reynolds_Mechanics_1972,
  title = {The Mechanics of an Organized Wave in Turbulent Shear Flow. {{Part}} 3. {{Theoretical}} Models and Comparisons with Experiments},
  author = {Reynolds, W. C. and Hussain, A. K. M. F.},
  year = 1972,
  month = jul,
  journal = {Journal of Fluid Mechanics},
  volume = {54},
  number = {2},
  pages = {263--288},
  issn = {1469-7645, 0022-1120},
  doi = {10.1017/S0022112072000679},
  urldate = {2025-04-24},
  langid = {english}
}

@article{rodriguez_Two_2013,
  title = {The Two Classes of Primary Modal Instability in Laminar Separation Bubbles},
  author = {Rodr{\'i}guez, Daniel and Gennaro, Elmer M. and Juniper, Matthew P.},
  year = 2013,
  month = nov,
  journal = {Journal of Fluid Mechanics},
  volume = {734},
  pages = {R4},
  issn = {0022-1120, 1469-7645},
  doi = {10.1017/jfm.2013.504},
  urldate = {2025-08-21},
  copyright = {https://www.cambridge.org/core/terms},
  langid = {english}
}

@article{rolandi_Invitation_2024,
  title = {An Invitation to Resolvent Analysis},
  author = {Rolandi, Laura Victoria and Ribeiro, Jean H{\'e}lder Marques and Yeh, Chi-An and Taira, Kunihiko},
  year = 2024,
  month = oct,
  journal = {Theoretical and Computational Fluid Dynamics},
  volume = {38},
  number = {5},
  pages = {603--639},
  issn = {0935-4964, 1432-2250},
  doi = {10.1007/s00162-024-00717-x},
  urldate = {2024-12-19},
  langid = {english}
}

@article{rukes_Assessment_2016,
  title = {An Assessment of Turbulence Models for Linear Hydrodynamic Stability Analysis of Strongly Swirling Jets},
  author = {Rukes, Lothar and Paschereit, Christian Oliver and Oberleithner, Kilian},
  year = 2016,
  month = sep,
  journal = {European Journal of Mechanics - B/Fluids},
  volume = {59},
  pages = {205--218},
  issn = {0997-7546},
  doi = {10.1016/j.euromechflu.2016.05.004},
  urldate = {2025-05-16}
}

@article{saldern_Role_2024,
  title = {On the Role of Eddy Viscosity in Resolvent Analysis of Turbulent Jets},
  author = {von Saldern, Jakob G. R. and Schmidt, Oliver T. and Jordan, Peter and Oberleithner, Kilian},
  year = 2024,
  month = dec,
  journal = {Journal of Fluid Mechanics},
  volume = {1000},
  pages = {A51},
  issn = {0022-1120, 1469-7645},
  doi = {10.1017/jfm.2024.922},
  urldate = {2025-04-24},
  langid = {english}
}

@article{sarras_Linear_2024,
  title = {Linear Stability Analysis of Turbulent Mean Flows Based on a Data-Consistent {{Reynolds-averaged Navier}}--{{Stokes}} Model: Prediction of Three-Dimensional Stall Cells around an Airfoil},
  shorttitle = {Linear Stability Analysis of Turbulent Mean Flows Based on a Data-Consistent {{Reynolds-averaged Navier}}--{{Stokes}} Model},
  author = {Sarras, K. and Tayeh, C. and Mons, V. and Marquet, O.},
  year = 2024,
  month = dec,
  journal = {Journal of Fluid Mechanics},
  volume = {1001},
  pages = {A41},
  issn = {0022-1120, 1469-7645},
  doi = {10.1017/jfm.2024.1009},
  urldate = {2025-05-16},
  langid = {english}
}

@phdthesis{sarwas_Experimental_2019,
  title = {Experimental Examination of New Separated Turbulent Flow Validation Test Geometry},
  author = {Sarwas, E. Sage},
  year = 2019
}

@inproceedings{savarino_Optimal_2024,
  title = {Optimal Transitional Mechanisms in Oblique Shock Wave-Boundary Layer Interaction Using Non-Linear Input/Output Analysis},
  booktitle = {{{AIAA SCITECH}} 2024 {{Forum}}},
  author = {Savarino, Flavio and Poulain, Arthur and Sipp, Denis and Rigas, Georgios},
  year = 2024,
  month = jan,
  publisher = {{American Institute of Aeronautics and Astronautics}},
  address = {Orlando, FL},
  doi = {10.2514/6.2024-2552},
  urldate = {2025-05-28},
  isbn = {978-1-62410-711-5},
  langid = {english}
}

@article{savarino_Optimal_2025,
  title = {Optimal Transitional Mechanisms of Incompressible Separated Shear Layers Subject to External Disturbances},
  author = {Savarino, Flavio and Sipp, Denis and Rigas, Georgios},
  year = 2025,
  month = aug,
  journal = {Journal of Fluid Mechanics},
  volume = {1016},
  pages = {A43},
  issn = {0022-1120, 1469-7645},
  doi = {10.1017/jfm.2025.10444},
  urldate = {2025-08-20},
  langid = {english}
}

@article{schmidt_Guide_2020,
  title = {Guide to {{Spectral Proper Orthogonal Decomposition}}},
  author = {Schmidt, Oliver T. and Colonius, Tim},
  year = 2020,
  month = mar,
  journal = {AIAA Journal},
  volume = {58},
  number = {3},
  pages = {1023--1033},
  publisher = {{American Institute of Aeronautics and Astronautics}},
  issn = {0001-1452},
  doi = {10.2514/1.J058809},
  urldate = {2025-04-24}
}

@article{simpson_Turbulent_1989,
  title = {Turbulent {{Boundary-Layer Separation}}},
  author = {Simpson, R L},
  year = 1989,
  month = jan,
  journal = {Annual Review of Fluid Mechanics},
  volume = {21},
  number = {1},
  pages = {205--232},
  issn = {0066-4189, 1545-4479},
  doi = {10.1146/annurev.fl.21.010189.001225},
  urldate = {2025-10-24},
  langid = {english}
}

@article{tammisola_Coherent_2016,
  title = {Coherent Structures in a Swirl Injector at {{Re}} = 4800 by Nonlinear Simulations and Linear Global Modes},
  author = {Tammisola, O. and Juniper, M. P.},
  year = 2016,
  month = apr,
  journal = {Journal of Fluid Mechanics},
  volume = {792},
  pages = {620--657},
  issn = {0022-1120, 1469-7645},
  doi = {10.1017/jfm.2016.86},
  urldate = {2025-04-24},
  langid = {english}
}

@article{tenaud_Wall_2016,
  title = {On Wall Pressure Fluctuations and Their Coupling with Vortex Dynamics in a Separated--Reattached Turbulent Flow over a Blunt Flat Plate},
  author = {Tenaud, C. and Podvin, B. and Fraigneau, Y. and Daru, V.},
  year = 2016,
  month = oct,
  journal = {International Journal of Heat and Fluid Flow},
  volume = {61},
  pages = {730--748},
  issn = {0142727X},
  doi = {10.1016/j.ijheatfluidflow.2016.08.002},
  urldate = {2025-07-29},
  langid = {english}
}

@article{towne_Spectral_2018,
  title = {Spectral Proper Orthogonal Decomposition and Its Relationship to Dynamic Mode Decomposition and Resolvent Analysis},
  author = {Towne, Aaron and Schmidt, Oliver T. and Colonius, Tim},
  year = 2018,
  month = jul,
  journal = {Journal of Fluid Mechanics},
  volume = {847},
  pages = {821--867},
  issn = {0022-1120, 1469-7645},
  doi = {10.1017/jfm.2018.283},
  urldate = {2025-01-12},
  copyright = {https://www.cambridge.org/core/terms},
  langid = {english}
}

@article{uzun_Direct_2025,
  title = {Direct Numerical Simulation of Flow Past a {{Gaussian}} Bump at a High {{Reynolds}} Number},
  author = {Uzun, Ali and Malik, Mujeeb R.},
  year = 2025,
  month = aug,
  journal = {Theoretical and Computational Fluid Dynamics},
  volume = {39},
  number = {4},
  pages = {30},
  issn = {0935-4964, 1432-2250},
  doi = {10.1007/s00162-025-00749-x},
  urldate = {2025-07-31},
  langid = {english}
}

@article{uzun_HighFidelity_2022,
  title = {High-{{Fidelity Simulation}} of {{Turbulent Flow Past Gaussian Bump}}},
  author = {Uzun, Ali and Malik, Mujeeb R.},
  year = 2022,
  month = apr,
  journal = {AIAA Journal},
  volume = {60},
  number = {4},
  pages = {2130--2149},
  issn = {0001-1452, 1533-385X},
  doi = {10.2514/1.J060760},
  urldate = {2024-08-30},
  langid = {english}
}

@article{uzun_Simulation_2021,
  title = {Simulation of a Turbulent Flow Subjected to Favorable and Adverse Pressure Gradients},
  author = {Uzun, Ali and Malik, Mujeeb R.},
  year = 2021,
  month = jun,
  journal = {Theoretical and Computational Fluid Dynamics},
  volume = {35},
  number = {3},
  pages = {293--329},
  issn = {0935-4964, 1432-2250},
  doi = {10.1007/s00162-020-00558-4},
  urldate = {2024-08-30},
  langid = {english}
}

@article{vonsaldern_Mean_2022,
  title = {Mean Flow Data Assimilation Based on Physics-Informed Neural Networks},
  author = {{von Saldern}, Jakob G. R. and Reumsch{\"u}ssel, Johann Moritz and Kaiser, Thomas L. and Sieber, Moritz and Oberleithner, Kilian},
  year = 2022,
  month = nov,
  journal = {Physics of Fluids},
  volume = {34},
  number = {11},
  pages = {115129},
  issn = {1070-6631},
  doi = {10.1063/5.0116218},
  urldate = {2025-04-24}
}

@article{wang_Unsteady_2022,
  title = {Unsteady Motions in the Turbulent Separation Bubble of a Two-Dimensional Wing},
  author = {Wang, Sen and Ghaemi, Sina},
  year = 2022,
  month = oct,
  journal = {Journal of Fluid Mechanics},
  volume = {948},
  pages = {A3},
  issn = {0022-1120, 1469-7645},
  doi = {10.1017/jfm.2022.603},
  urldate = {2025-07-28},
  langid = {english}
}

@article{weiss_Spectral_2022,
  title = {Spectral {{Proper Orthogonal Decomposition}} of {{Unsteady Wall Shear Stress Under}} a {{Turbulent Separation Bubble}}},
  author = {Weiss, Julien and Steinfurth, Ben and Chamard, L{\'e}o and Giani, Alain and Combette, Philippe},
  year = 2022,
  month = apr,
  journal = {AIAA Journal},
  volume = {60},
  number = {4},
  pages = {2150--2159},
  issn = {0001-1452, 1533-385X},
  doi = {10.2514/1.J061087},
  urldate = {2025-07-28},
  langid = {english}
}

@article{weiss_Unsteady_2015,
  title = {Unsteady {{Behavior}} of a {{Pressure-Induced Turbulent Separation Bubble}}},
  author = {Weiss, Julien and {Mohammed-Taifour}, Abdelouahab and Schwaab, Quentin},
  year = 2015,
  month = sep,
  journal = {AIAA Journal},
  volume = {53},
  number = {9},
  pages = {2634--2645},
  issn = {0001-1452, 1533-385X},
  doi = {10.2514/1.J053778},
  urldate = {2025-01-23},
  langid = {english}
}

@article{welch_Use_1967,
  title = {The Use of Fast {{Fourier}} Transform for the Estimation of Power Spectra: {{A}} Method Based on Time Averaging over Short, Modified Periodograms},
  shorttitle = {The Use of Fast {{Fourier}} Transform for the Estimation of Power Spectra},
  author = {Welch, P.},
  year = 1967,
  month = jun,
  journal = {IEEE Transactions on Audio and Electroacoustics},
  volume = {15},
  number = {2},
  pages = {70--73},
  issn = {0018-9278},
  doi = {10.1109/TAU.1967.1161901},
  urldate = {2025-01-02},
  copyright = {https://ieeexplore.ieee.org/Xplorehelp/downloads/license-information/IEEE.html},
  langid = {english}
}

@inproceedings{williams_Experimental_2020,
  title = {Experimental {{Study}} of a {{CFD Validation Test Case}} for {{Turbulent Separated Flows}}},
  booktitle = {{{AIAA Scitech}} 2020 {{Forum}}},
  author = {Williams, Owen and Samuell, Madeline and Sarwas, E. Sage and Robbins, Matthew and Ferrante, Antonino},
  year = 2020,
  month = jan,
  publisher = {{American Institute of Aeronautics and Astronautics}},
  address = {Orlando, FL},
  doi = {10.2514/6.2020-0092},
  urldate = {2024-12-19},
  isbn = {978-1-62410-595-1},
  langid = {english}
}

@article{winkelman_Flowfield_1980,
  title = {Flowfield {{Model}} for a {{Rectangular Planform Wing}} beyond {{Stall}}},
  author = {Winkelman, Allen E. and Barlow, Jewell B.},
  year = 1980,
  month = aug,
  journal = {AIAA Journal},
  volume = {18},
  number = {8},
  pages = {1006--1008},
  issn = {0001-1452, 1533-385X},
  doi = {10.2514/3.50846},
  urldate = {2025-07-29},
  langid = {english}
}

@article{wu_Spatiotemporal_2020,
  title = {Spatio-Temporal Dynamics of Turbulent Separation Bubbles},
  author = {Wu, Wen and Meneveau, Charles and Mittal, Rajat},
  year = 2020,
  month = jan,
  journal = {Journal of Fluid Mechanics},
  volume = {883},
  pages = {A45},
  issn = {0022-1120, 1469-7645},
  doi = {10.1017/jfm.2019.911},
  urldate = {2025-07-29},
  copyright = {https://www.cambridge.org/core/terms},
  langid = {english}
}

@article{zaman_Natural_1989,
  title = {A Natural Low-Frequency Oscillation of the Flow over an Airfoil near Stalling Conditions},
  author = {Zaman, K. B. M. Q. and Mckinzie, D. J. and Rumsey, C. L.},
  year = 1989,
  month = may,
  journal = {Journal of Fluid Mechanics},
  volume = {202},
  pages = {403--442},
  issn = {0022-1120, 1469-7645},
  doi = {10.1017/S0022112089001230},
  urldate = {2025-07-29},
  copyright = {https://www.cambridge.org/core/terms},
  langid = {english}
}

@article{zhou_Sensitivity_2024,
  title = {Sensitivity Analysis of Wall-Modeled Large-Eddy Simulation for Separated Turbulent Flow},
  author = {Zhou, Di and Bae, H. Jane},
  year = 2024,
  month = jun,
  journal = {Journal of Computational Physics},
  volume = {506},
  pages = {112948},
  issn = {00219991},
  doi = {10.1016/j.jcp.2024.112948},
  urldate = {2025-10-24},
  langid = {english}
}

@article{Bodony2006_sponge,
    title = {{Analysis of sponge zones for computational fluid mechanics}},
    year = {2006},
    journal = {J. Comput. Phys.},
    author = {Bodony, D},
    pages = {681--702},
    volume = {212},
    url = {https://consensus.app/papers/analysis-of-sponge-zones-for-computational-fluid-bodony/f9771a5ffc165799915a7e87f2ff709b/},
    doi = {10.1016/j.jcp.2005.07.014}
}

\end{document}